\documentclass[%
preprint,
amsmath,amssymb,
aps,
]{revtex4-1}

\usepackage{graphicx, bm}
\usepackage{amsmath}
\usepackage{mathrsfs}
\usepackage{amsmath}
\usepackage{amssymb}
\usepackage{bbold}
\usepackage[usenames]{color}
\usepackage{float}

\usepackage{subfigure}
\usepackage{graphicx}
\usepackage{subcaption}

\usepackage[unicode=true]{hyperref}   
\usepackage{slashed}

\usepackage[version=4]{mhchem}

\usepackage{graphicx,color}
\usepackage{amsmath,amssymb}
\usepackage{url}
\usepackage{epstopdf}

\newcommand{\hs}{\hspace*{0.5cm}}

\newcommand{\be}{\begin{equation}}
	\newcommand{\ee}{\end{equation}}
\newcommand{\bea}{\begin{eqnarray}}
	\newcommand{\eea}{\end{eqnarray}}
\newcommand{\bit}{\begin{itemize}}
	\newcommand{\eit}{\end{itemize}}

\newcommand{\al}{\alpha}

\newcommand{\ga}{\gamma}

\newcommand{\bc}{\begin{center}}
	\newcommand{\ec}{\end{center}}
\newcommand{\Ga}{\Gamma}

\newcommand{\La}{\Lambda}

\newcommand {\ba}{\begin{array}}
	\newcommand {\ea}{\end{array}}
\newcommand{\ben}{\begin{enumerate}}
	\newcommand{\een}{\end{enumerate}}

\newcommand{\IAST}{ Laboratory of Advanced Materials and Natural Resources, Institute for
	Advanced Study in Technology, Ton Duc Thang University, Ho Chi Minh City,
	Vietnam}
\newcommand{\DAS}{ Faculty of Applied Sciences, Ton Duc Thang University, Ho Chi Minh City,
	Vietnam}

\newcommand{\IOP}{Institute of Physics,  Vietnam Academy of Science and Technology, 10 Dao Tan, Ba
	Dinh, Hanoi 10000, Vietnam}

\newcommand{\SPtwo}{Department of Physics, Hanoi Padegogical 2, Xuan Hoa, Phu Tho Vietnam}

\newcommand{\HNNU}{VNU University of Science,334 Nguyen Trai Road, Thanh Xuan, Hanoi}

\newcommand{\AdrHEPC}{Phenikaa Institute for Advanced Study, Phenikaa University, Nguyen Trac, Duong Noi, Hanoi 12116, Vietnam}

\begin{document}
\markboth{Authors' Names}{Instructions for typing manuscripts (paper's title)}

%

%

	\title{ \boldmath   
	Lepton Flavor Physics in the flipped 3-3-1-1 Model: Non-Universality and Violation }

\author{ D.T.Huong \footnote{dthuong@iop.vast.vn}}
\affiliation{\IOP}

\author{ V. H. Binh \footnote{vhbinh@iop.vast.vn}}
\affiliation{\IOP}

\author{N.T.Huong \footnote{nguyenthuhuong@hus.edu.vn}}
\affiliation{\HNNU}

\author{H.T.Hung \footnote{hathanhhung@hpu2.edu.vn}}
\affiliation{\SPtwo}

\author{Duong Van Loi \footnote{loi.duongvan@phenikaa-uni.edu.vn}}
\affiliation{\AdrHEPC}

\author{D. T. Binh \footnote{dinhthanhbinh@tdtu.edu.vn}}
\affiliation{\small \IAST }\affiliation{\DAS}

\date{\today}


\begin{abstract}
	We investigate the flavor violation (FV) of Z decays to leptons at tree level and flavor conserving Z decays to leptons in the frame work of the flipped $	SU(3)_C\otimes SU(3)_L \otimes U(1)_X\otimes U(1)_N$,(F3311) model. In addition, we analyze the processes $l_i\rightarrow l_j \gamma$ and the leptonic three-body decay. Using  the experimental bounds on these decays we set the constraint on  $\sin \phi$ which represents the mixing between Z-Z' boson. The most stringent limits arises  from $\mu \rightarrow e \gamma$ decay where $\sin \phi \sim \mathcal{ O}(10^{-3})$. The leptonic three-body decay set lower bound on the mass of the new neural gauge boson $m_{Z'} \geq 3.2TeV$. Using the LUX-ZEPLIN (LZ) experiment data we set bounds to the mass of the dark matter candidates.
	Subsequently, we investigate  the lepton non-universality in B decays within the $F3311$ model  by calculating the generic one-loop contribution to the process $u_i\rightarrow d_j e_b \bar{\nu}_a$ in the unitary gauge as well as numerical evaluating the branching ratio $R_D, R_{D^{(*)}},R(X_c)$. We demonstrate that the $F3311$ model can address the $3.3 \sigma$ discrepancies between Standard Model and experimental data. To reaffirm our results, we also analyze the $d \to u$ transitions and  $s\to u$ transitions.  These two transitions also give consistent result with experiment data. Combine all experiment dat a we obtain the  operating region for the mass of the model  
	specifically $m_E \in [6.5, 9 ]TeV$, $m_Q \in [6,11]TeV$  and the dark matter candidate $m_\xi \in [1.5,2 ]TeV$.   
\end{abstract}

\maketitle


\section{Introduction}
\label{intro}

Although the Standard Model is a remarkable
achievement, it still leaves several fundamental questions unanswered, including the origin of neutrino masses \cite{RevModPhys.88.030501, RevModPhys.88.030502}, the existence of dark matter\cite{Bertone:2004pz,Arcadi:2017kky}, the matter-antimatter asymmetry of the universe \cite{Canetti:2012zc}, and the flavor puzzles. These unresolved puzzles strongly imply  that the SM is an effective theory and need to be extended  at higher energy scales. 
Many extended models of the SM have been proposed to address experimental data and predict new physics (NP) phenomenology.

To search for new particles and interactions, current experiments are implemented via direct and indirect searches.  While direct searches often involve investigations 
at high energy collider, indirect searches focus on detecting  quantum loop effects in precise observables such as 
the violation of (approximate) symmetries of the SM. These effects are highly  sensitive to NP especially in scenarios where the symmetries of the SM are not preserved. 
Consequently, even if the new physics's mass scale is quite high, these effects are still significant.   Recently, several measurements in flavor physics have shown deviations  from the SM  prediction, particularly
the anomalies associated to the charged current processes.
The discrepancies found in charged current involving $b \rightarrow c l^- \nu_l$ decays have been observed at  BARBAR \cite{BaBar:2012obs, BaBar:2013mob}, Belle \cite{Belle:2016dyj, Belle:2015qfa, Belle:2019rba} and LHCb\cite{ LHCb:2015gmp, LHCb:2017smo, LHCb:2017rln}.
In particular, measurements of the ratios of branching ratios (BRs) are
\begin{equation}
	\mathcal{R}_{D}\equiv  \frac{\mathcal{B}(B\rightarrow D\tau \nu_\tau)}{\mathcal{B}(B\rightarrow D l\nu_l)}, 
\end{equation}
and 
\begin{equation}
	\mathcal{R}_{D^*}\equiv  \frac{\mathcal{B}(B\rightarrow D^* \tau \nu_\tau)}{\mathcal{B}(B\rightarrow D^* l\nu_l)}, 
\end{equation}
with $l=e, \mu $.

The latest 
\begin{align}
	\mathcal{R}(D)_{exp} &=0.339 \pm 0.026 \pm 0.014, \\
	\mathcal{R}(D^*)_{exp} &=0.295 \pm 0.010 \pm 0.008, 
\end{align}
where the first and second uncertainty is the statistical and systematic error, respectively. These measurements are 
at level of $3.3 \sigma$ tension with SM predictions \cite{HFLAV:2022esi}
\begin{align}
	\mathcal{R}_{SM}(D) &=0.298 \pm 0.004 , \\
	\mathcal{R}_{SM}(D^*) &=0.254 \pm 0.005 , 
\end{align}
Even though these decays have imperfect cancellations of the form factor dependence since the mass of tau 
is quite big. However, these measurement are supported by $\mathcal{R}(J/\psi)$ \cite{LHCb:2017vlu} which is observed to be larger than SM prediction. 
These deviations  hint a violation of the universality of interactions of leptons. In SM, the gauge bosons interact with the same strength with all  lepton flavors. The lepton flavor universality (LFU) 
is only broken by the Higgs Yukawa couplings. However, these couplings are very small (1\% for the $\tau$ lepton), hence the LFU is an approximate symmetry of the SM. Therefore, the LFU violation (LFUV) 
represent sign of NP. 

Among the various proposed extensions to the SM, gauge extensions offer a
compelling framework for addressing some of these shortcomings. Models based on the $SU(3)_C \otimes SU(3)_L \otimes U(1)_X$ (3-3-1) gauge group are particularly intriguing. These models inherently incorporate three generations of fermions \cite{Singer:1980sw, Valle:1983dk, Pisano:1992bxx, Frampton:1992wt, PhysRevD.50.R34}, which naturally leads to explanations for the number of observed families and often provides solutions for neutrino masses through mechanisms like the seesaw \cite{Tully:2000kk,Chang:2006aa, Dong:2006mt, Dong:2008sw, Dong:2010gk, Dong:2010zu, Dong:2011vb, Boucenna:2014ela, Boucenna:2014dia, Dias:2005yh}. Furthermore, the stability of dark matter in some of these models is often interpreted via global or discrete symmetries \cite{Long:2003hht, Filippi:2005mt, Mizukoshi:2010ky, Ruiz-Alvarez:2012nvg, Dong:2013ioa, Profumo:2013sca}. Given typical fermion contents, charge quantization is a direct consequence of the 3-3-1 model \cite{Pisano:1996ht,Doff:1998we, deSousaPires:1998jc,deSousaPires:1999ca,Dong:2005ebq}.

The $[SU(3)_L]^3$  anomaly cancellation condition  \cite{Gross:1972pv,Georgi:1972bb,Banks:1976yg,Okubo:1977sc} requires the number of triplets equal to the number of anti-triplets. As a consequence, there are different arrangements of particle content of the models. Traditionally, quarks and leptons are arranged into triplets or anti-triplets. However, in recent works \cite{Fonseca:2016tbn,VanLoi:2020xcq}, a distinct arrangement is explored where one generation of leptons is placed in a sextet, while the other two are arranged in triplets. This setup is often referred to as the flipped 3-3-1 (F331) model.
With this type of arrangement, one lepton generation transforms differently from two others to result  
the different interactions of leptons with gauge bosons. As a result, the F331 incorporates lepton-flavor-violating interactions at tree-level. This feature directly impacts lepton number violating decay processes of charged leptons and is expected to give rise to observable effects related to lepton flavor non-universality \cite{Duy:2022qhy,Huong:2019vej, Dinh:2019jdg,Thu:2023xai}. Additionally, in the F331 model, there is a mixing between the Z and Z' bosons. Therefore, the model not only contains neutral currents that change lepton flavor associated with both Z and Z', but these interactions also directly predict lepton-flavor-violating decay processes of the Z boson, which can thus be very stringently constrained by current experimental limits \cite{Abu-Ajamieh:2025vxw}. Furthermore,   in the F331 model, the charge Q and the quantum number 
$B-L$  charge neither close nor commute with the $SU(3)_L$ \cite{Dong:2013wca}. To close the algebra, the  electroweak symmetry must be extended to $SU(3)_L \otimes U(1)_X \otimes U(1)_N$ where the Abelian charges X and N are related to $Q$ and $B-L$ through generators of $SU(3)_L$ respectively (often referred to as the F3311 model). This specific setup naturally reveals matter parity as a residual gauge symmetry \cite{Dong:2013wca,VanLoi:2020xcq}, which plays a vital role in stabilize symmetry various dark matter candidates within the model, alongside its implications for other related phenomena, as thoroughly investigated in prior studies  \cite{Dong:2013wca,Dong:2015yra,Dong:2014wsa,Alves:2016fqe}.

This paper focuses on investigating the implications of the F3311 model for both lepton number violating processes in the decays of the Z-boson and the observed $R_D/R_{D^*}$ anomalies. The Z-boson, being a fundamental electroweak mediator, provides a clean environment to search for rare decay modes that could be sensitive to new physics scales. Furthermore, the model's ability to explain the $R_D/R_{D^*}$ anomalies presents an additional avenue for its experimental verification. We specifically examine the rates for various Z-boson decay channels involving different lepton flavors, and delve into the F3311 model's contributions to the $R_D/R_{D^*}$ ratios, as discussed in detail in our previous work. Our analysis leverages the extended scalar and fermion content of the F3311 model, which introduces new interactions mediating these exotic decays and lepton flavor non-universality. By calculating the branching ratios of these processes and evaluating the theoretical predictions for $R_D/R_{D^*}$, we aim to constrain the parameter space of the F3311 model using current experimental limits.

The remainder of this paper is organized as follows. In Section \ref{model}, we provide a concise review of the F3311 model, highlighting its particle content and relevant interaction Lagrangian. Section \ref{LFPheno} details the consideration of Z-boson decay rates involving lepton flavor and lepton number violation. The flavor non-universality of the effective Hamiltonian for $u_i-d_j$ transitions is derived in Section \ref{Eff}. We then proceed with a numerical study of lepton flavor non-universality processes in Section \ref{nonuni2}. Finally, Section \ref{Conclu} summarizes our findings and offers concluding remarks.

\section{F3311  model \label{model}}

In this section, we present a concise overview of the F3311 model which forms the theoretical basis of our analysis. We begin by detailing the model's particle content, followed by a discussion of its relevant interaction Lagrangian \cite{VanLoi:2020xcq}.

\subsection{Gauge Group and Charge Relations}
The F3311 model extends the SM gauge symmetry by incorporating an additional $U(1)_N$ factor, resulting in the full gauge group:
\begin{equation}
SU(3)_C\otimes SU(3)_L \otimes U(1)_X\otimes U(1)_N\,. \label{eq:gauge_group}
\end{equation}
This extended symmetry naturally leads to novel charge assignments and particle content. The electric charge operator $\mathcal{Q}$ and the baryon minus lepton number $B-L$ are fundamentally related to the diagonal generators of the $SU(3)_L$ group, $T_3$ and $T_8$, and the $U(1)_X$ and $U(1)_N$ charges as follows:
\begin{align}
\mathcal{Q}&=T_3+\frac{1}{\sqrt{3}}T_8 +X, \hs B-L=\frac{2}{\sqrt{3}}T_8 +N\,. \label{eq:charge_relations}
\end{align}
These relations are crucial for understanding the embedding of SM fields and the properties of new exotic particles within this framework.
\subsection{Particle Content of the Model}
The particle spectrum of the F3311 model is designed to ensure anomaly cancellation and to naturally accommodate three fermion generations, along with new exotic states. The unique "flipped" arrangement, as introduced in Ref. \cite{VanLoi:2020xcq}, dictates that one generation of leptons transforms as a sextet under $SU(3)_L$, while the remaining two transform as anti-triplets. Specifically, the left-handed fermion content is arranged as:
\begin{align}
\psi_{1L}&=
\begin{pmatrix}
	\xi_L^+ &\frac{1}{\sqrt{2}}\xi_L^0 &\frac{1}{\sqrt{2}} \nu_L \\
	\frac{1}{\sqrt{2}}\xi_L^0  & \xi_L^- & \frac{1}{\sqrt{2}}e_{1L} \\
	\frac{1}{\sqrt{2}}\nu_{1L}  & \frac{1}{\sqrt{2}}e_{1L}  & E_{1L}  \\
\end{pmatrix}\sim (1,6,-1/3,-2/3)\,, \label{eq:lepton_sextet} \\
\psi_{\alpha L}&=
\begin{pmatrix}
	\nu_{\alpha L} & e_{\alpha L} & E_{\al L}
\end{pmatrix}^T \sim  (1,3,-2/3,-4/3)\,, \label{eq:lepton_triplet}\\
Q_{a L}&=
\begin{pmatrix}
	d_{a L } & -u_{aL} & U_{aL}
\end{pmatrix}^T \sim (3,3^*,1/3,2/3)\,, \label{eq:quark_triplet}
\end{align}
where $\alpha=2,3$ denotes the generation indices for the lepton anti-triplets. The right-handed fermion fields  are given by:
\begin{align}
\nu_{aR}&\sim (1,1,0,-1), \hs e_{aR} \sim (1,1,-1,-1), \hs E_{aR}\sim (1,1,-1,-2)\,, \label{eq:right_leptons}\\
u_{aR}&\sim(3,1,2/3,1/3), \hs d_{aR} \sim (3,1,-1/3,1/3), \hs U_{aR}\sim (3,1,2/3,4/3)\,, \label{eq:right_quarks}
\end{align}
with $a=1,2,3$ as the full generation index for all right-handed fermions. This particular arrangement of fermion fields is key to the model's unique phenomenology, especially regarding lepton flavor violation and lepton flavor non-universality. 

\subsection{Scalar Sector and Electroweak Symmetry Breaking}
The scalar fields, along with their corresponding representations and VEVs are defined as follows:
\begin{align}
\eta&=
\begin{pmatrix}
	\eta_1^0 & \eta_2^- & \eta_3^-
\end{pmatrix}^T \sim(1,3,-2/3,-1/3)\,, \label{eq:eta_scalar}\\
\rho&=
\begin{pmatrix}
	\rho_1^+ & \rho_2^0 & \rho_3^0
\end{pmatrix}^T\sim (1,3,1/3,-1/3)\,, \label{eq:rho_scalar} \\
\chi&= 
\begin{pmatrix}
	\chi_1^+ & \chi_2^0 & \chi_3^0
\end{pmatrix}^T \sim (1,3,1/3,2/3)\,, \label{eq:chi_scalar} \\
S&= 
\begin{pmatrix}
	S_{11}^{++} & \frac{1}{\sqrt{2}}S_{12}^+ & \frac{1}{\sqrt{2}}S_{13}^+ \\
	\frac{1}{\sqrt{2}}S_{12}^+ & S_{22}^0 & \frac{1}{\sqrt{2}}S_{23}^0 \\
	\frac{1}{\sqrt{2}}S_{13}^+ & \frac{1}{\sqrt{2}}S_{23}^0 &S_{33}^0 \\	
\end{pmatrix} \sim (1,6,2/3,4/3)\,, \label{eq:S_scalar}\\
\Phi&\sim (1,1,0,2)\,, \label{eq:Phi_scalar}
\end{align}
These scalar fields acquire non-zero vacuum expectation values (VEVs), which are crucial for electroweak symmetry breaking and mass generation:
\begin{eqnarray}
\langle\eta \rangle	&&=\begin{pmatrix}
	\frac{u}{\sqrt{2}}\\
	0\\
	0 \\
\end{pmatrix}, 
\hs 
\langle\rho \rangle=\begin{pmatrix}
	0\\
	\frac{v}{\sqrt{2}}\\
	0 \\
\end{pmatrix},  
\hs \langle\chi \rangle=\begin{pmatrix}
	0\\
	0\\
	\frac{\omega}{\sqrt{2}} \\
\end{pmatrix}\,,\hs
\langle S \rangle=\begin{pmatrix}
	0&0&0\\
	0&\frac{\kappa}{\sqrt{2}}&0\\
	0
	&0
	&\frac{\Lambda}{\sqrt{2}}\\
\end{pmatrix},   \nonumber \\
\langle\Phi \rangle &&=\frac{\Delta}{\sqrt{2}}\,. \label{eq:VEVs}
\end{eqnarray}
The initial breaking from $SU(3)_C \otimes SU(3)_L \otimes U(1)_X \otimes U(1)_N$ to the $SU(3)_C \otimes SU(2)_L \otimes U(1)_Y \otimes W_P$ symmetry is triggered by the higher-scale VEVs, $\omega$ and $\Delta$. Here, $W_P$ represents a residual discrete symmetry defined as $W_P=(-1)^{3(B-L)+2s}$, where $B-L=\frac{2}{\sqrt{3}}T_8+N$. Subsequently, the electroweak symmetry of the SM is further broken to $U(1)_{EM}$ by the VEVs $u, v,$ and $\kappa$.

The explicit hierarchy of these VEVs ($\kappa \ll u, v \ll \omega$ and $\Lambda \ll \Delta$) not only ensures the correct breaking pattern but also dictates the mass scales of the new gauge bosons and exotic fermions, directly influencing the phenomenological predictions of the model.

\subsection{Gauge sector}
Building upon the scalar sector, the masses of the gauge bosons in the F3311 model fundamentally arise from the kinetic terms of the scalar multiplets, which acquire non-zero VEVs. Specifically, these masses are generated when the scalar fields are expanded around their VEVs within the covariant derivative term,
\bea
\mathcal{L}  \owns \sum_{\phi= \chi, \rho, \eta, S, \varphi} \left( D_\mu <\phi>\right)^\dag \left( D^\mu <\phi>\right).
\eea
The non-Hermitian gauge boson states and their corresponding masses are presented in the following format:
\begin{itemize}
\item $W^\pm_\mu = \frac{1}{\sqrt{2}}(A_{1\mu} \mp iA_{2\mu})$ with $m_W^2 \approx \frac{g^2}{4}(u^2+v^2)$
\item $X^\pm_\mu = \frac{1}{\sqrt{2}}(A_{4\mu} \mp iA_{5\mu})$ with $m_X^2 = \frac{g^2}{4}(u^2+\omega^2 +2\Lambda^2)$
\item $Y^{0,0*}_\mu = \frac{1}{\sqrt{2}}(A_{6\mu} \mp iA_{7\mu})$ with $m_Y^2 \approx \frac{g^2}{4}(u^2+\omega^2 +2\Lambda^2)$
\end{itemize}
Assumming that $\Delta \gg \La,w \gg u,v$, we integrate out the boson $B_N$, three remaining neutral gauge bosons $(A_3, A_8, B)$ mix, yielding:
\begin{itemize}
\item The massless photon, denoted as $A$, is defined as:

\begin{equation}
	A = s_W A_3 + c_W \left( \frac{t_W}{\sqrt{3}} A_8 + \sqrt{1 - \frac{t^2_W}{3}} B \right)
\end{equation}

where $s_W = \sin \theta_W = \frac{\sqrt{3}t_W}{\sqrt{3+4t_W^2}}$, $c_W = \cos \theta_W$, and $t_W = \frac{g_X}{g}$.

\item Orthogonal fields: $Z, Z'$
\begin{align}
	Z &= c_W A_3 - s_W \left( \frac{t_W}{\sqrt{3}} A_8 + \sqrt{1 - \frac{t^2_W}{3}} B \right) \\
	Z' &= \sqrt{1 - \frac{t^2_W}{3}} A_8 - \frac{t_W}{\sqrt{3}} B
\end{align}
\item $Z$-$Z'$ mixing produces mass eigenstates:
\begin{align}
	Z_1 &= c_\phi Z - s_\phi Z'  \label{Zengenstate1}\\
	Z_2 &= s_\phi Z + c_\phi Z'   \label{Zengenstate2}
\end{align}
with $c_\phi = \cos \phi,$  $s_\phi =\sin  \phi$, and  mixing angle $\phi$ is defined as
\begin{align}
	\tan(2\phi)=	t_{2\phi} \approx \frac{\sqrt{3-t_W^2}}{2c_W^3} \frac{u^2-v^2 c_{2W}}{\omega^2+4\Lambda^2}\,.
	\label{tan2phi}
\end{align}
\item Mass squared: \begin{align}
	m_{Z_1}^2 \approx \frac{m_W^2}{c_W^2}, \hs  m_{Z_2}^2 \approx \frac{g^2}{(3-t_W^2)}(\omega^2+4\Lambda^2)\,.
\end{align}
	\end{itemize}
	The physical and original neutral gauge bosons are related by:
	$$
	(A_{3}, A_{8}, B)^T = U (A, Z_{1}, Z_{2})^T
	$$
	where the mixing matrix $U$ is given by:
	
	\begin{align*}
U &= \begin{pmatrix}
	s_W & c_W c_\phi & c_W s_\phi \\
	\frac{1}{\sqrt{3}}s_W & -\frac{1}{\sqrt{3}}s_W t_W c_\phi - \sqrt{1-\frac{1}{3}t_W^2} s_\phi & -\frac{1}{\sqrt{3}}s_W t_W s_\phi + \sqrt{1-\frac{1}{3}t_W^2} c_\phi \\
	\sqrt{1-\frac{4}{3}s_W^2} & \frac{t_W}{\sqrt{3}}(s_\phi + \sqrt{3-4 s_W^2} c_\phi) & \frac{t_W}{\sqrt{3}}(c_\phi - \sqrt{3-4 s_W^2} s_\phi).
\end{pmatrix}
\end{align*}
We identify $Z_1$  as the SM Z boson, and $Z_2$ as a new gauge boson.
\subsection{Fermion masses}
Having established the particle content and the mechanism of symmetry breaking, we now present the full Lagrangian of the F3311 model, which governs the dynamics and interactions of all fields. The total Lagrangian for the model can be broadly categorized into kinetic, Yukawa, and potential terms:
The total Lagrangian for the model is
\begin{equation}
\mathcal{L}=\mathcal{L}_{kinetic} + \mathcal{L}_{Yukawa} -V\,.
\end{equation}
For the purpose of this work, our primary focus lies on the Yukawa interactions. The relevant Yukawa Lagrangian terms, separated for quarks and leptons, are given as follows:
\begin{equation}
\mathcal {L}^{Quark}_{Yukawa}=h^u_{ab} \bar{Q}_{aL}\rho^*u_{bR}+h^d_{ab}\bar{Q}_{aL}\eta^* d_{bR} + h^U_{ab}\bar{Q}_{aL}\chi^*U_{bR}  \label{LYquark} \,,
\end{equation}
\begin{align}
\mathcal{L}^{lepton}_{Yukawa}&=h^e_{\alpha a} \overline{\psi}_{\al L} \rho e_{aR} 
+  h^E_{\alpha a} \overline{\psi}_{\al L} \chi E_{aR}
+  h^E_{1 e} \overline{\psi}_{1 L} S E_{aR}
+  h^E_{\xi} \overline{\psi}_{1 L}^c \psi_{1L} S \nonumber \\
&+   h^\nu_{\alpha b} \overline{\psi}_{\al L} \eta \nu_{b R} 
+  h^R_{a b} \overline{\nu}_{\al R}^c \nu_{bR} \Phi + H.c	\,.
\label{LYlept}
\end{align}
From the Yukawa Lagrangian of quarks in Eq.~(\ref{LYquark}), the quark 
mass mixing is given as followings
\begin{align}
[m_u]_{ab}&= \frac{h^u_{ab}}{\sqrt{2}}v, \hs  [m_d]_{ab}= -\frac{h^d_{ab}}{\sqrt{2}}u, \hs [m_U]_{ab}= -\frac{h^U_{ab}}{\sqrt{2}}\omega \,. 
\end{align}
and the exotic  lepton mass mixing matrices 
are obtained from (\ref{LYlept}), which have a following form:
\begin{equation}
[m_E]_{1 b} = -\frac{h^E_{1b}}{\sqrt{2}}\Lambda,  \hs  [m_E]_{\alpha b}=-\frac{h^E_{\alpha b}}{\sqrt{2}}\omega, \hs \alpha=2,3; b=1,2,3\,,
\end{equation}
\begin{align}
m_\xi^0 &=-\sqrt{2}h_\xi^E\Lambda, \hs m_\xi= -\sqrt{2}h_\xi^E\Lambda, 
\end{align}
where $\xi $ and  $E_a$ are mixing with $[m_{\xi E_a}]=-\frac{h^E_{1a}\kappa}{\sqrt{2}} $. However this mixing is small and can be neglected since $\kappa<< \omega, \Lambda$. The SM  leptons gain masses
\begin{align}
[m_e]_{ab}&= -\left(
\begin{array}{ccc}
	0 & 0 & 0 \\
	\frac{v h^e{}_{21}}{\sqrt{2}} & \frac{v h^e{}_{22}}{\sqrt{2}} & \frac{v h^e{}_{23}}{\sqrt{2}} \\
	\frac{v h^e{}_{31}}{\sqrt{2}} & \frac{v h^e{}_{32}}{\sqrt{2}} & \frac{vh^e{}_{33}}{\sqrt{2}} \\
\end{array}
\right)\,,	\hs \text{or} \hs [m_e]_{\alpha b}= -\frac{ h^e_{\alpha b}}{\sqrt{2}}v\,,\\
[m_\nu]_{11}&= -\sqrt{2}h_\xi^E\kappa  \,.
\end{align}
The electron mass can be  radiative induced  as in original study \cite{Fonseca:2016tbn}.

The  mass eigenstate and the gauge state are related by
\begin{align}
u'_{L,R}&=\left(u_1',u_2',u_3' \right)^T_{L,R} =V^u_{L,R}\left(u_1,u_2,u_3 \right)^T_{L,R} =V^u_{L,R}u_{L,R} \,,\\
d'_{L,R}&=\left(d_1',d_2',d_3' \right)^T_{L,R} =V^d_{L,R}\left(d_1,d_2,d_3 \right)^T_{L,R} =V^d_{L,R}d_{L,R}\,, \\
U'_{L,R}&=\left(U_1',U_2',U_3' \right)^T_{L,R} =V^U_{L,R}\left(U_1,U_2,U_3 \right)^T_{L,R} = V^U_{L,R}U_{L,R}\,,\\
e'_{L,R}&=\left(e_1',e_2',e_3' \right)^T_{L,R} =V^e_{L,R}\left(e_1,e_2,e_3 \right)^T_{L,R} =V^e_{L,R}e_{L,R}\,, \\
\nu'_{L,R}&=\left(\nu_1',\nu_2',\nu_3' \right)^T_{L,R} =V^\nu_{L,R}\left(\nu_1,\nu_2,\nu_3 \right)^T_{L,R} =V^\nu_{L,R}\nu_{L,R}\,, \\	
E'_{L,R}&=\left(E_1',E_2',E_3' \right)^T_{L,R} =V^E_{L,R}\left(E_1,E_2,E_3 \right)^T_{L,R} =V^E_{L,R}E_{L,R} \,.
\end{align}
A crucial consequence of $W_P$ parity conservation in this model is the absence of mixing mass terms between SM fermions and new exotic fermions. This stands in contrast to the F331 model \cite{Duy:2022qhy, Huong:2019vej, Dinh:2019jdg, Thu:2023xai}.
\subsection{Charged and neutral currents}
\subsubsection{Charged currents}
The Lagrangian for the charged currents are given as \cite{VanLoi:2020xcq}
\begin{align}
\mathcal{L}_f^C=J^{-\mu}_W W^+_\mu + J^{-\mu}_X X^+_\mu + J^{0*\mu}_Y Y^0_\mu  + H.c\,, 
\end{align}
where the charged currents are given as
\begin{align}
J^{-\mu}_W &= -\frac{g}{\sqrt{2}}\left( \overline{\nu}_{aL}\gamma^\mu e_{aL}  + \overline{u}_{aL}\gamma^\mu d_{aL} \right) \label{Wcurrent}\,, \\
J^{-\mu}_X &= -\frac{g}{\sqrt{2}} \left( \sqrt{2} \overline{\nu}_{1L}\gamma^\mu E_{1L} +\overline{\nu}_{\alpha L} \gamma^\mu E_{\alpha L} -\overline{U}_{aL} \gamma^\mu d_{aL} + \sqrt{2} \overline{\xi}^+_{L} \gamma^\mu \nu_{1L} + \overline{\xi^0}_L \gamma^\mu e_{1L}   \right) \,, \label{Xcurrent} \\
J^{0*\mu}_Y &= -\frac{g}{\sqrt{2}} \left(  \sqrt{2} \overline{e}_{1L}\gamma^\mu E_{1L} + \overline{e}_{\alpha L}\gamma^\mu E_{\alpha L} + \overline{U}_{aL}\gamma^\mu u_{aL} + \sqrt{2}\overline{\xi}^-_L \gamma^\mu e_{1L} +\overline{\xi^0}_L\gamma^\mu \nu_{1L} \label{Ycurrent} \right)\,,
\end{align}
In these mass eigenstates,
the Lagrangian describing the interaction has form
\begin{align}
\mathcal{L}^{CC} & \supset -\frac{g}{\sqrt{2}} \bigg[   \left( U_{PMNS} \right)_{ab} W'_\mu \overline{\nu^{'a}}_{aL}\gamma^\mu e^{'b}_{aL}  + \left(V_{CKM}\right)_{ij} W'_\mu \overline{u^{'i}}_{aL}\gamma^\mu d^{'j}_{aL} \bigg]  \nonumber \\
& -\frac{g}{\sqrt{2}} \bigg[    \bigg( V^\nu_L (V^E_L)^\dagger \bigg)_{ac}  X'_\mu \overline{\nu^{'a}}_{L}\gamma^\mu E^{'c}_{L}  - \bigg( V^U_L (V^d_L)^\dagger \bigg)_{ij}  X'_\mu \overline{U^{'i}}_{L}\gamma^\mu d^{'j}_{L}  \nonumber \\
&+ (V^\nu_L)_{1a} X'^\mu \overline{\xi'}^+_{L} \gamma^\mu \nu^{'a}_{L} + (V^e_L)_{1a} X'^\mu \overline{\xi^{0'}}_L \gamma^\mu e^{'a}_{L} \bigg]  \nonumber \\
&  -\frac{g}{\sqrt{2}} \bigg[    \bigg( V^e_L (V^E_L)^\dagger \bigg)_{bc}  Y^{0'}_\mu \overline{e^{'b}}_{L}\gamma^\mu E^{'c}_{L}  + \bigg( V^U_L (V^u_L)^\dagger \bigg)_{ij}  Y^{0'}_\mu \overline{U^{'i}}_{L}\gamma^\mu u^{'j}_{L} \bigg] \nonumber \\
&+ (V^e_L)_{1a} Y^{0'}_\mu \overline{\xi'}^-_{L} \gamma^\mu e^{'a}_{L} + (V^\nu_L)_{1a} Y^{0'}_\mu \overline{\xi^{0'}}_L \gamma^\mu \nu^{'a}_{L} \bigg]  + H.c \,,
\end{align}

where  $V_{CKM}=V^u_L (V^d_L)^\dagger$ and $U_{PMNS}=V^\nu_L (V^l_L)^\dagger$.
Because of the conservation of $W_P$ parity, it forbids mixing terms between the $W^\pm$ bosons and new charged gauge bosons ($X^\pm$). This property ensures the absence of anomalous flavor-changing couplings  associated with charged currents ($W^\pm$) at the tree level, which significantly differentiates this work from previous studies \cite{Thu:2023xai}. 
\subsubsection{Neutral Currents \label{neutral01}}
This model exhibits a distinctive feature: tree-level Flavor Changing Neutral Currents (FCNCs) exclusively in the lepton sector. This arises as a direct consequence of the unique transformation properties of the first lepton generation under the $SU(3)_L$ gauge group, which fundamentally differs from the behavior of the other two generations. Crucially, quark flavors remain conserved at tree-level within this framework, providing a clear distinction from other extended models.

The nature of these neutral currents, particularly their dependence on the Cartan charges, is central to understanding the model's phenomenology. The interactions are defined by the kinetic term of the fermions, expanded around the gauge fields. Specifically, the coupling of fermions to the neutral gauge bosons can be extracted from the covariant derivative term in the Lagrangian:
\bea
\mathcal{L} \supset \bar{F}i \ga^\mu D_\mu F 	\supset -g \bar{F}\ga^\mu \left\{ T_3A_{3\mu}+T_8 A_{8\mu}+t_X \left(Q-T_3-\frac{T_8}{\sqrt{3}}\right)B_\mu\right\}F,
\eea
where $F$ denotes the various fermion multiplets within the model.

Regarding the properties of these neutral currents, it is important to emphasize that the model inherently ensures flavor conservation for quarks and right-handed leptons at tree-level, as previously stated. In contrast, the distinct lepton flavor-changing neutral currents originate specifically from interactions involving the $T_8$ generator. These FCNCs are predominantly mediated by the mixing of the $Z_1$ (SM-like) and $Z_2$ (new exotic) neutral gauge bosons. Building on these characteristics, the relevant Lagrangian terms that give rise to these specific lepton flavor-changing neutral current interactions are expressed as:
\begin{equation}
\mathcal{L} \supset \left(g_L^{Z_1}\right)_{ij} \bar{l}'_{iL}\gamma^\mu l'_{jL}Z^\mu_1 + \left(g_L^{Z_2}\right)_{ij} \bar{l}'_{iL}\gamma^\mu l'_{jL}Z^\mu_2, \label{eq:LFCNC_Lagrangian}
\end{equation}
where the coupling strengths $\left(g_L^{Z_1}\right)_{ij}$ and $\left(g_L^{Z_2}\right)_{ij}$ are explicitly given by:
\begin{equation}
\left(g_L^{Z_1}\right)_{ij}  = s_\phi \frac{g}{\sqrt{3-t^2_W}}(V^*_{lL})_{1i}(V_{lL})_{1j}, \quad  \left(g_L^{Z_2}\right)_{ij}  = c_\phi \frac{g}{\sqrt{3-t^2_W}}(V^*_{lL})_{1i}(V_{lL})_{1j}.
\label{gLZ1Z2}
\end{equation}
The matrix $V_{lL}$ is a $3 \times 3$ unitary charged lepton mixing matrix. While its individual entries are undetermined, its product $V_{\nu_L}^\dag V_{eL}$ is constrained by neutrino oscillation data. This matrix plays a crucial role in characterizing the flavor-changing interactions. Analogous to the Cabibbo-Kobayashi-Maskawa (CKM) matrix in the quark sector, $V_{lL}$ can be parameterized using three mixing angles $\theta_{ij}^l$ and a single CP-violating phase $\delta^l$ as:
\begin{equation}
V_{lL}=\left(
\begin{array}{ccc}
	c_{12} c_{13} & c_{13} s_{12} & s_{13} e^{-i\delta^l} \\
	-c_{12} s_{13} s_{23} e^{i\delta^l}-c_{23} s_{12} & c_{12} c_{23}-s_{12} s_{13} s_{23} e^{i\delta^l} & c_{13} s_{23} \\
	s_{12} s_{23}-c_{12} c_{23} s_{13} e^{i\delta^l} & -e^{i\delta^l}c_{23} s_{12} s_{13} -c_{12} s_{23} & c_{13} c_{23} \\
\end{array}
\right),
\label{VlL}
\end{equation}
where $c_{ij} = \cos \theta_{ij}^l, s_{ij}=\sin \theta_{ij}l^l$.
In contrast to the lepton sector, the uniform transformation properties of the quark generations under the $SU(3)_L$ gauge group ensure that flavor changes are absent in $Z'$ interactions at tree-level. This leads to purely flavor-conserving neutral current interactions involving the neutral gauge bosons, which are expressed as:
\begin{align}
\mathcal{L}_f^N = & -eQ(f)\bar{f} \gamma^\mu f A_\mu  \nonumber \\
&- \frac{g}{2c_W} \left\{ \bar{f}\gamma^\mu \left[ g_V^{Z_1}(f) - g_A^{Z_1}(f) \gamma_5 \right] f Z_{1\mu} + \bar{f}\gamma^\mu \left[ g_V^{Z_2}(f) - g_A^{Z_2}(f) \gamma_5 \right] f Z_{2\mu} \right\}
\end{align}
where $e = gs_W$. The couplings $g_A^{Z_{1,2}}$ and $g_V^{Z_{1,2}}$ are detailed in Table \ref{coupling1}.
Having established the full Lagrangian and the mass spectrum of the gauge bosons, we can now derive the effective Lagrangian relevant for specific low-energy processes  \cite{Dinh:2019jdg}.

\section{Lepton FCNC Phenomenology \label{LFPheno}}
\subsection{Constraints from  SM Z boson decay data}

The model features $Z-Z'$ mixing, controlled by a small mixing angle $\phi $. This leads to additional contributions to the SM  boson decays, including both flavor-conserving and flavor-violating modes.

\subsubsection{Flavor-Violating Decays} 
The model predicts the FCNCs in the lepton sector at the tree level. Experimental limits on $Z$ boson decays with lepton flavor violation (e.g., $Z \to e\mu$, $Z \to e\tau$, $Z \to \mu\tau$) provide strong constraints on the mixing parameters and the mass of the SM $Z$ boson. The branching ratios for the flavor-violating $Z\to f_af_b$ decays is \cite{Abu-Ajamieh:2025vxw}

\begin{equation}
\Gamma(Z_1 \to l_i^+ l_j^-) = \frac{M_{Z_1}}{12 \pi } \left( |(g_L^{Z_1})_{ij}|^2 + |(g_R^{Z_1})_{ij}|^2 \right)
\end{equation}

\begin{figure}[ht!]
\centering
\includegraphics[width=0.45\linewidth]{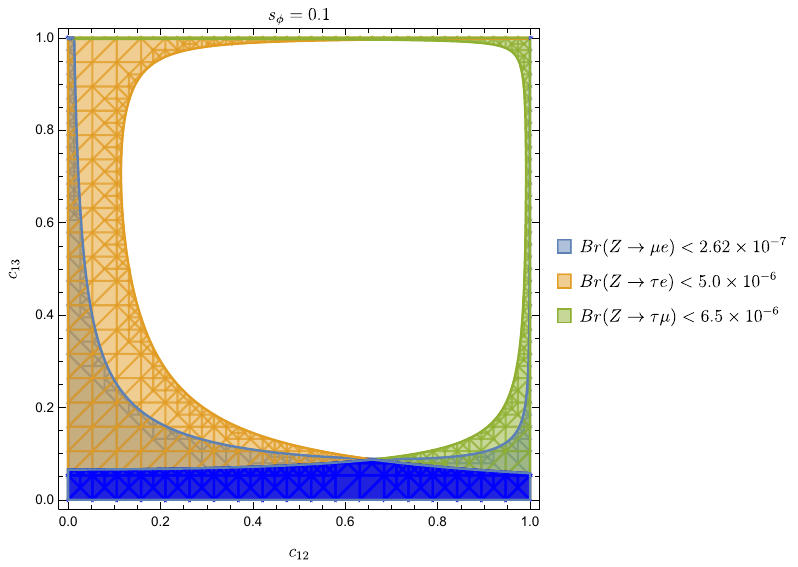} 
\includegraphics[width=0.45\linewidth]{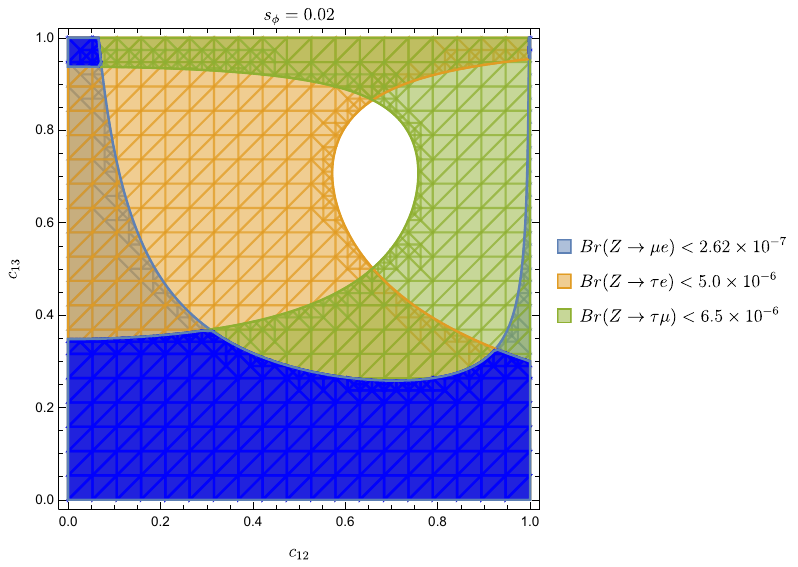} 
\caption{Allowed region of $Br(Z\rightarrow l_i l_j)$. Blue region is the combined region restricted by experimental values.}
\label{ZliljFig} 
\end{figure}

The experimental upper limits on lepton flavor violating decays of the SM $Z$ boson, as reported in \cite{ParticleDataGroup:2022pth}, are:

\begin{align}
\text{Br}(Z \to e^\pm \mu^\mp) &< 2.62 \times 10^{-7}, 		\label{ExpBrZlilj1} \\
\text{Br}(Z \to e^\pm \tau^\mp) &< 5 \times 10^{-6}, 		\label{ExpBrZlilj2} \\
\text{Br}(Z \to \tau^\pm \mu^\mp) &< 6.5 \times 10^{-6}.
\label{ExpBrZlilj3}
\end{align}
Clarification on how these $Z$ boson decay are used to restrict the model's parameter space and is essential for a comprehensive analysis. The strength of the Z-Z' mixing depends on the mixing angle $\sin \phi$ (\ref{gLZ1Z2}). In the limit where the new scale approaches the electroweak scale (EW), the order of $\tan 2\phi$ (\ref{tan2phi}) is of order $10^{-1}$ hence $\sin \phi$ is of order $10^{-1}$. The branching of decay $Z_1\rightarrow l_i l_j$ will be proportional to the mixing between Z-Z' and the lepton mixing parameters as in (\ref{VlL}). In FIG.(\ref{ZliljFig}) we investigate the $Br(Z_1\rightarrow l_i l_j)$ for the $F 3311$ model. The allowed regions for lepton mixing are restricted by combined  the experimental value (\ref{ExpBrZlilj1}),(\ref{ExpBrZlilj2}),(\ref{ExpBrZlilj3}). We investigate for different value of mixing angle $\sin \phi$. With maximal value of $\sin \phi=0.1$ the value of $c_{12}$ is in the range $c_{12}\in [0,1]$ while the range for the $c_{13}$ is $c_{13}\in [0,0.1]$. In case $\sin \phi =0.02$ then $c_{12}\in [0,1]$ and $c_{13}\in [0, 0.4]$

\subsubsection{Flavor-Conserving Decays}
\begin{figure}[ht!]
\centering
\includegraphics[width=0.25\linewidth]{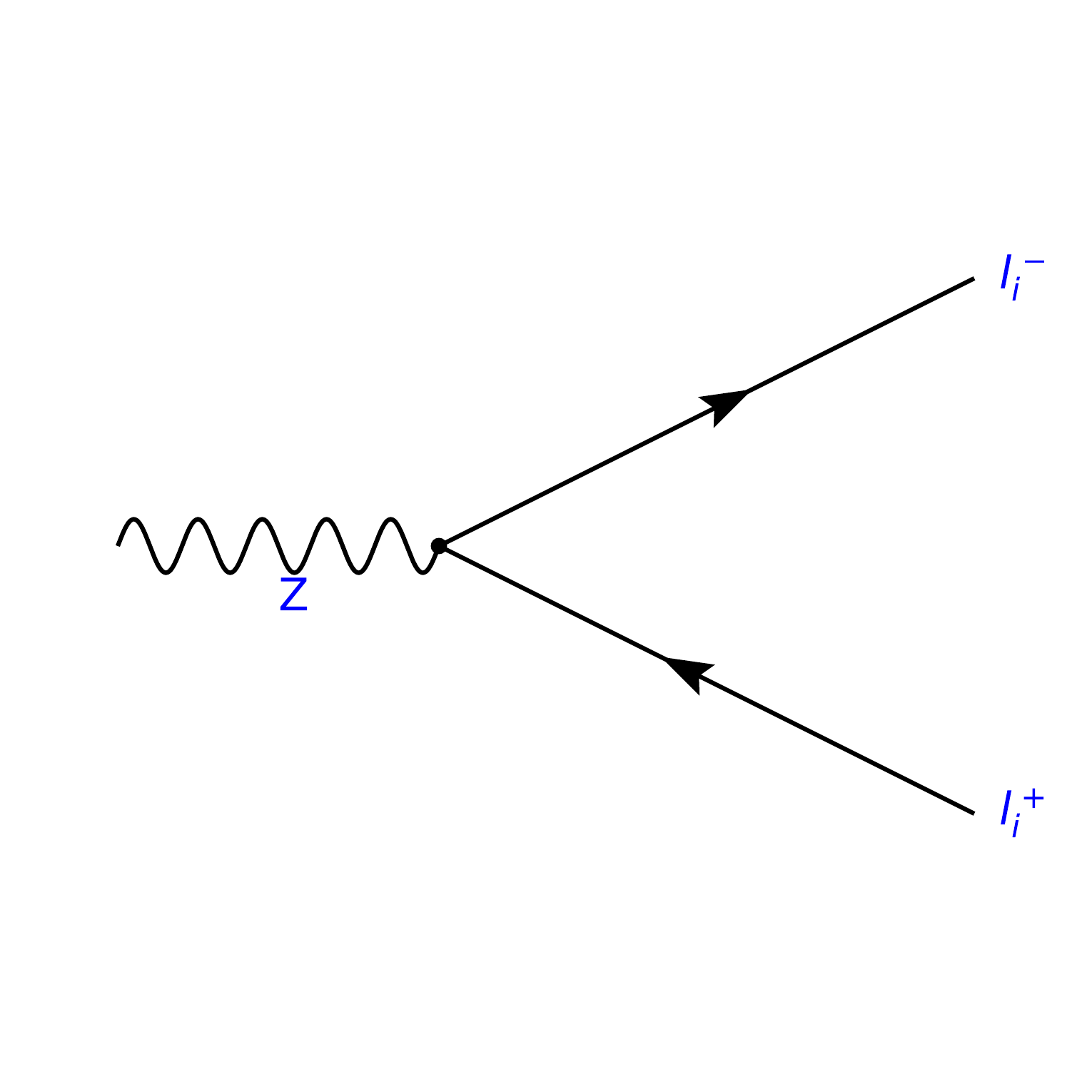} 
\includegraphics[width=0.25\linewidth]{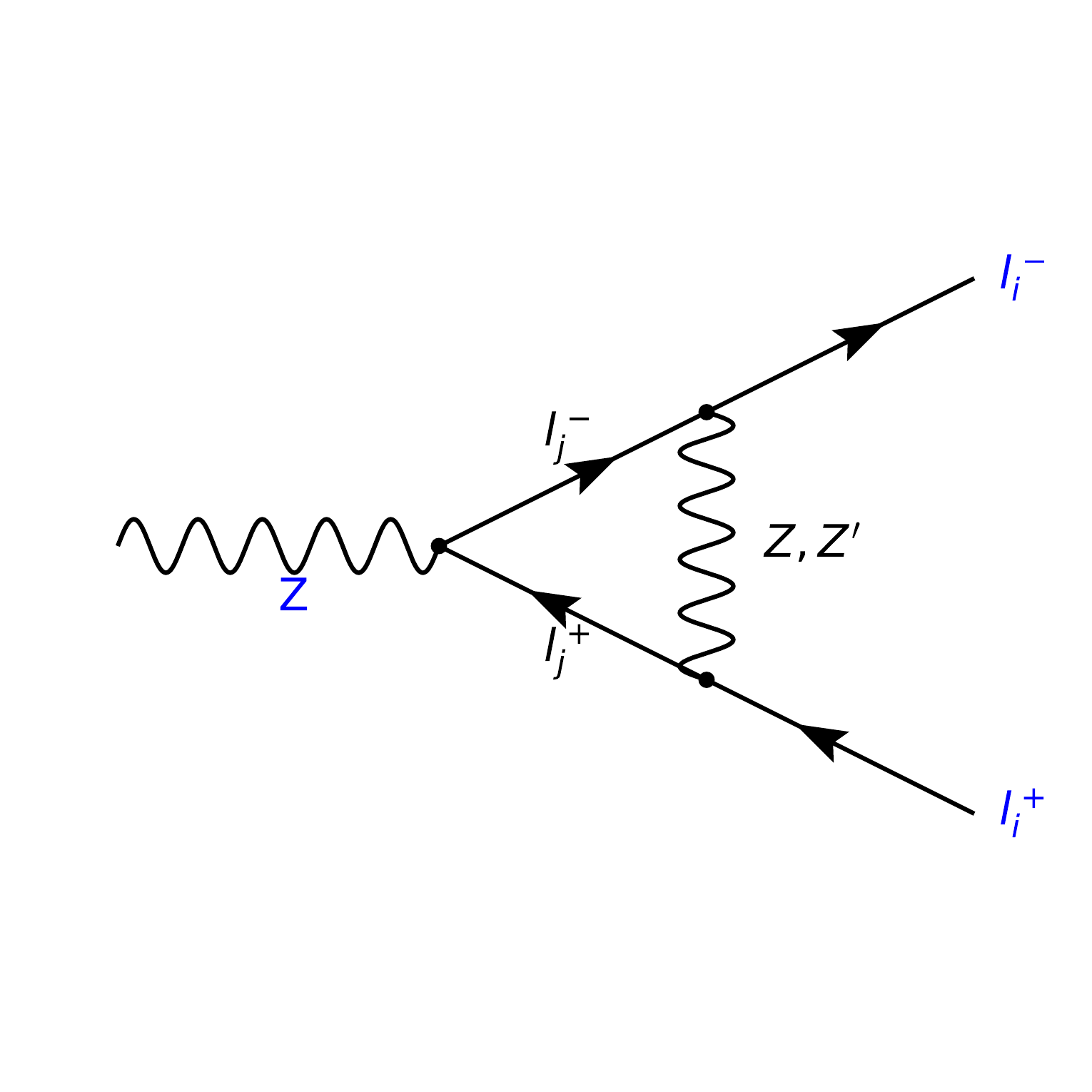} 
\includegraphics[width=0.25\linewidth]{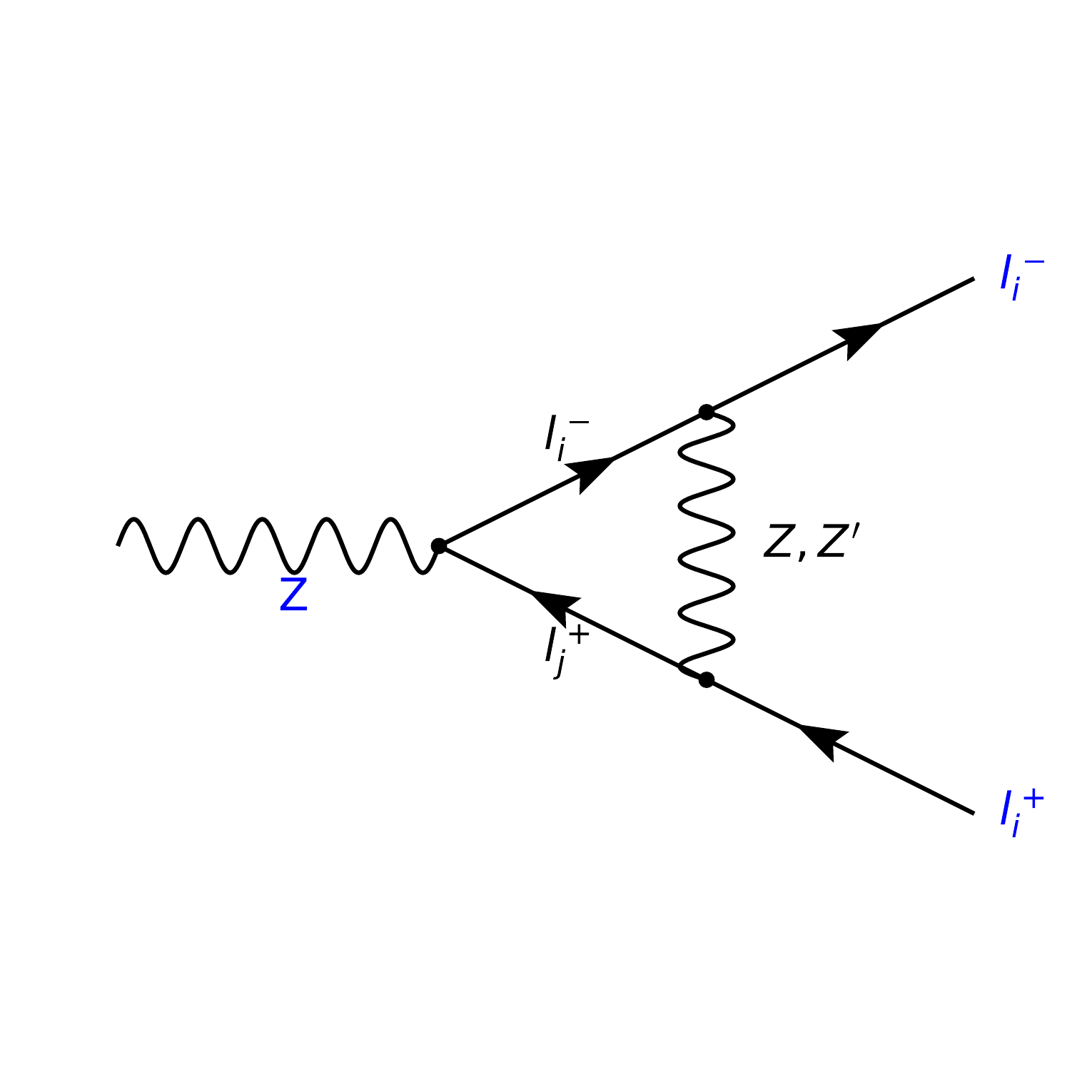} 
\includegraphics[width=0.25\linewidth]{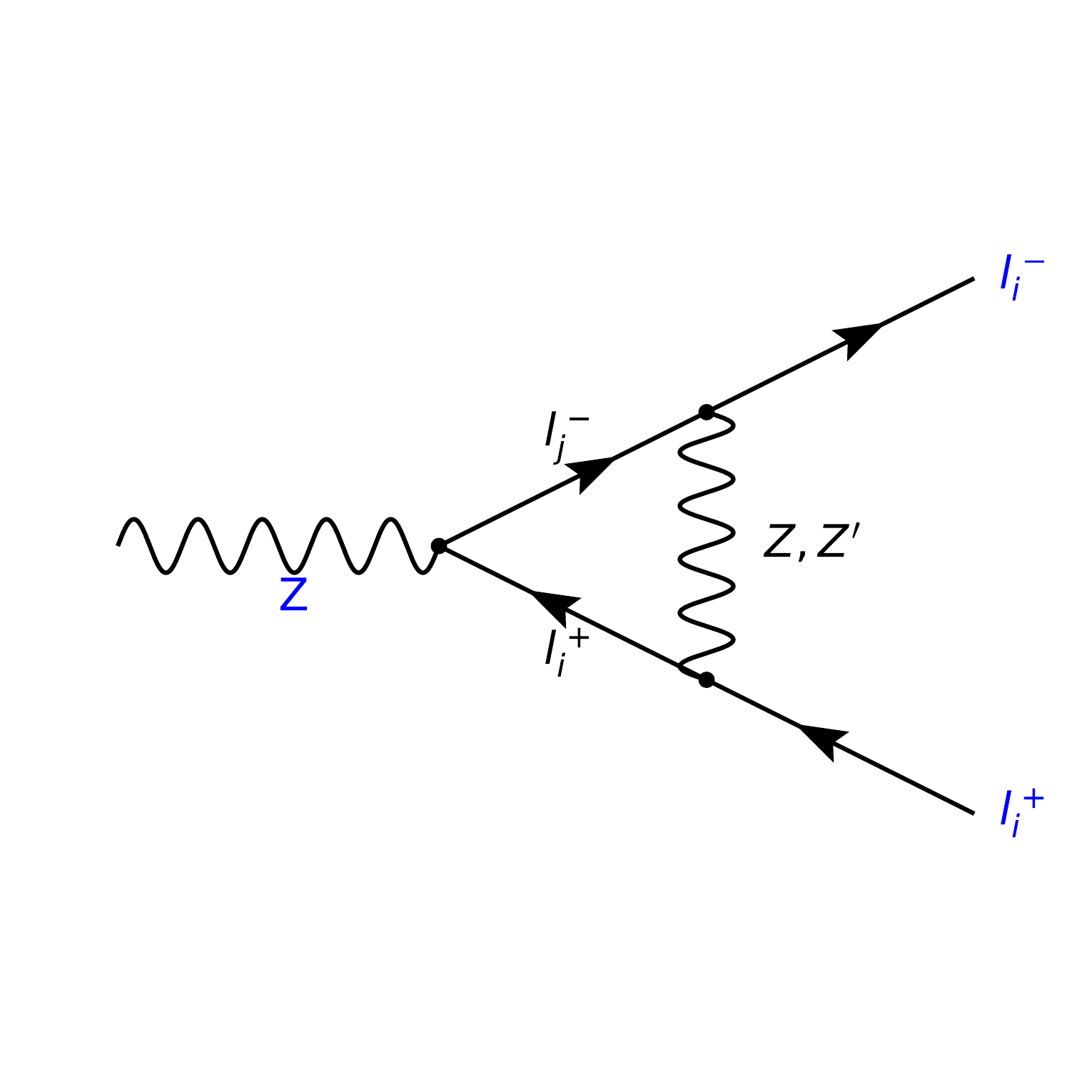} 
\includegraphics[width=0.25\linewidth]{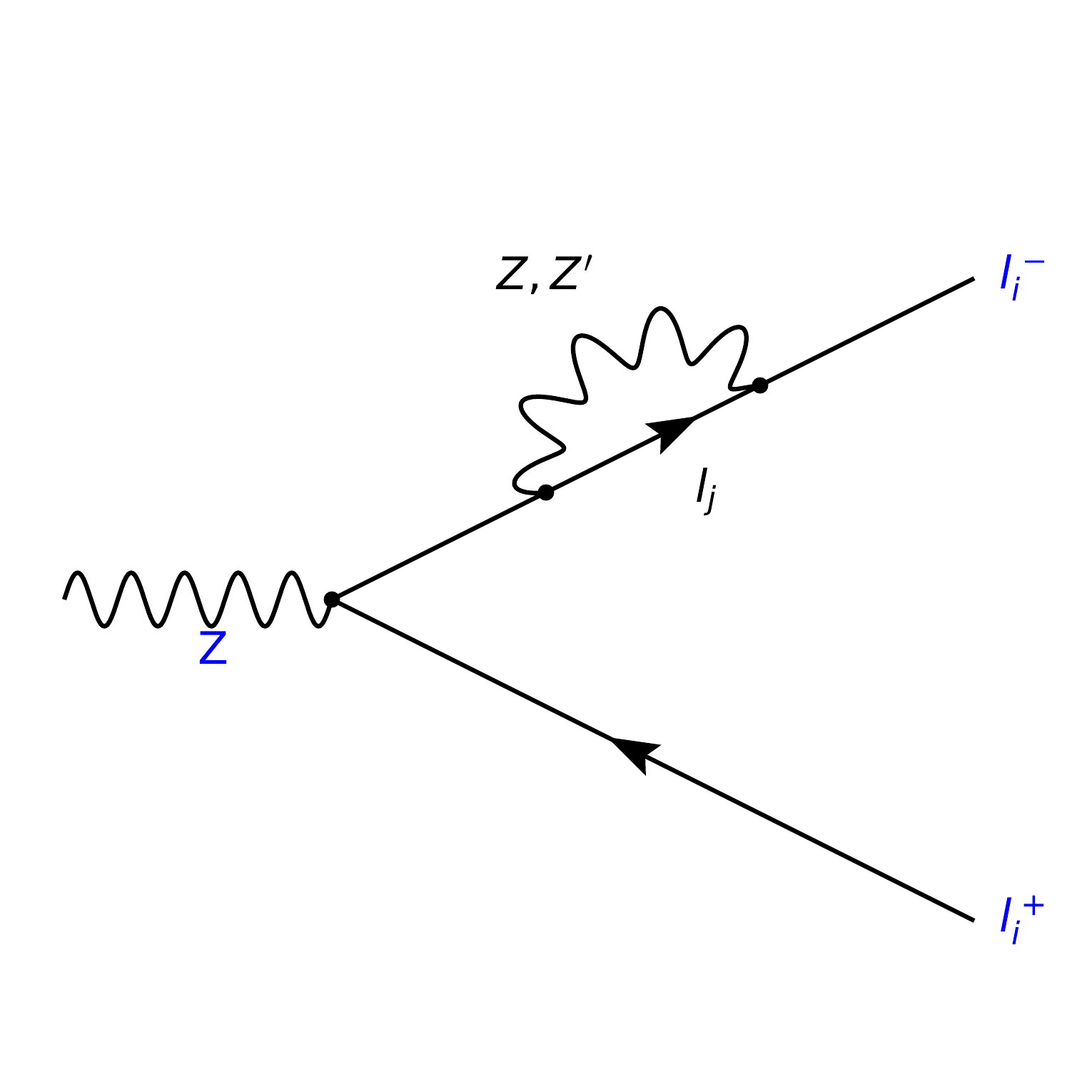} 
\includegraphics[width=0.25\linewidth]{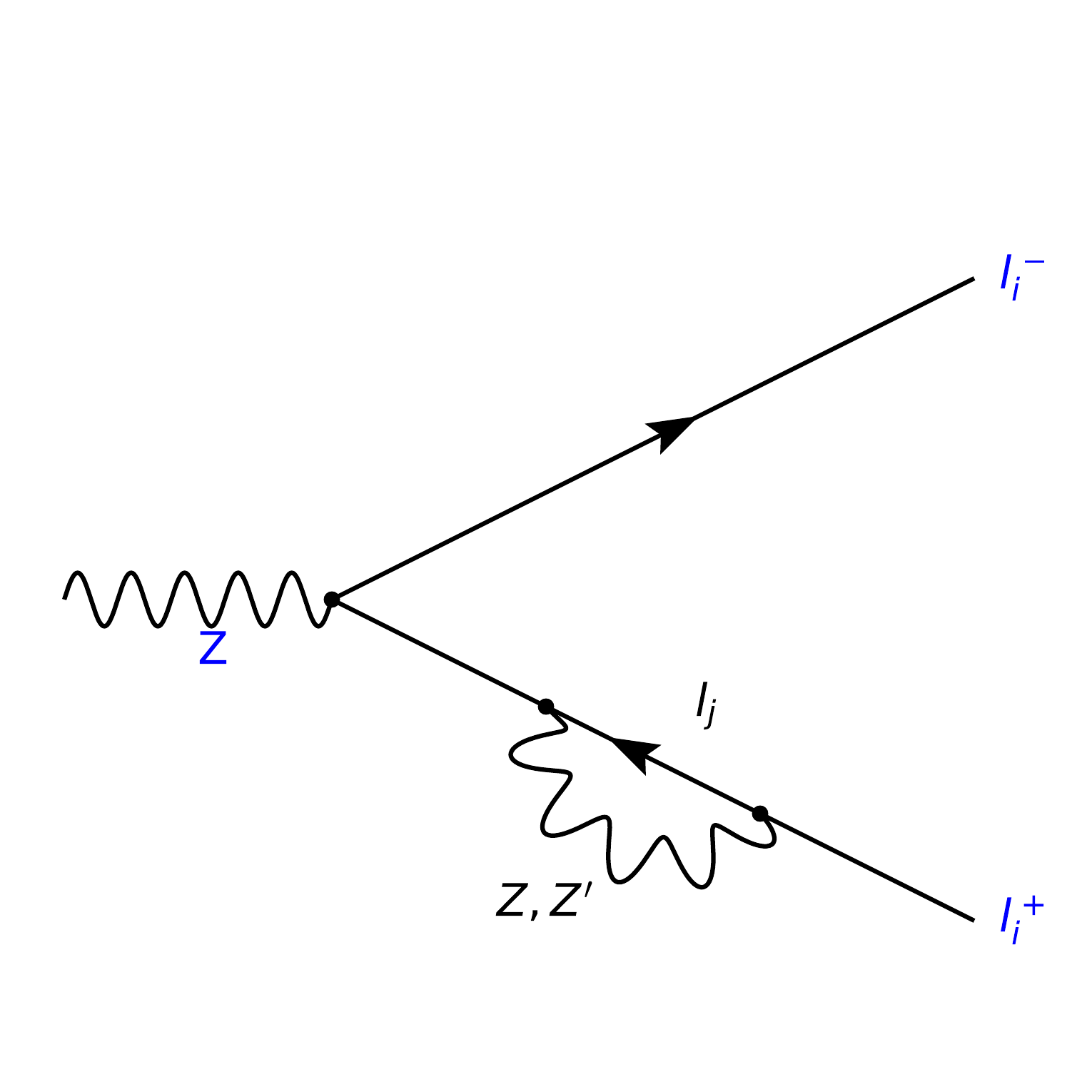} 
\caption{1-loop contribution to $Z_1\rightarrow l_i l_i$ decay.}
\label{Zlili1loop} 
\end{figure}

The $Z-Z'$ mixing modifies the couplings of the SM $Z$ boson to lepton pairs, and lepton FCNC interactions also contribute to these couplings through one-loop diagrams, as illustrated in FIG \ref{Zlili1loop}. Including both new contributions to the SM Z boson couplings, the decay width for $Z_1 \to l_i^+ l_i^-$ can be approximated as:

\bea
\Gamma(Z_1 \to l_i^+ l_i^-) && \simeq \Gamma_0 + \frac{1}{12 \pi} \left\{(g_L^0)_{ii} (\delta g_L^{Z_1})_{ii} + (g_R^0)_{ii} (\delta g_R^{Z_1})_{ii}\right\}  m_{Z_1}\\ \nonumber  & & +\frac{16 \pi^2 - 171}{1536 \pi^3} \left\{ \left(g_L^0 \right)_{ii}^2 \left [\left( g_L^{Z_1}\right)_{ij}^2 +\frac{m_{Z_1}^2}{m_{Z_2}^2}\left( g_L^{Z_1}\right)_{ij}^2 \right] \right\} m_{Z_1}
\eea
where $\Gamma_0=  \frac{1}{24 \pi} \left( (g_L^0)_{ii}^2+(g_R^0)_{ii}\right) m_{Z_1}$ represents the tree level SM contribution to the decay width; $(g_L^0)_{ii}$ and $(g_R^0)_{ii}$ are the SM left-handed and right-handed couplings, respectively; $(\delta g_L)_{ii}^{Z_1}$ and $(\delta g_R)_{ii}^{Z_1}$ represent the deviations from the SM couplings due to new physics at the tree level; $m_{Z_1}$ is the mass of the SM $Z$ boson. The experimental bounds on leptonic $Z$ decays are given in Table \ref{Zlilimodes}. To satisfy the $90\%$ Confidence Level (CL) constraint, the following condition must hold:

\begin{equation}
\frac{\delta \Gamma(Z_1 \to l_i^+ l_i^-)}{\Gamma_{Z_1}} < 1.7 \, \delta \Gamma_{Z_1}^{\text{Exp}},
\end{equation}

\begin{figure}[ht!]
\centering
\includegraphics[width=0.6\textwidth]{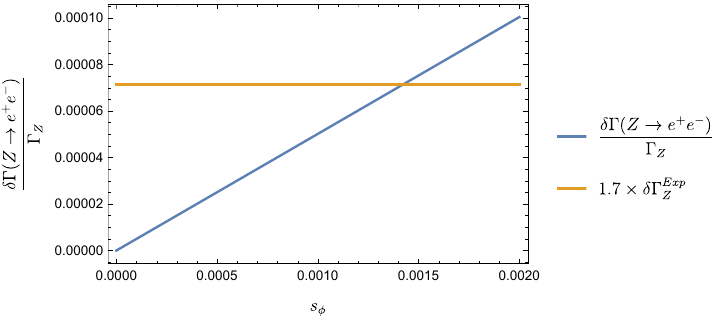} 
\caption{Condition for $90\% C.L$ for $Br(Z\rightarrow e^+e^-)$}
\label{ZeeFig90CL}
\end{figure}

We will find the parameter space for more strict mode which is $Z\rightarrow e^+e^-$. The bound for $\sin \phi$ is obtained by setting $c_{12}=1, c_{13}=0.3$. In this set up, the condition for the mixing angle between Z-Z' is $\sin \phi \leq 0.0015$ as in FIG.\ref{ZeeFig90CL}.

\begin{table}[ht!]
\caption{Flavor-conserving $Z$ decay modes \cite{ParticleDataGroup:2022pth}}
{\begin{tabular}{lc}
		\hline
		Decay Mode & Branching Ratio (\%) \\
		\hline
		$Br(Z \to e^+e^-)$       & $(3.3632 \pm 0.0042)$ \\
		$Br(Z \to \mu^+\mu^-)$     & $(3.3662 \pm 0.0066)$ \\
		$Br(Z \to \tau^+\tau^-)$   & $(3.3696 \pm 0.0083)$ \\
		$Br(Z \to c\bar{c})$       & $(12.03 \pm 0.21)$ \\
		$Br(Z \to b\bar{b})$       & $(15.12 \pm 0.05)$ \\
		\hline
	\end{tabular}
	\label{Zlilimodes}
}
\end{table}

\subsection{Lepton FCNCs and the \texorpdfstring{$\mu \rightarrow e \gamma$} Process}
\begin{figure}[ht!]
\centering
\includegraphics[width=0.4\linewidth]{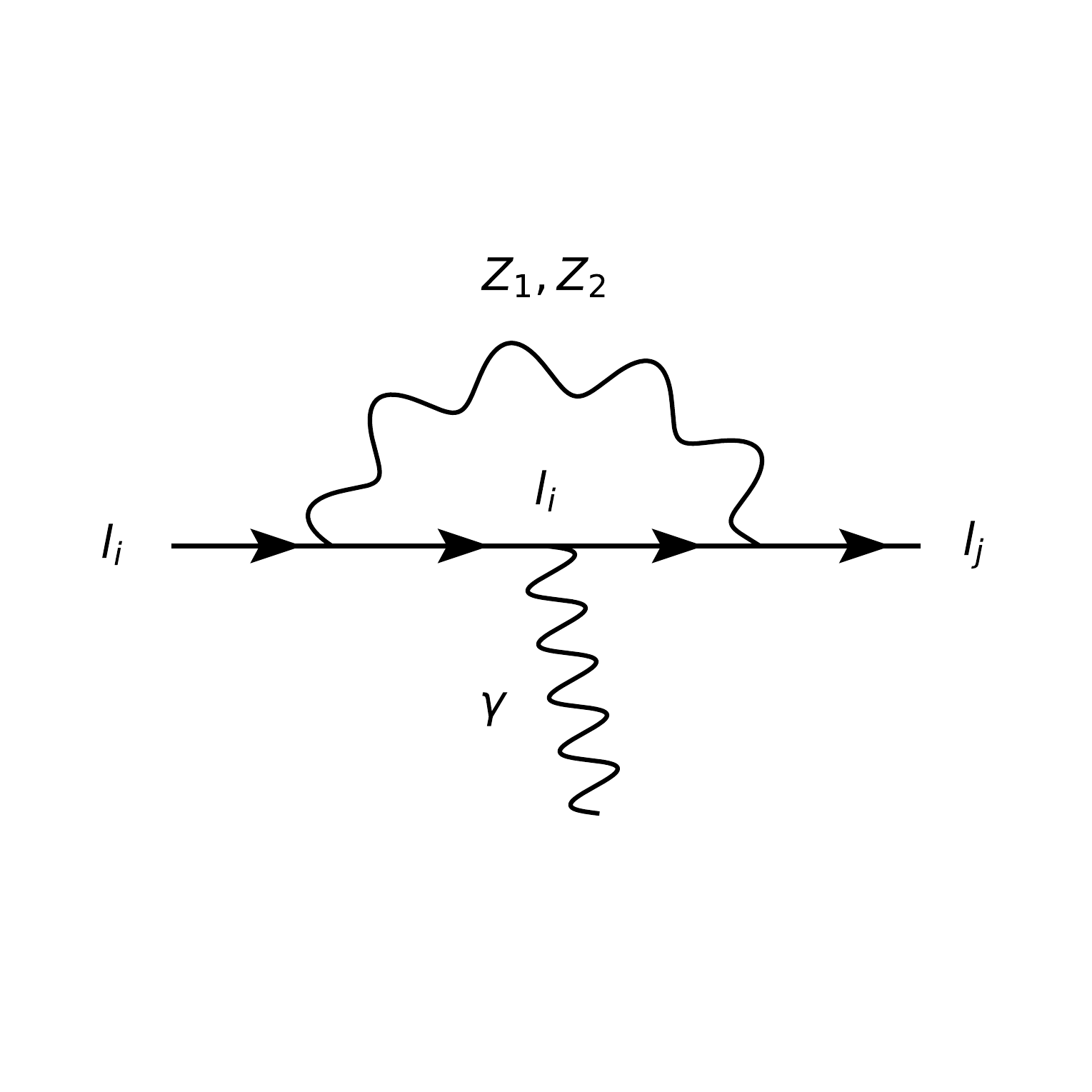} 
\includegraphics[width=0.4\linewidth]{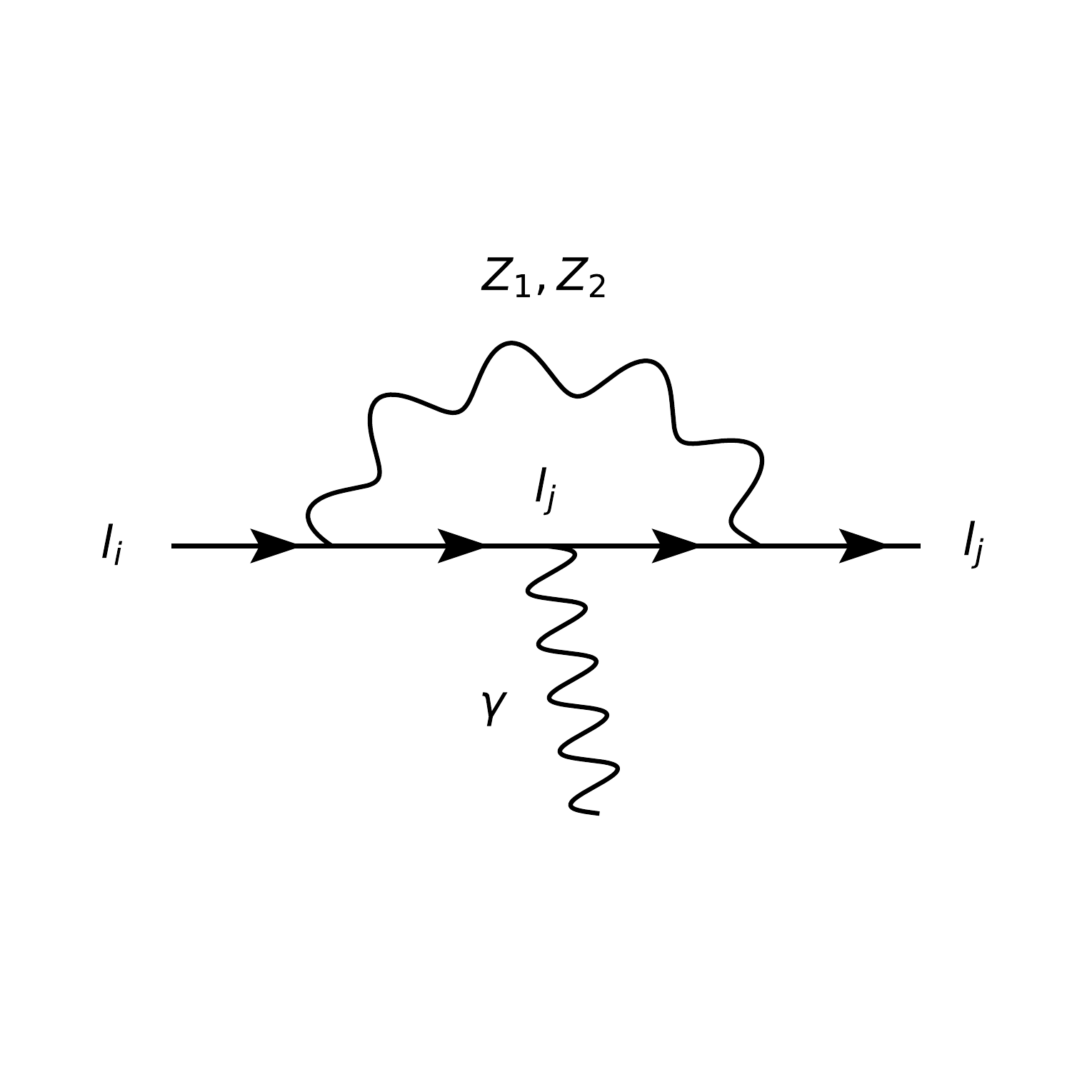} 
\caption{Loop diagrams for process $l_i \rightarrow l_j \gamma$}
\label{liljgammaFig} 
\end{figure}

Lepton FCNCs can significantly influence the radiative decay process $\mu \to e\gamma$. In this work, we focus specifically on the contributions to the general $l_i \rightarrow l_j \gamma$ decay arising from FCNCs, neglecting other potential contributions. The relevant Feynman diagrams are illustrated in FIG.\ref{liljgammaFig}.

It is convenient to introduce 

\begin{align}
(\Gamma^{Z_1,Z_2}_L)_{ij}=g^{Z_1,Z_2}_L \delta_{ij} + (g^{Z_1,Z_2}_L)_{ij}
\end{align}

in which the diagonal is the flavor conserving couplings and the non-diagonal is the flavor violating couplings. 

In the limit where $\frac{m_{\mu, e}}{m_{Z_1 , Z_2}} \simeq 0$ and $m_\tau >> m_\mu >>m_e$, we obtain the following expression for the decay width \cite{Kim:2016bdu}:

\begin{align}
\Gamma(l_i \to l_j \gamma) & =\frac{\al m_{l_i} }{9(4\pi)^4}\times 
\Big[\Big(\frac{m_{l_i}}{m_{Z_1}}\Big)^4 \left\vert\sum_k (\Gamma_L^{Z_1})_{ik}(\Gamma_L^{Z_1})_{kj}\right\vert^2 \nonumber \\
&+ \Big(\frac{m_{l_i}}{m_{Z_2}}\Big)^4 \left\vert \sum_k (\Gamma_L^{Z_2})_{ik}(\Gamma_L^{Z_2})_{kj} \right\vert^2\nonumber \\
&+ \Big(\frac{m_{l_i}}{m_{Z_1}}\Big)^2 \Big(\frac{m_{l_i}}{m_{Z_2}}\Big)^2 \sum_k (\Gamma_L^{Z_1})_{ik}(\Gamma_L^{Z_1})_{kj} \sum_k (\Gamma_L^{Z_2})_{il}(\Gamma_L^{Z_1})_{lj}\Big]	
\end{align}

The branching ration of the process $l_i \rightarrow l_j \gamma $ can be written as:

\begin{align}
Br(l_i\rightarrow l_j \gamma)&= \frac{\Gamma(l_i \rightarrow l_j \gamma)}{\Gamma(l_i \rightarrow l_j \bar{\nu}_j \nu_i)}\times Br(l_i \rightarrow l_j \bar{\nu}_j \nu_i)
\end{align}

where 
\begin{align}
\Gamma(l_i \rightarrow l_j \bar{\nu}_j \nu_i)&=\frac{G_F^2 m_i^5}{192 \pi^3}
\end{align}

The most rencent bounds on the $\mu \to e \gamma$ is given in \cite{ParticleDataGroup:2022pth} with the branching ratio $Br(\mu \rightarrow e \bar{\nu}_e \nu_\mu) \sim 100\%$
\bea
Br(\mu \to e \gamma) < 4.2 \times 10^{-13}.
\eea

\begin{figure}[ht!] 
\centering
\includegraphics[width=0.4\linewidth]{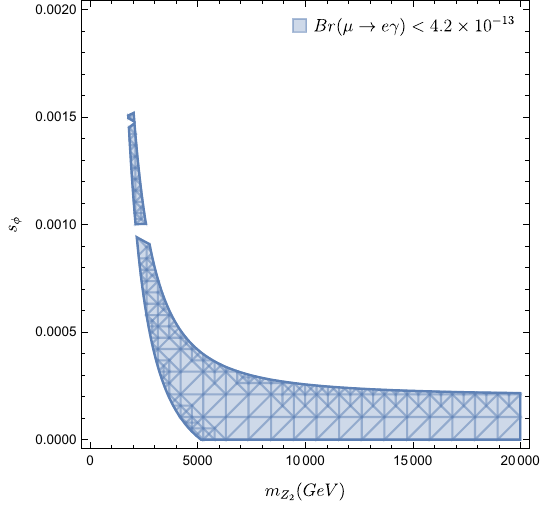} 
\caption{Allowed region of parameter space ($m_{Z_2}, s_{\phi}$) of $\mu \rightarrow  e \gamma $ }
\label{muegFig} 
\end{figure}

We investigate the allowed parameter space $(\sin \phi ,m_{Z_2})$ and illustrate as in Fig.(\ref{muegFig}). The mass of the new Z' boson of  the model is then $m_{Z_2}\geq 2.5 TeV$ when $\sin\phi  \leq 0.0015$ which is consistent with the constraint  $\sin \phi \leq 0.0015$ from the decay $Z \rightarrow e^+ e^-$.

\subsection{ Leptonic three-body decay}
Building upon the proposed model of lepton flavor violation (LFV) mediated by $Z_1$ and $Z_2$ bosons, this section systematically surveys various leptonic three-body decay channels. Our analysis focuses on scenarios where these LFV interactions occur at the tree-level approximation. We will specifically examine decays of the form $l_i \to l_j  l_\beta l_\beta $, where $l_i$ represents a heavier charged lepton decaying into lighter leptons. Intergrating out the neutral bosons, $Z_1,Z_2$, from neutral currents, we obtain an effective Lagrangian  for these processes 

\begin{equation}
\mathcal{L}_{\text{eff}}^{l_i \to l_j l_\beta l_\beta} \simeq -\frac{4G_F}{\sqrt{2}} \left\{ g_{LL}^{ij\beta\beta}\bar{l_i}\gamma_\mu P_L l_j \bar{l}_\beta \gamma^\mu P_L l_\beta +g_{LR}^{ij\beta\beta}\bar{l_i}\gamma_\mu P_L l_j \bar{l}_\beta \gamma^\mu P_R l_\beta\right\} +H.c.
\label{eff3lep}\end{equation}
where the effective couplings are defined as:
\begin{equation}
g_{LL(R)}^{ij\beta\beta} = \frac{\sqrt{2}}{4G_F} \left(\frac{(g_L^{Z_1})_{ij}(g_{L(R)}^{Z_1})_{\beta \beta}}{m_{Z_1}^2}+\frac{(g_L^{Z_2})_{ij}(g_{L(R)}^{Z_2})_{\beta \beta}}{m_{Z_2}^2}\right),
\end{equation}
with the lepton flavor conserving couplings $(g_{L(R)}^{Z_{1,2}})_{\beta \beta}$ given by:
\begin{equation}
(g_R^{Z_{1,2}})_{\beta \beta} = -\frac{g}{4c_W}\left(g_V^{Z_{1,2}}(f_\beta) +g_A^{Z_{1,2}}(f_\beta)\right),
\end{equation}
\begin{equation}
(g_L^{Z_{1,2}})_{\beta \beta} = -\frac{g}{4c_W}\left(g_V^{Z_{1,2}}(f_\beta) -g_A^{Z_{1,2}}(f_\beta)\right).
\end{equation}
The effective Lagrangian presented in Eq. (\ref{eff3lep}) directly describes various charged LFV processes, including $\mu \to 3e$, $\tau \to 3\mu$, $\tau \to 2e\mu$, and $\tau \to 2\mu e$.  In particular, the branching ratio for the $\tau^+ \to \mu^+ \mu^+ \mu^-$ decay channel is given by:
\begin{equation}
Br\left( \tau^+ \to \mu^+ \mu^+ \mu^-\right) =\left( |g_{LL}^{\tau \mu \mu \mu}|^2+ |g_{LR}^{\tau \mu \mu \mu}|^2 \right) Br(\tau^+ \to \bar{\nu}_\tau e^+ \nu_e). \label{tauto3mu}
\end{equation}
To obtain the branching ratios for other analogous decay modes, such as $\tau^+ \to e^+e^-e^+$ and $\mu^+ \to e^+ e^- e^+$, we can adapt Eq. (\ref{tauto3mu}). Specifically, for $\tau^+ \to e^+e^-e^+$, one replaces all $\mu$ indices with $e$ indices. For $\mu^+ \to e^+ e^- e^+$, the initial lepton $\tau$ is replaced by $\mu$, and all final state $\mu$ indices are replaced by $e$ indices. 

\begin{figure}[ht!] 
\centering
\includegraphics[width=0.65\linewidth]{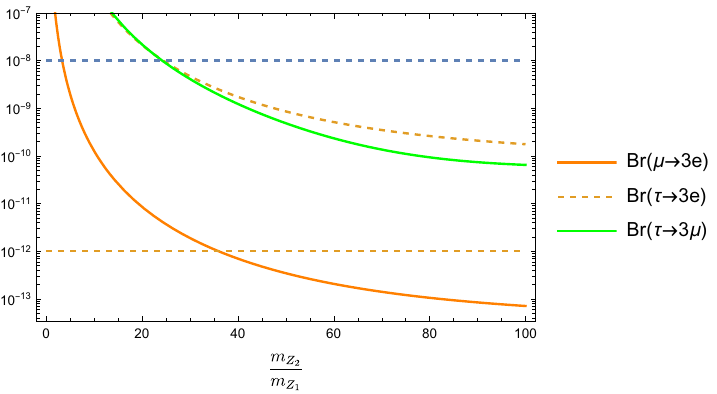}
\caption{The branching ratio $Br(l_i\rightarrow l_j l_\beta l_\beta)$ as a function of the mass ratio $\frac{m_{Z_2}}{m_{Z_1}}$.}
\label{mm} 
\end{figure}

In previous investigations of these processes \cite{Dinh:2019jdg}, we primarily focused on the impact of lepton flavor violating (LFV) neutral currents associated with the new gauge boson, often neglecting the contribution from those mediated by the neutral SM gauge boson. However, our current analysis reveals that the proportional contributions from both neutral gauge bosons are comparable, a crucial finding that necessitates their inclusion in comprehensive evaluations. Specifically, we've found that the coupling strengths of the lepton flavor conserving (LFC) neutral currents for both the $Z_1$ and $Z_2$ gauge bosons are of the same order of gauge magnitude. Consequently, their proportional contributions to the branching ratio are determined as follows:
\bea
\frac{(|g_L^{Z_1}|^2)_{ij}}{(|g_L^{Z_2}|^2)_{ij}} \times \frac{m_{Z_2}^2}{m_{Z_1}^2} =\left( \tan  \phi \right)^2 \frac{m_{Z_2}^2}{m_{Z_1}^2} \simeq  \mathcal{O}(1).
\eea
Under the assumption of maximal mixing among charged lepton generations, the branching ratio $Br(l_i \to l_j l_\beta l_\beta)$ reaches its largest predicted value. Additionally, by assuming a mixing angle between the $Z$ and $Z'$ bosons given by $\phi \simeq  \frac{u^2,v^2}{w^2, \La^2} \simeq \frac{m_{Z_1}^2}{m_{Z_2}^2}$, we derive the branching ratio $Br(l_i \to l_j l_\beta l_\beta)$ as a function of the mass ratio $\frac{m_{Z_2}}{m_{Z_1}}$, as illustrated in Fig.(\ref{mm}).

Our theoretical predictions for the branching ratios of charged LFV decays, namely $Br(\mu \to 3e)$, $Br(\tau \to 3e)$, and $Br(\tau \to 3\mu)$, are presented as a function of the mass ratio $m_{Z_2}/m_{Z_1}$ in Fig.~(\ref{mm}). These predictions are crucial for assessing the viability of our proposed model against stringent experimental constraints.

The current experimental upper limits on these charged LFV processes are highly restrictive. For the muon decay, $Br(\mu \to 3e) < 1.0 \times 10^{-12}$ at 90\% C.L. \cite{ParticleDataGroup:2022pth}. In contrast, the limits for tau decays are comparatively less stringent, with $Br(\tau \to 3e) < 2.7 \times 10^{-8}$ at 90\% C.L.  \cite{ParticleDataGroup:2022pth} and $Br(\tau \to 3\mu) < 2.1 \times 10^{-8}$ at 90\% C.L.  \cite{ParticleDataGroup:2022pth}.

\texorpdfstring{}{}

The most constraining channel for our model proves to be $\mu \to 3e$, due to its exceptionally tight experimental limit. To ensure consistency with the current experimental bound of $1.0 \times 10^{-12}$ for $\mu \to 3e$, our model necessitates that the mass ratio $m_{Z_2}/m_{Z_1}$ falls within the region where the predicted $Br(\mu \to 3e)$ curve drops below this limit. Visually inspecting Fig.~(\ref{mm}), this consistency is achieved for values of $\frac{m_{Z_2}}{m_{Z_1}} >35$ and the mass of new neural gauge boson $m_{Z'}\geq 3.2TeV $ which  compliment the  LHC constraints \cite{ParticleDataGroup:2022pth}.

In summary, the $F3311$ model, by design, has the FLV at tree level in the lepton sector. This feature will contribute to physical processes involve neural current.We have investigated the flavor changing in neural current in some decay processes and set constraints on the mixing angle between Z-Z bosons and the mixing in the lepton sector and the mass of the new neural gauge boson Z'. The process $\mu \to e \gamma $ gives the most strict condition on the mixing parameters which  $ \sin\phi <0.0015$ while  three body leptonic decay set bound  $m_{Z'}\geq 3.2 TeV$. In next section we will study the non-universality effects in the $u^i-d^j$ transitions using these constraints on Z-Z' mixing parameters $\sin \phi$ and the mixing parameters among lepton generations.

\section{The flavor non-universality  in \texorpdfstring{$u^i - d^j$} transitions  \label{Eff}}
The flavor non-universality  in $u^i - d^j$ transitions has been previously investigated within the framework of the F331 model \cite{Thu:2023xai}. However, as highlighted, crucial distinctions of the F3311 model compared to the F331 model include the absence of direct mixing between SM fermions and newly introduced heavy fermions, as well as the lack of mixing between the SM $W^\pm$ gauge boson and new $X^\pm$ gauge boson. These fundamental differences imply that the F3311 model does not induce any modifications to the Wilson coefficients of the relevant SM operators at the tree-level approximation; instead, all new contributions to these transitions arise exclusively at the one-loop level (or higher orders). Given these fundamental differences, a re-evaluation of these processes within the F3311 model is warranted, which is the focus of the present study.

\subsection{The effective Hamiltonian in \texorpdfstring{$u^i - d^j$} transitions}
In the framework of the model, the effective contributions of charged-currents, as presented in Eqs.~(\ref{Wcurrent}), (\ref{Xcurrent}), and (\ref{Ycurrent}), to lepton-flavor non-universal processes such as $u_i \to d_j e_b \bar{\nu}_a$ transition are described by the effective Hamiltonian:
\begin{align}
\mathcal{H}_{eff}
&=\left[ \mathcal{C}_{\nu_a e_b}^{u_id_j}\right]( \bar{d}^\prime_{j} \ga^\mu P_L u_{i}^{'} )  ( \bar{e}_{b}^\prime \ga_\mu P_L \nu^\prime_{a} ).
\end{align}
Including one loop correction, the Wilson Coefficients (WCs), $\mathcal{C}^{u_id_j}_{\nu_a e_b}$, can be separated as followings
\begin{align}
\mathcal{C}^{u_id_j}_{\nu_a e_b}&=\left[\mathcal{C}^{u_id_j}_{\nu_a e_b}\right]_{\text{tree}} +\left[\mathcal{C}^{u_id_j}_{\nu_a e_b}\right]_{\text{penguin}}+\left[\mathcal{C}^{u_id_j}_{\nu_a e_b}\right]_{\text{box}}.
\end{align}
where the tree level Wilson Coefficient is
\begin{align}
\left[\mathcal{C}^{u_i d_j}_{\nu_a e_b}\right]_{\text{tree}}&=\frac{4 G_F}{\sqrt{2}}(U_{PMNS})^\dagger_{ab} (V_{CKM})_{ij}\,.
\end{align}

\begin{figure}[ht!]
\centering
\includegraphics[width=0.2\linewidth]{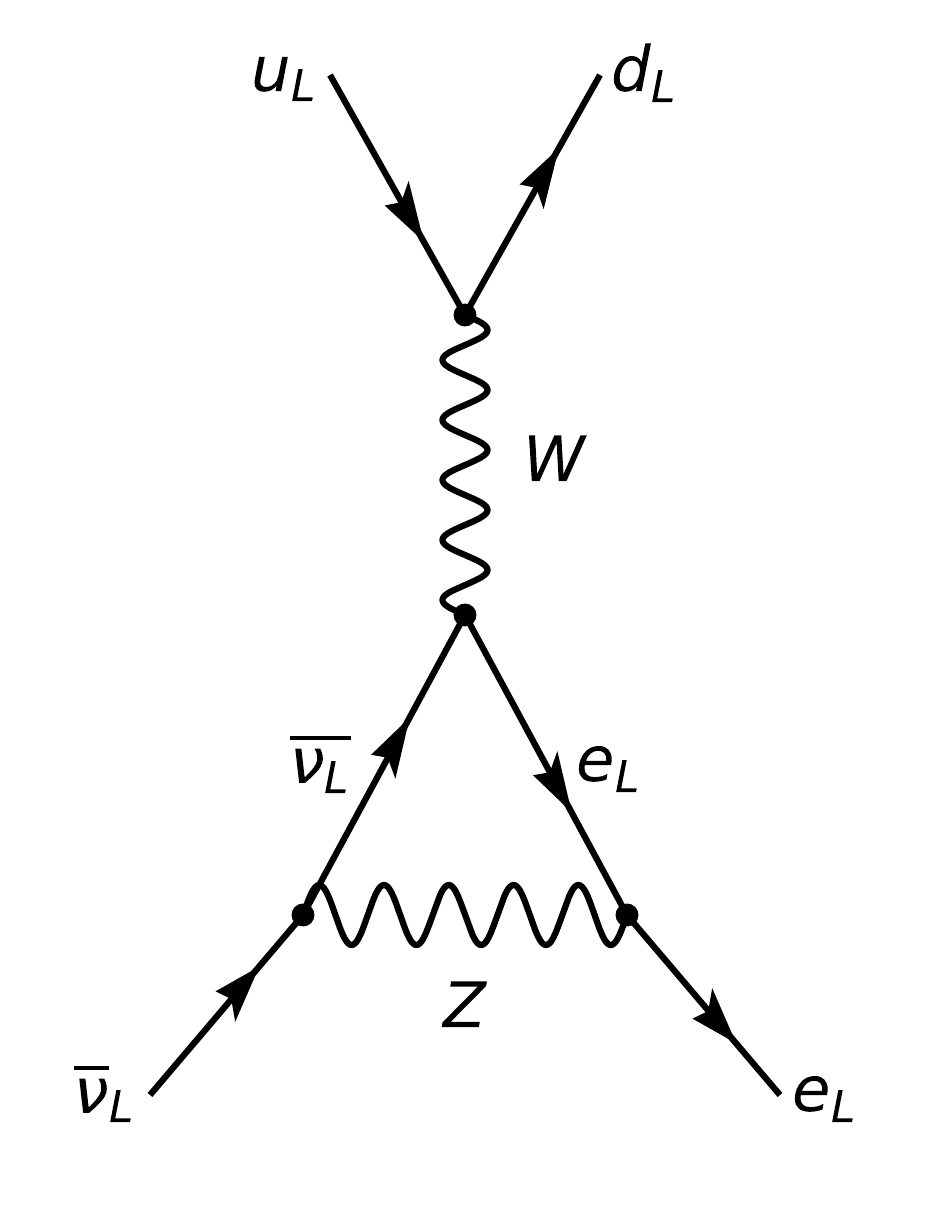} 
\includegraphics[width=0.2\linewidth]{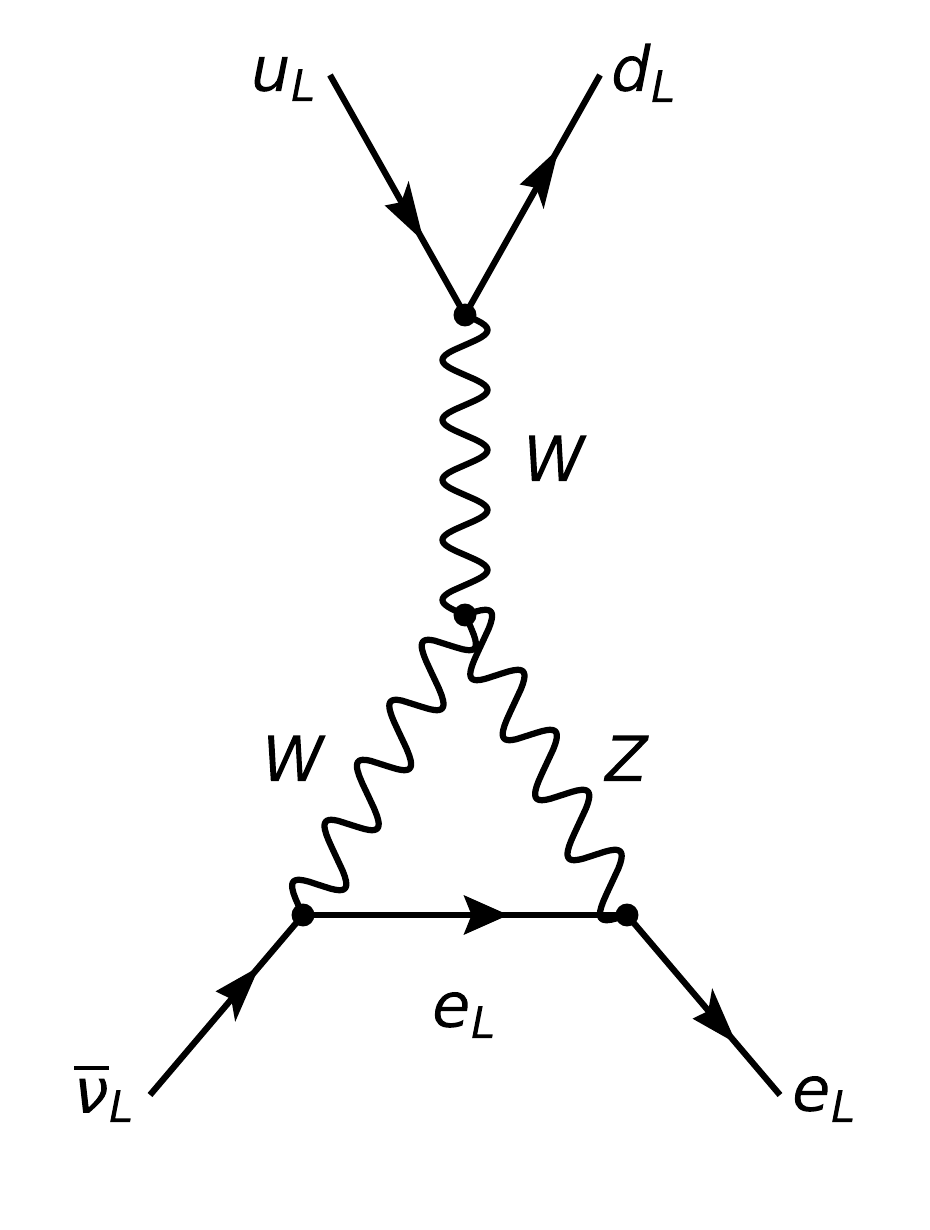} 
\includegraphics[width=0.2\linewidth]{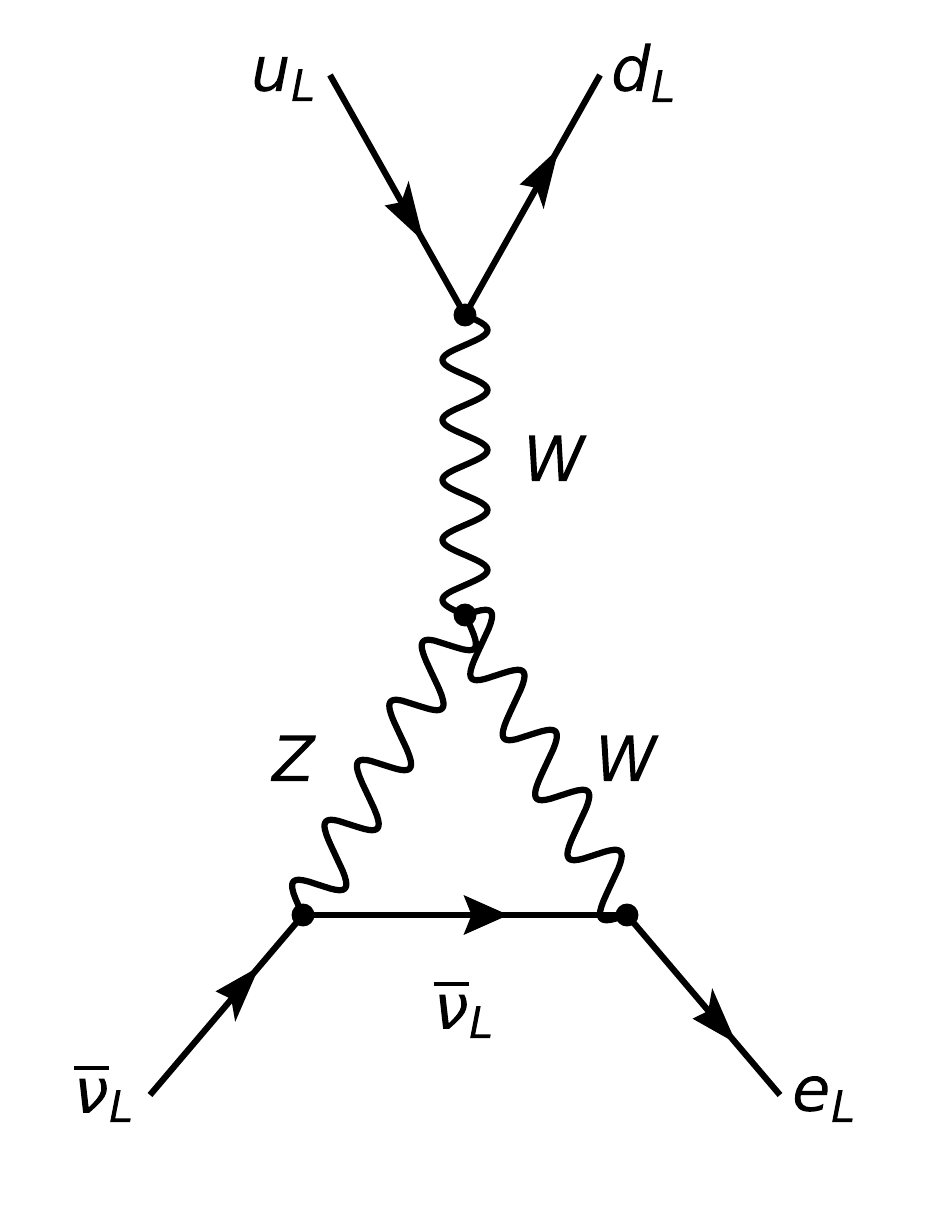} 
\includegraphics[width=0.2\linewidth]{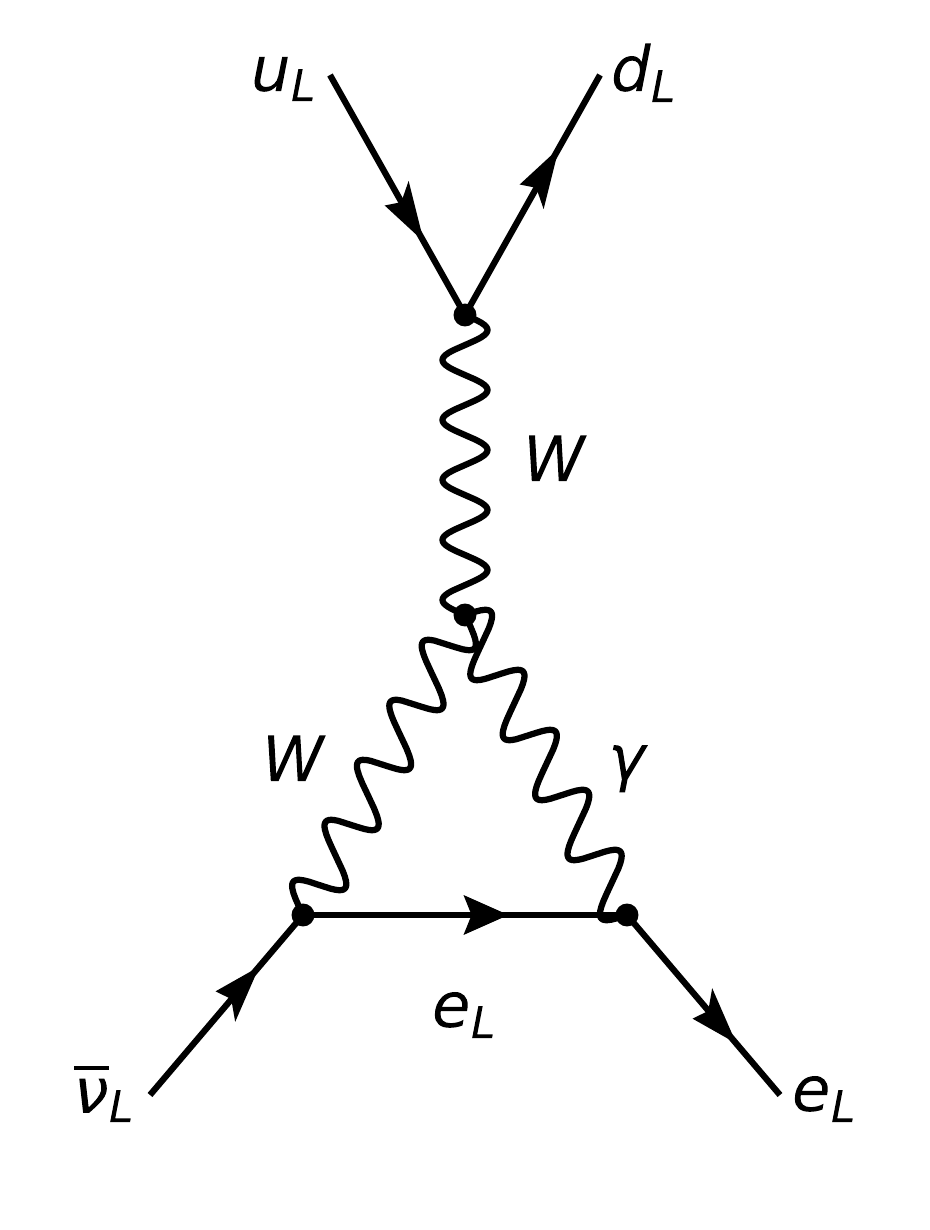} 
\caption{SM  contributions-penguin diagrams.}
\label{fig1} 
\end{figure}

\begin{figure}[ht!]
\centering
\includegraphics[width=0.16\linewidth]{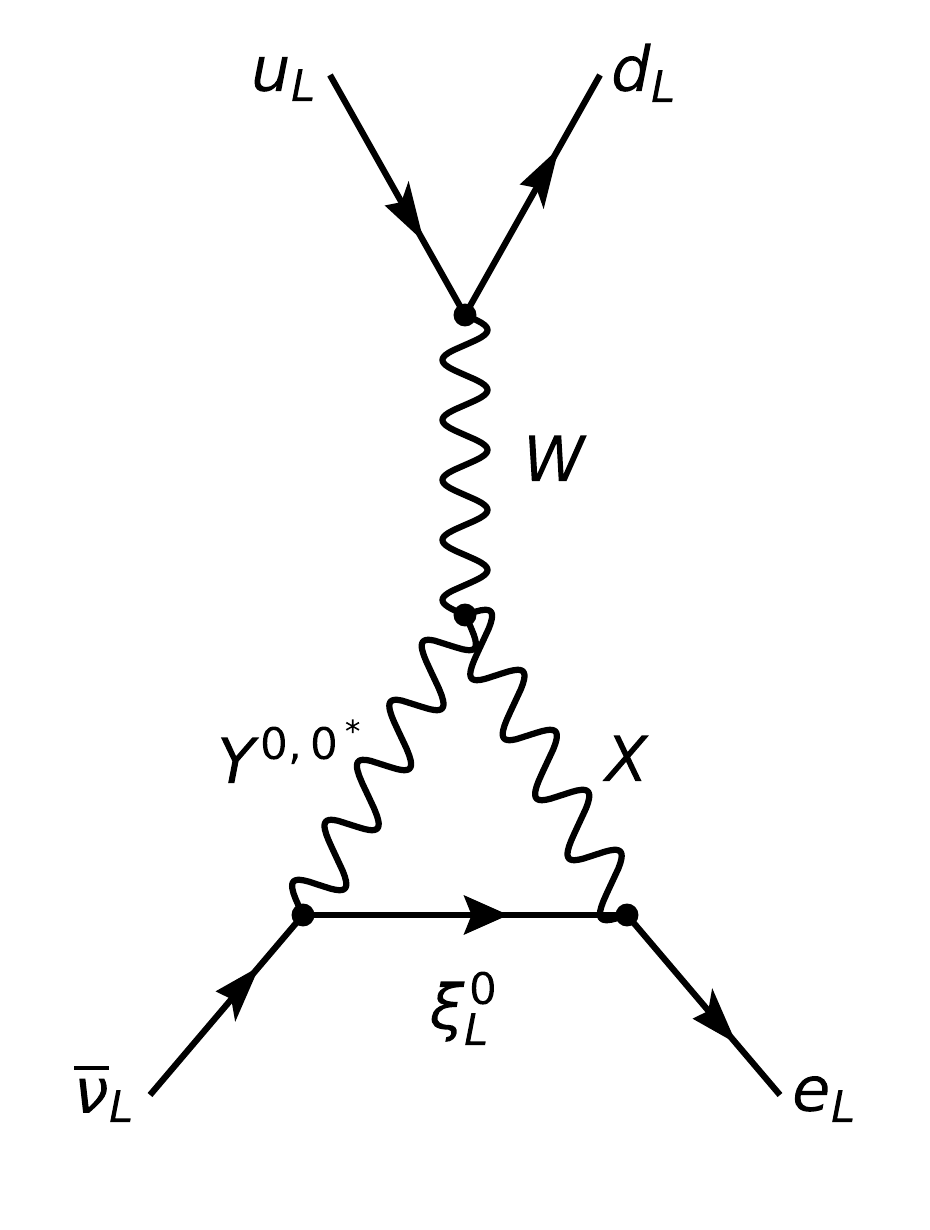} 
\includegraphics[width=0.16\linewidth]{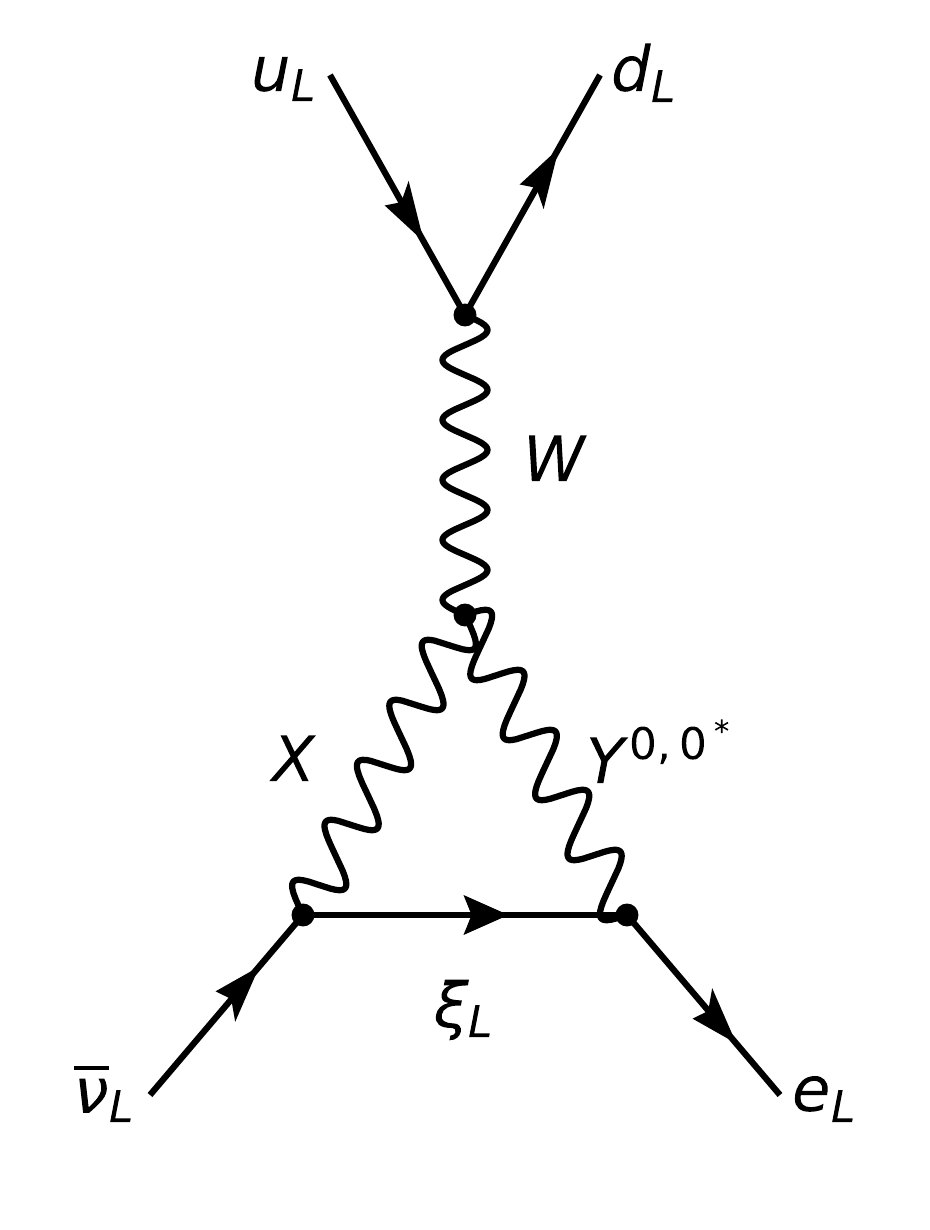} 
\includegraphics[width=0.16\linewidth]{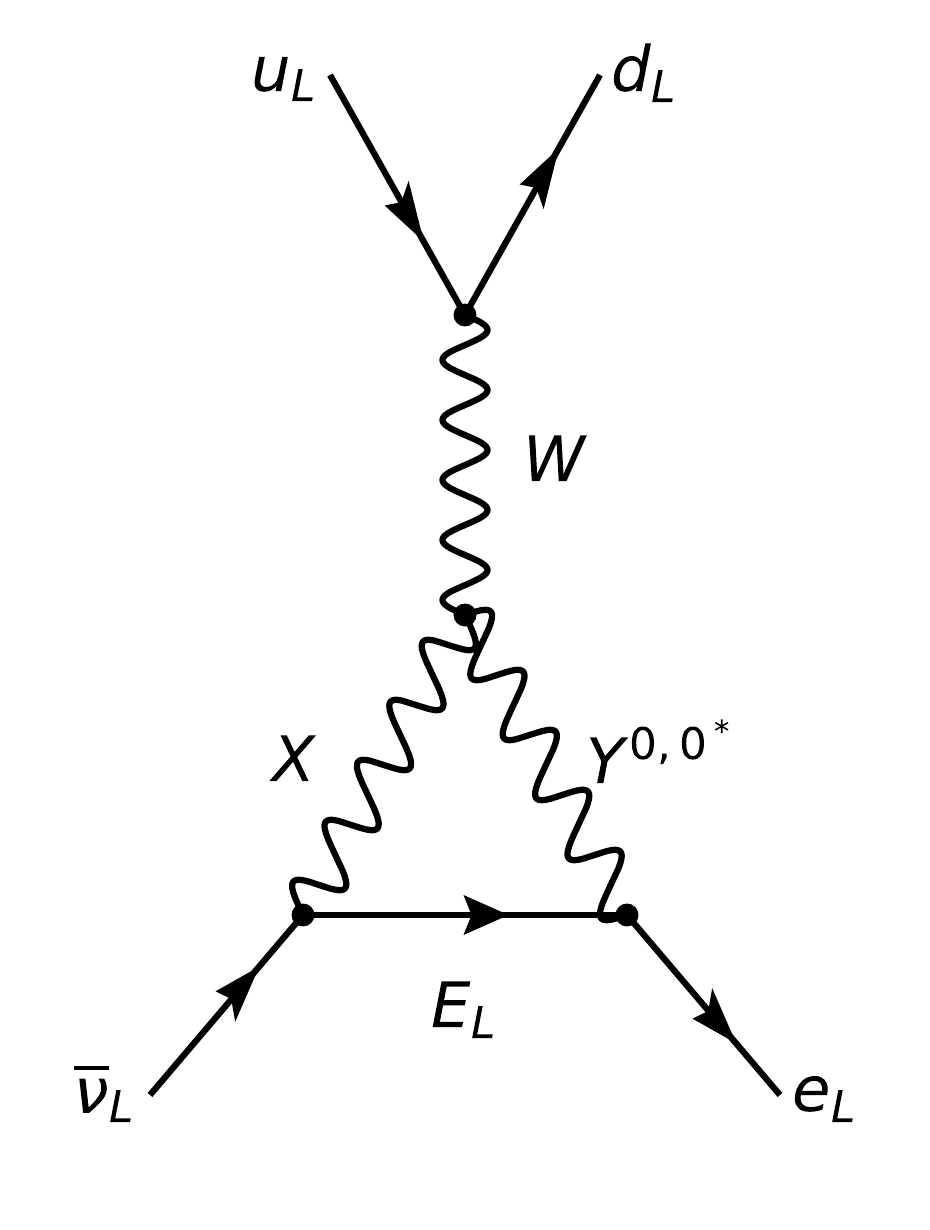} 
\includegraphics[width=0.16\linewidth]{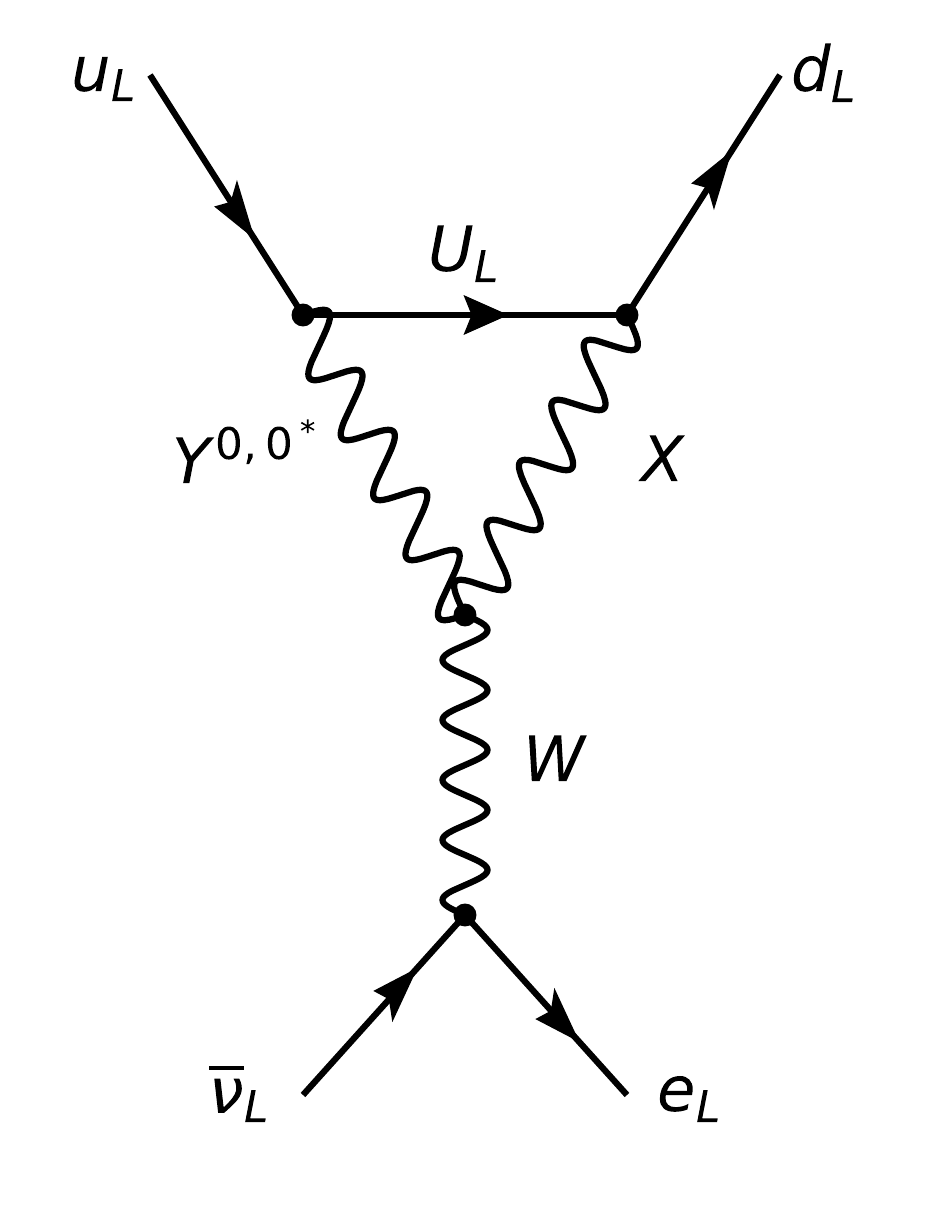} 
\includegraphics[width=0.16\linewidth]{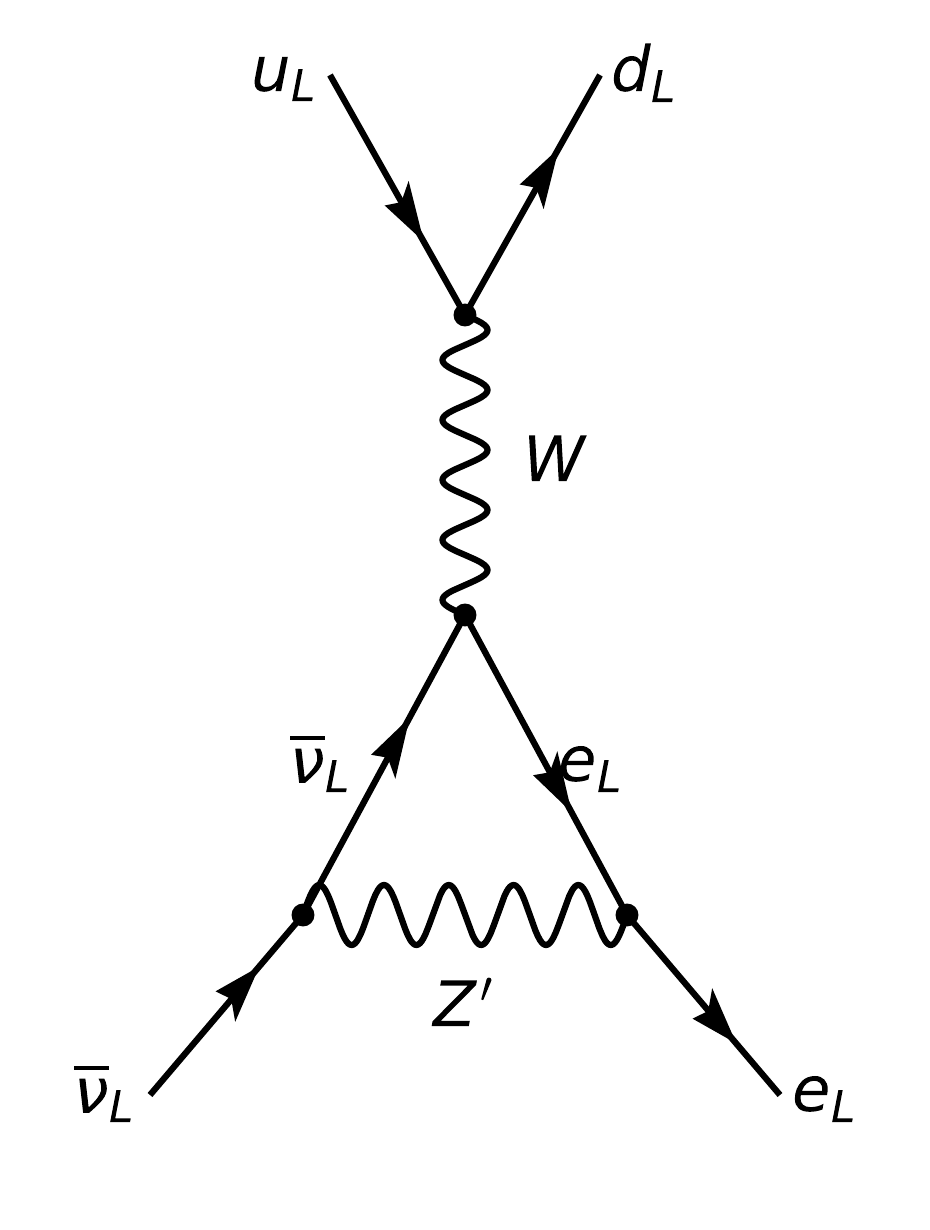} 
\caption{$F3311$ model contributions-penguin diagrams.}
\label{fig2} 
\end{figure}

The penguin diagrams' contribution can be split
into two parts, SM contribution Fig. \ref{fig1} and NP contribution Fig \ref{fig2}. 
\begin{align}
\left[\mathcal{C}^{u_id_j}_{\nu_a e_b}\right]_\text{penguin}&=\left[\mathcal{C}^{u_id_j}_{\nu_a e_b}\right]_\text{penguin}^{\text{SM}}+\left[\mathcal{C}^{u_id_j}_{\nu_a e_b}\right]_\text{penguin}^{\text{NP}}.
\end{align}

The SM contribution, $\left[\mathcal{C}^{u_id_j}_{\nu_a e_b}\right]_\text{penguin}^{\text{SM}}$, is written as
\begin{align}
\left[\mathcal{C}^{u_id_j}_{\nu_a e_b}\right]_\text{penguin}^{\text{SM}}&= 
\mathcal{C}^{u_id_j}_{\nu_a e_b}(\nu e Z) 
+\mathcal{C}^{u_id_j}_{\nu_a e_b}(WZe)  
+\mathcal{C}^{u_id_j}_{\nu_a e_b}(WZ\nu)
+\mathcal{C}^{u_id_j}_{\nu_a e_b}(W\gamma e) \,.
\end{align}

The contribution of NP 
of penguin diagrams are given as
\begin{align}
\left[\mathcal{C}^{u_id_j}_{\nu_a e_b}\right]_\text{penguin}^{\text{F3311}}&=
\mathcal{C}^{u_id_j}_{\nu_a e_b}(Y^0  X^\pm  \xi^0)+ \mathcal{C}^{u_id_j}_{\nu_a e_b}(Y^0  X^\pm  \xi) +
\mathcal{C}^{u_id_j}_{\nu_a e_b}(Y^0  X^\pm  E) \nonumber \\
&+\mathcal{C}^{u_id_j}_{\nu_a e_b}(Y^0  X^\pm  U) + \mathcal{C}^{u_id_j}_{\nu_a e_b}(\nu  e  Z')\,. 
\end{align}

where the WCs 
are give as in Eqs.~(\ref{penWC1}) , (\ref{penWC2}):
\begin{align}
\mathcal{C}^{u_i d_j}_{e_a \nu_b}(f_1f_2V)&= \frac{g_4^L}{m_{W}^2}\Big( V_{CKM}\Big)_{ij}   \Big(U_{PMNS}\Big)^\dagger_{ab} \Ga^{f_1^a f_2^b V} 
\label{penWill1} \,, \\
\mathcal{C}^{u_i d_j}_{e_a \nu_b}(V_1 V_2 f)&= \frac{g_4^L}{m_{W}^2}\Big( V_{CKM}\Big)_{ij}   \Big(U_{PMNS}\Big)^\dagger_{ab} \Ga^{V_1 V_2 f}    
\label{WC}
\end{align}

with $g_4^L=\frac{g}{\sqrt{2}}$.

\begin{figure}[ht!] 
\centering
\includegraphics[width=0.3\linewidth]{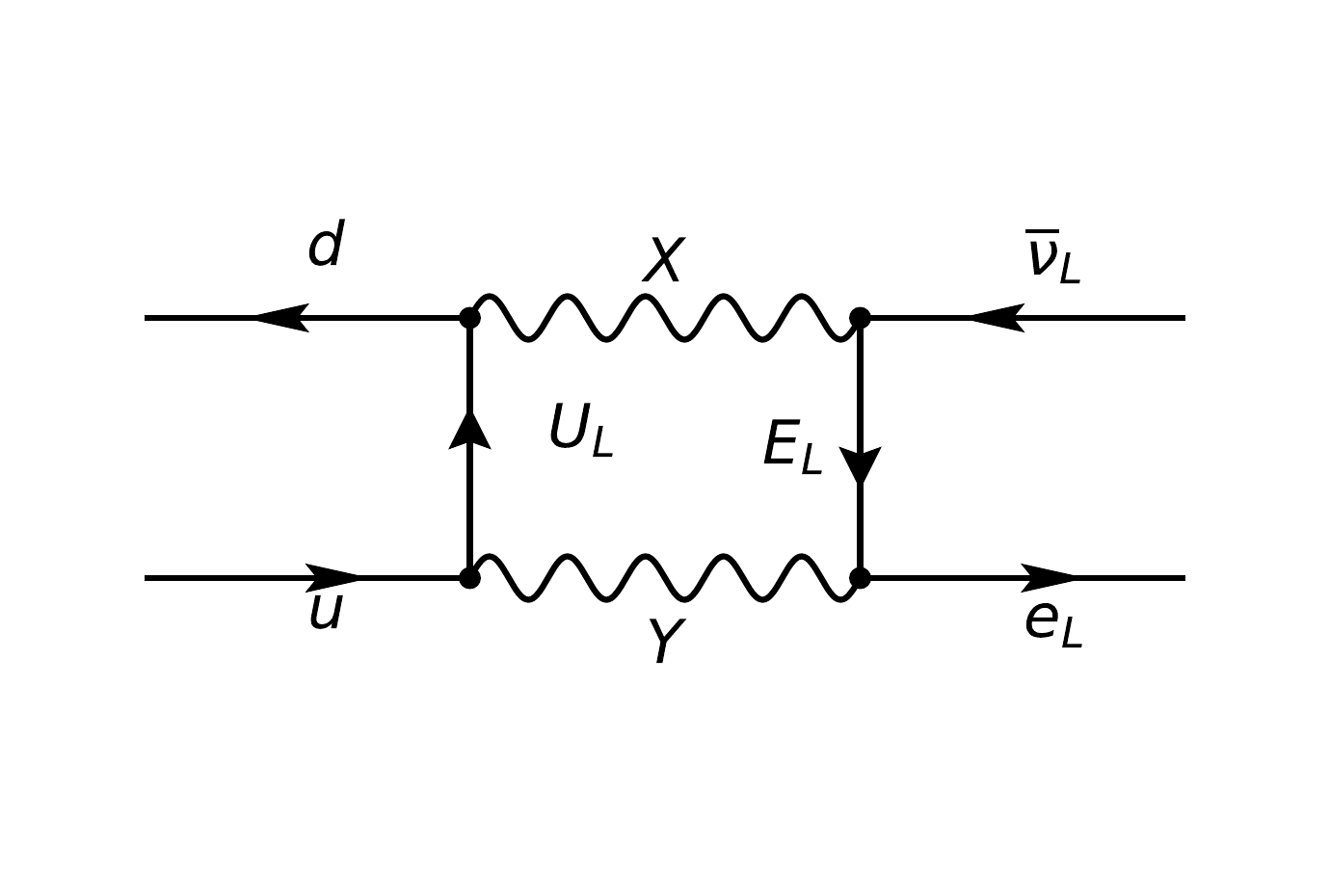} 
\includegraphics[width=0.3\linewidth]{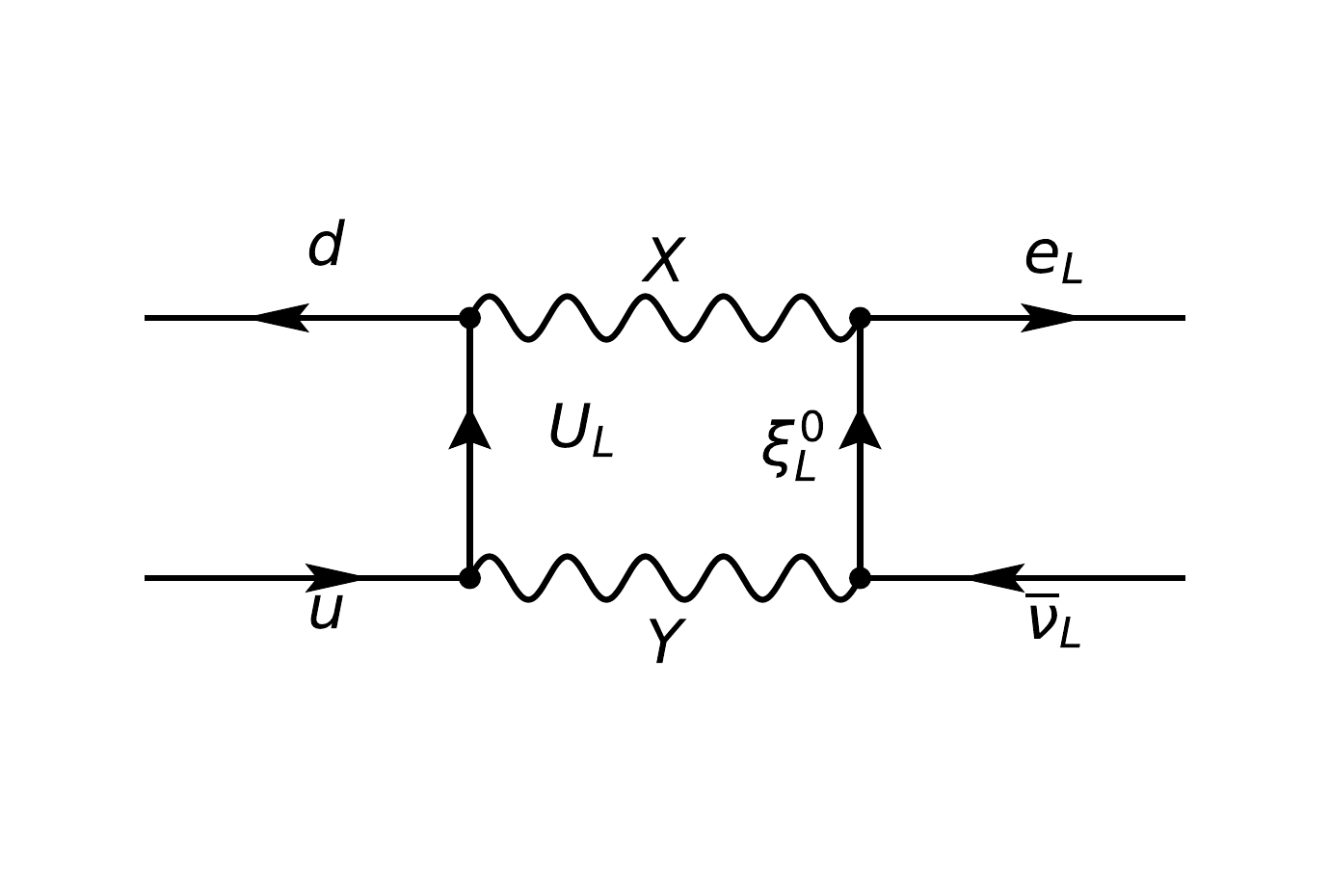} 
\includegraphics[width=0.3\linewidth]{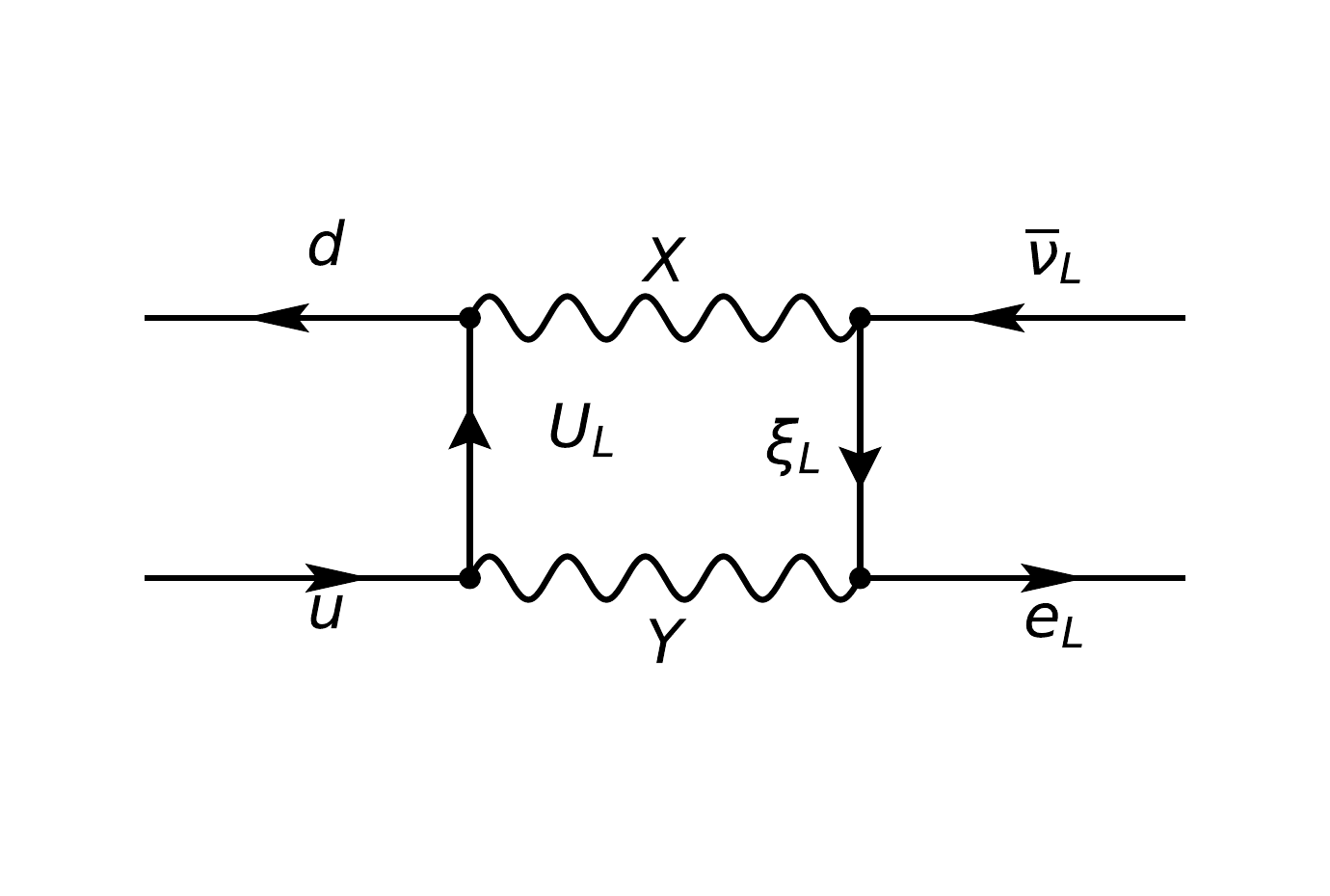} 
\caption{ $F3311$ contributions-box diagrams.}
\label{fig3} 
\end{figure}

The  WCs  of  general box diagram Fig. \ref{Boxdiagrams} are given as in \ref{BoxWC}. Hence The contributions of box diagrams Fig \ref{fig3} are given as 
\begin{align}
\left[\mathcal{C}^{u_id_j}_{\nu_a e_b}\right]_\text{box}^{\text{F3311}}&=\mathcal{C}^{u_id_j}_{\nu_a e_b}(U X^\pm E Y^0 ) + \mathcal{C}^{u_id_j}_{\nu_a e_b}(U X^\pm \xi^0 Y^0 ) + \mathcal{C}^{u_id_j}_{\nu_a e_b}(U X^\pm \xi Y^0 ) \,,
\end{align}

In summary,  we have the contribution to WCs 
of each diagram as followings
\begingroup
\allowdisplaybreaks
\begin{align}
\left[\mathcal{C}^{u_id_j}_{\nu_a e_b}\right]_\text{(1a)}^{e \nu Z} &= \frac{g}{\sqrt{2} m_{W}^2}\Big( V_{CKM}\Big)_{ij}   \Big(U_{PMNS}\Big)^\dagger_{ab}  \nonumber \\
&\times \Ga^{e_b \nu_a Z}(g_{eeZ}^L,g_{\nu \nu Z}^L, g_{\nu e W}^L,g_{\nu e W}^R,   m_{e_a}, m_{\nu_b},m_Z)\,,  \\
\left[\mathcal{C}^{u_id_j}_{\nu_a e_b}\right]_\text{(1b)}^{Z W e} &=\frac{g}{\sqrt{2} m_{W}^2}\Big( V_{CKM}\Big)_{ij}   \Big(U_{PMNS}\Big)^\dagger_{ab}  \nonumber \\
&\times	\Ga^{Z W e}(g_{eeZ}^L,g_{\nu e W}^L,g_{WWZ},m_e,m_{Z}, m_{W}) \,, \\
\left[\mathcal{C}^{u_id_j}_{\nu_a e_b}\right]_\text{(1c)}^{W Z \nu} &=\frac{g}{\sqrt{2} m_{W}^2}\Big( V_{CKM}\Big)_{ij}   \Big(U_{PMNS}\Big)^\dagger_{ab} \nonumber \\ 
&\times	\Ga^{WZ\nu}(g_{\nu e W}^L,g_{\nu \nu Z}^L,g_{WWZ},m_{\nu},m_{Z}, m_{W}) \,, \\
\left[\mathcal{C}^{u_id_j}_{\nu_a e_b}\right]_\text{(1d)}^{\gamma W e} &=\frac{g}{\sqrt{2} m_{W}^2}\Big( V_{CKM}\Big)_{ij}   \Big(U_{PMNS}\Big)^\dagger_{ab}  \nonumber \\
&\times 	\Ga^{\gamma W e}(g_{e e \gamma}^L,g_{\nu e W}^L,g_{WW\gamma},m_{\nu},0, m_{W})\,,  \\
\left[\mathcal{C}^{u_id_j}_{\nu_a e_b}\right]_\text{(2a)}^{X Y \xi^0} &=\frac{g}{\sqrt{2} m_{W}^2}\Big( V_{CKM}\Big)_{ij}   (V_L^\nu)_{1a} (V^l_L)^\dagger_{1b} \nonumber \\ 
&\times	\Ga^{X Y \xi^0}(g_{e \xi^0 X}^L,g_{\xi^0 \nu Y}^L,g_{X Y W},m_{\xi^0},m_X, m_{Y})\,,  \\
\left[\mathcal{C}^{u_id_j}_{\nu_a e_b}\right]_\text{(2b)}^{Y X \xi} &=\frac{g}{\sqrt{2} m_{W}^2}\Big( V_{CKM}\Big)_{ij}   (V_L^\nu)_{1a} (V^l_L)^\dagger_{1b}  \nonumber \\
&\times	\Ga^{Y X \xi}(g_{e \xi Y}^L,g_{\xi \nu X}^L,g_{Y X W},m_{\xi},m_Y, m_{X}) \,, \\
\left[\mathcal{C}^{u_id_j}_{\nu_a e_b}\right]_\text{(2c)}^{Y X E} &=\frac{g}{\sqrt{2} m_{W}^2}\Big( V_{CKM}\Big)_{ij}   \Big(U_{PMNS}\Big)^\dagger_{ab} \nonumber \\
&\times \sum_{c=1}^3	\Ga^{Y X E_c}(g_{e E_c Y}^L,g_{E_c \nu X}^L,g_{Y X W},m_{E_c},m_Y, m_{X}) \,, \\
\left[\mathcal{C}^{u_id_j}_{\nu_a e_b}\right]_\text{(2d)}^{Y X U} &=\frac{g}{\sqrt{2} m_{W}^2} \Big(U_{PMNS}\Big)^\dagger_{ab}
\sum_{k=1}^3   \big(V_L^u (V_L^U)^\dagger \big)_{ik}\big(V_L^U (V_L^d)^\dagger \big)_{kj}   \notag \\
&\times \Ga^{Y X U_k}(g_{u U_k Y}^L,g_{U_k d X}^L,g_{Y X W},m_{U_k},m_Y, m_{X}) \,, \\
\left[\mathcal{C}^{u_id_j}_{\nu_a e_b}\right]_\text{(2e)}^{e \nu Z'} &= \frac{g}{\sqrt{2} m_{W}^2}\Big( V_{CKM}\Big)_{ij}   \Big(U_{PMNS}\Big)^\dagger_{ab}  \nonumber \\
&\times \Ga^{e_b \nu_a Z'}(g_{eeZ'}^L,g_{\nu \nu Z'}^L, g_{\nu e W}^L,g_{\nu e W}^R,   m_{e_a}, m_{\nu_b},m_{Z'})\,, \\
\left[\mathcal{C}^{u_id_j}_{\nu_a e_b}\right]_\text{(3a)}^{EXUY} &=  \sum_{k=1}^3 \sum_{c=1}^3 \big(V^u_L (V_L^U)^\dagger \big)_{ik}
\big(V^U_L (V_L^d)^\dagger \big)_{kj}   
\big(V^l_L (V_L^E)^\dagger \big)_{bc} \big(V^E_L (V_L^\nu)^\dagger \big)_{ca} \nonumber \\
&\times \Ga^{E_c X U_k Y}( g_{e_b E_c Y}^L,g_{E_c \nu_b X}^L, g_{U_k d_j X}^L, g_{u_i U_k Y}^L,   m_{E_c}, m_{X},m_{U_k},m_Y )\,,  \\
\left[\mathcal{C}^{u_id_j}_{\nu_a e_b}\right]_\text{(3b)}^{\xi^0 Y U X} &=  \sum_{k=1}^3  \big(V^u_L (V_L^U)^\dagger \big)_{ik}
\big(V^U_L (V_L^d)^\dagger \big)_{kj}   
\big(V^\nu_L \big)_{a1} \big((V_L^l)^\dagger \big)_{1b} \nonumber \\
&\times \Ga^{\xi^0 X U_k Y}( g_{e_b \xi^0 X}^L,g_{\xi^0 \nu_a Y}^L, g_{U_k d_j X}^L, g_{u_i U_k Y}^L, m_{\xi^0}, m_{X},m_{U_k},m_Y )\,,  \\
\left[\mathcal{C}^{u_id_j}_{\nu_a e_b}\right]_\text{(3c)}^{\xi X U Y} &=  \sum_{k=1}^3  \big(V^u_L (V_L^U)^\dagger \big)_{ik}
\big(V^U_L (V_L^d)^\dagger \big)_{kj}   
\big(V^\nu_L \big)_{a1} \big((V_L^l)^\dagger \big)_{1b} \nonumber \\
&\times \Ga^{\xi X U_k Y}( g_{e_b \xi Y}^L,g_{\xi \nu_a X}^L, g_{U_k d_j X}^L, g_{u_i U_k Y}^L, m_{\xi}, m_{X},m_{U_k},m_Y )\,.
\end{align}
\endgroup

\section{Numerical study of flavor non-universality interactions \label{nonuni2}}
In this section we will numerically investigate  observables 
connected to the flavor non-universality interactions.
\subsection{Parameters setup}

\begin{figure}[ht] 
\centering
\includegraphics[width=.6\linewidth]{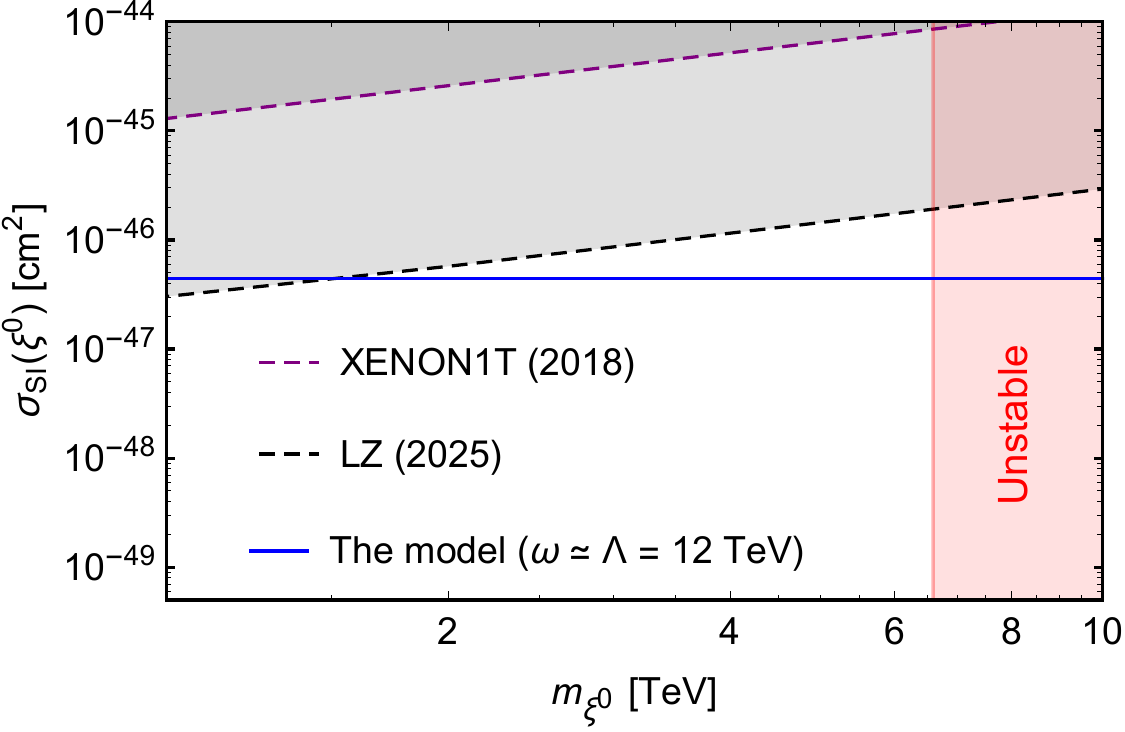} 
\caption{SI DM--nucleon scattering cross-section limit versus DM mass.}
\label{figmxi0} 
\end{figure}

The WCs 
depend on the parameters of the model which are 
the  parameters of the embedded SM model and new mixing parameters  $( \sin \phi,  V_L^{u,d}, V_L^{l,\nu}, V_L^{U,E} ) $ as well as 
new particles masses   $(m_X,m_Y,m_{Z'},m_{E_i},m_{\xi},m_{\xi^0}, m_{U_i} )$. The SM parameters are used as in \cite{ParticleDataGroup:2022pth}. Here, we have used the  recent results on the upper bound of neutrino masses  
$ m_\nu \leq 0.8 eV c^{-2} $ at $90\% CL$  \cite{KATRIN:2021uub}. The numerical values of the CKM matrix are 
taken as in \cite{ParticleDataGroup:2022pth}. The update fit of the numerical values of the PMNS matrix are 
\cite{Esteban:2020cvm}. The new parameters without loss of generality is:

\begin{itemize}
\item $ V_L^u \equiv V_L^U  \equiv V_L^E \equiv Diag(1,1,1)$, \hs $ V_L^\nu  (V_L^l)^\dag \equiv U_{PMNS}$, \hs   $V_L^d \equiv V_{CKM}$\,.
Note that in this work we use the paramterization of $V^l_L$ as in (\ref{VlL}) hence $V^\nu_L=U_{PMNS} \times V^l_L $. 
\item Since the new exotic leptons and quarks are assigned to the same sextet and triplet representations as SM leptons and quarks, it is reasonable to assume that they either follow a similar mass hierarchy (normal or inverse) or exhibit degenerate masses compared to their SM counterparts.
\item The triplet fermion $\xi^0$ is regarded as a DM candidate~\cite{VanLoi:2020xcq}. At tree level, the $\xi$ triplet has a degenerate mass, $m_\xi = -\sqrt{2}\,h_\xi^E \Lambda$. However, loop correction can make the mass of $\xi^\pm $ larger than $\xi^0$ as $m_{\xi^\pm} = m_{\xi^0} + 0.168~\text{GeV}$. Therefore, $\xi^0$ is expected to be the lightest odd particle and hence a viable DM candidate. As shown in Ref.~\cite{VanLoi:2020xcq}, $\xi^0$ with a mass in the TeV range can reproduce the observed relic abundance through annihilation channels mediated by the new Higgs and gauge bosons. Moreover, the mass of $\xi^0$ is constrained by direct detection experiments, as illustrated in Fig.~\ref{figmxi0}. 
In particular, the most stringent bound comes from the LUX-ZEPLIN (LZ) experiment \cite{LZ:2024zvo}, where the excluded parameter space is shaded, while the DM unstable region is shown in light red. From this figure, the allowed mass range of $\xi^0$ is found to be $1.53~\text{TeV} \lesssim m_{\xi^0} \lesssim 6.59~\text{TeV}$, assuming $\omega \simeq \Lambda = 12~\text{TeV}$.
\item The new neutral boson masses are imposed to satisfy the constrained by recent LHC data $m_{Z'}\geq 4500GeV$ \cite{ParticleDataGroup:2022pth}
and the 
model constraints e.g $m_Y^2=m_W^2+m_X^2$.
\end{itemize} 

\subsection{Numerical Analysis}

In our work, the numerical evaluation of Veltman Passarino functions  is  performed using  Collier \cite{Denner:2016kdg} which is a Fortran library to evaluate complex one-loop in Extended Regularizations.

In the following, we will investigate the parameter space according to two cases:

\begin{itemize}
\item Case 1: Degenerate mass hierarchy, where $m_{U_1}=m_{U_2}=m_{U_3}=m_{U}$ and $m_{E_1}=m_{E_2}=m_{E_3}=m_{E}$.
\item Case 2: Normal hierarchy, with $(m_{U_1} \ll m_{U_2} \ll m_{U_3})$ and $(m_{E_1} \ll m_{E_2} \ll m_{E_3})$. While an inverse hierarchy could generally be considered, we focus solely on the normal hierarchy to simplify the problem.
\end{itemize}
\subsection{ b to c  transitions}
The ratios $R(D^{(*)}),R(D)$ for the $F3311$  model  are presented in the form of WCs 
\cite{Boucenna:2016qad} as
\begin{align}
R(D^{(*)})&=\frac{\Gamma(B\rightarrow D^{(*)}\tau \bar{\nu})}{\Gamma(B\rightarrow D^{(*)}l \bar{\nu})}=\frac{\sum_i |\mathcal{C}_{3i}^{cb}|^2}{\sum_i (|\mathcal{C}_{1i}^{cb}|^2+|\mathcal{C}_{2i}^{cb}|^2)} \times \left[ \frac{\sum_i (|\mathcal{C}_{1i}^{cb}|^2+|\mathcal{C}_{2i}^{cb}|^2)}{\sum_i |\mathcal{C}_{3i}^{cb}|^2} \right]_{SM} \nonumber \\
&\times R(D^{(*)})_{SM}\,,\\
R(X_c)&=\frac{\Gamma(B\rightarrow X_c\tau \bar{\nu})}{\Gamma(B\rightarrow X_c l\bar{\nu})}=\frac{\sum_i |\mathcal{C}_{3i}^{cb}|^2}{\sum_i |\mathcal{C}_{1i}^{cb}|^2} \times \left[ \frac{\sum_i |\mathcal{C}_{1i}^{cb}|^2}{\sum_i |\mathcal{C}_{3i}^{cb}|^2} \right]_{SM} \times R(X_c)_{SM}\,, 
\end{align}

where $i$ is the generation index of 
leptons.

\begin{figure}
\centering     
\subfigure[]{\label{RDFig}\includegraphics[width=60mm]{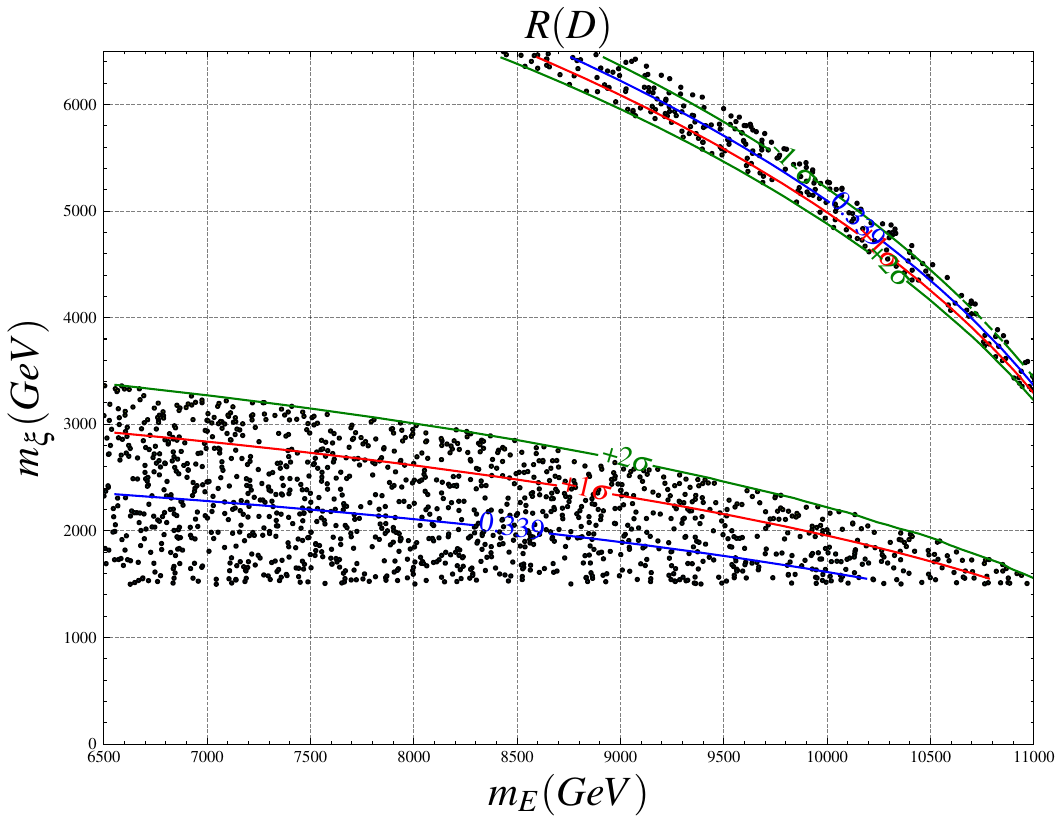}}
\subfigure[]{\label{RDSFig}\includegraphics[width=60mm]{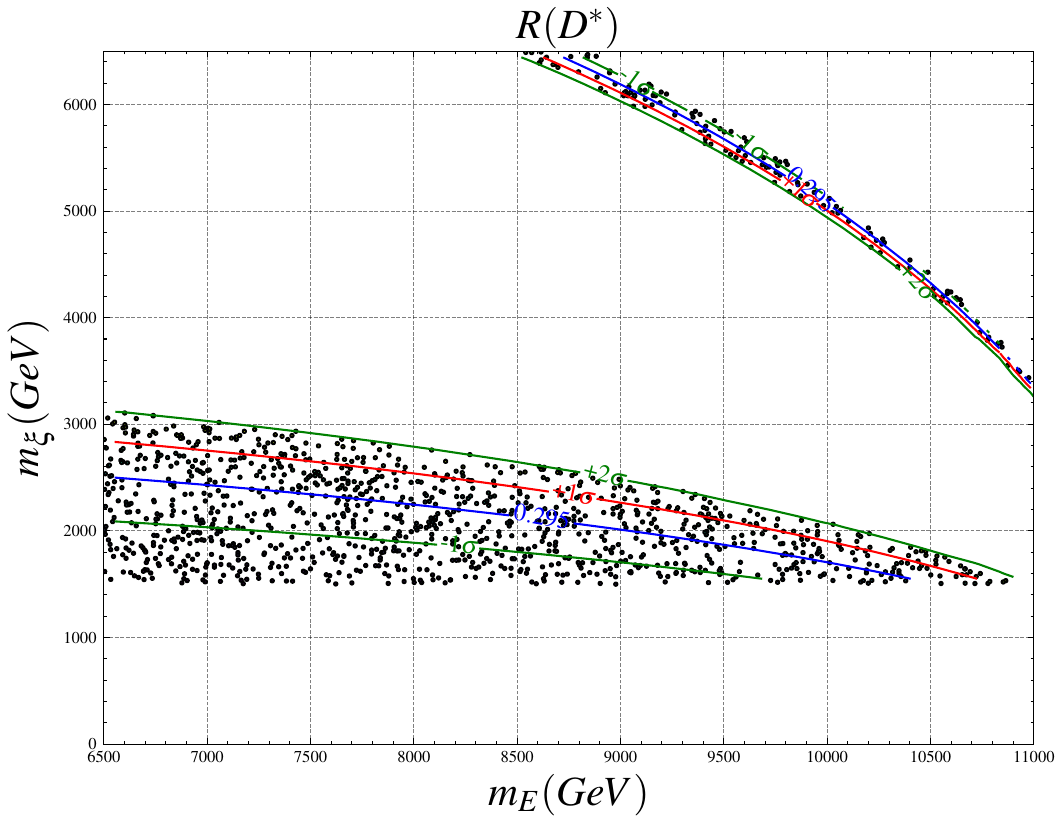}}
\caption{The ratio $R(D),R(D^{(*)})$ as function of exotic mass, contour lines (red, green ) are $1\sigma , 2\sigma$ of the average value blue line }
\label{figRDRDS} 
\end{figure}

We recall the conditions of the lepton $\xi$ imposed by dark matter constraint \cite{VanLoi:2020xcq}, specify that  $\xi$ is considered  as  dark matter candidate and LUX-ZEPLIN (LZ) experiment data limits  the mass of $\xi$: $ 1.53 \leq m_\xi \leq 6.59 $ TeV. To ensure that $\xi$ is stable and does not decay into SM particles, the masses of the exotic leptons ($E_i$) and exotic quarks ($U_i$) must be greater than that of $\xi$.

In FIG.\ref{RDFig} we investigate $R(D)$ as the function of  the mass of $(m_E, m_\xi)$ with     the mass of the exotic quark fixed at the scale of the model. The experiment bounds  $\mathcal{R}(D)^\pm_{exp} =0.339 \pm 0.02952$ are used to find the allowed region for the mass of the exotic leptons within $\pm 2\sigma$ of experiment average value. In the absence of a mass hierarchy, the operating region for the masses of the exotic leptons is primarily in the ranges $m_E \in [6.5,11]TeV$ and $m_\xi \in [1.53, 3.4]TeV$.

It is noteworthy that no parameter region exists where both the mass hierarchy and the condition  $m_\xi \leq m_E$ are simultaneously satisfied for  $R(D)$.  This is because, to preserve the mass hierarchy, the mass of one generation must be at least ten times that of another. Let $m_E$ represent the mass of the first generation, and assume the mass of the third generation is $m_{E_3}=10^2 \times m_E$. Since the lightest exotic lepton has to be heavier than $m_\xi$ ( for example $m_E\geq m_\xi$), then the mass hierarchy implies $m_{E_3}\geq 10^2 m_\xi=10^2 \times 1TeV$. Given $m_\xi$ is on the order of 1 TeV, this results in $m_{E_3}\geq 10^2 TeV$,  which is significantly above the model scale of 12 TeV.

Similarly, in FIG.\ref{RDSFig} we analyze the ratio $R(D^{(*)})$ in the model $F3311$ and compare it with the experimental value  $\mathcal{R}(D^*)_{exp} =0.295 \pm 0.0148661 $. To determine the experimental bounds,  we consider region $\pm 2\sigma$ of experimental mean value. Using these values, we  identify the allowed parameter space for the model, specifically the mass of the exotic leptons $m_E$ and  $m_\xi$. The allowed region are approximately $m_\xi \in [1.53,3]TeV$ and $m_E \in [6.5,11]TeV$.

\begin{figure}
\centering     
\includegraphics[width=80mm]{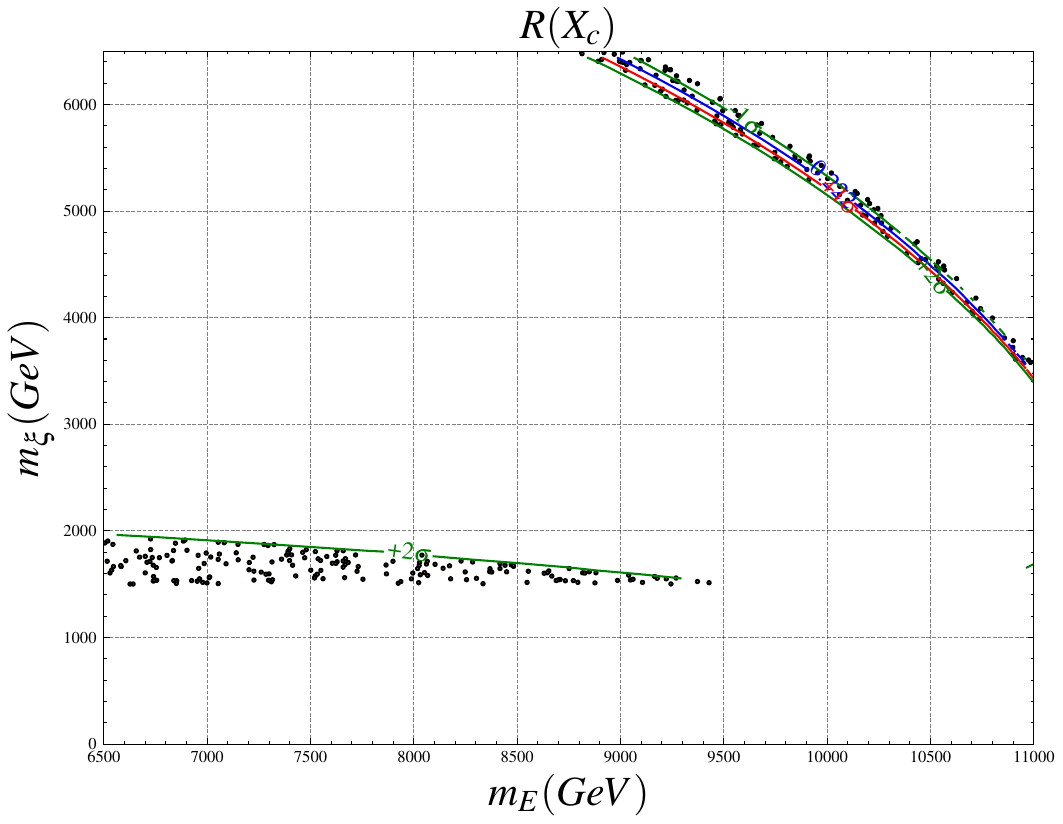}
\caption{The ratio $R(X_c)$ in $F3311$ model with experiment value  $R(X_c)=0.22\pm 0.022$, contour lines (red, green ) are $1\sigma , 2\sigma$ of the average value blue line.}
\label{RXcFig} 
\end{figure}

Finally, to further analyze the   $b\rightarrow c$ transitions, we evaluate the observable $R(X_c)$ within  the model and compare it to the experimental data. In FIG.\ref{RXcFig}, the region where the mass $m_E\in [6.5,11]TeV$ and $m_\xi \in [1.53,2]TeV$, the predicted  the value of $R(X_c)$  is consistent with the  experiment value $R(X_c)=0.22\pm 0.022$.

\subsection{d to u  transitions}
It is important to examine the  flavor non-universality of the $F3311$ model across different quark transitions. In this section, we focus on  the flavor non-universality in the decay process $d \rightarrow u l \bar{\nu}$ corresponding to $\pi \rightarrow l \bar{\nu}$. To minimize the dependence of the combination of $G_F|V_{ud}|$, we analyze the ratios $\frac{\Ga
(\pi \rightarrow \mu \bar{\nu})}{\Ga(\pi \rightarrow e \bar{\nu})}$ and $\frac{\Ga
(\tau \rightarrow \pi \nu)}{\Ga(\pi \rightarrow e \bar{\nu})}$. The experiment values for these two ratios  are taken from  \cite{Boucenna:2016qad}.
\begin{equation}
\Big[ \frac{\Ga
	(\pi \rightarrow \mu \bar{\nu})}{\Ga(\pi \rightarrow e \bar{\nu})}\Big]_{exp}=8.13(3)\times 10^3, \hs
\Big[ \frac{\Ga
	(\tau \rightarrow \pi \nu)}{\Ga(\pi \rightarrow e \bar{\nu})}\Big]_{exp}=7.90(5)\times 10^7 
	\end{equation}
	
	The SM's predictive values of these two ratios are given as \cite{Boucenna:2016qad,ParticleDataGroup:2022pth}
	\begin{equation}
\Big[ \frac{\Ga
	(\pi \rightarrow \mu \bar{\nu})}{\Ga(\pi \rightarrow e \bar{\nu})}\Big]_{SM}=8.096(1)\times 10^3, \hs
\Big[ \frac{\Ga
	(\tau \rightarrow \pi \nu)}{\Ga(\pi \rightarrow e \bar{\nu})}\Big]_{SM}=7.91(1)\times 10^7 
	\end{equation}
	The two ratios, expressed in terms of Wilson coefficient are:
	\begin{align}
\frac{\Ga
	(\pi \rightarrow \mu \bar{\nu})}{\Ga(\pi \rightarrow e \bar{\nu})}&=\frac{\sum_i |\mathcal{C}^{ud}_{2i}|^2}{\sum_i |\mathcal{C}^{ud}_{1i}|^2} \times
\left[ \frac{\sum_i |\mathcal{C}^{ud}_{1i}|^2}{\sum_i |\mathcal{C}^{ud}_{2i}|^2}\right]_{SM} \times \Big[ \frac{\Ga
	(\pi \rightarrow \mu \bar{\nu})}{\Ga(\pi \rightarrow e \bar{\nu})}\Big]_{SM} \\
\frac{\Ga
	(\tau \rightarrow \pi \nu)}{\Ga(\pi \rightarrow e \bar{\nu})}&=\frac{\sum_i |\mathcal{C}^{ud}_{3i}|^2}{\sum_i |\mathcal{C}^{ud}_{1i}|^2} \times
\left[ \frac{\sum_i |\mathcal{C}^{ud}_{1i}|^2}{\sum_i |\mathcal{C}^{ud}_{3i}|^2}\right]_{SM} \times	\Big[ \frac{\Ga
	(\tau \rightarrow \pi \nu)}{\Ga(\pi \rightarrow e \bar{\nu})}\Big]_{SM}
	\end{align}

\begin{figure}
\centering     
\subfigure[]{\label{pimunubarFig}\includegraphics[width=60mm]{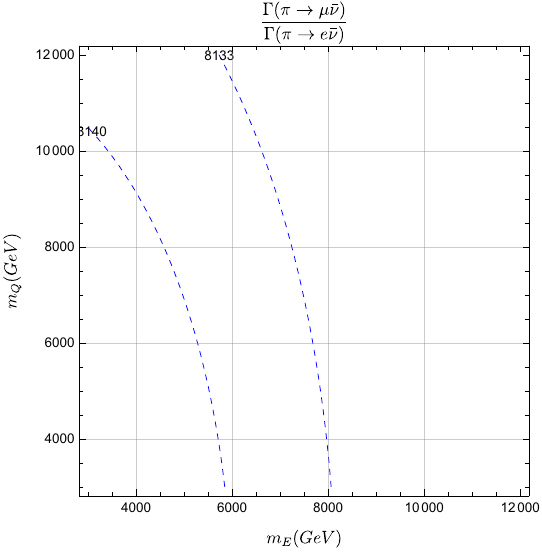} }
\subfigure[]{\label{taupinuFig}\includegraphics[width=60mm]{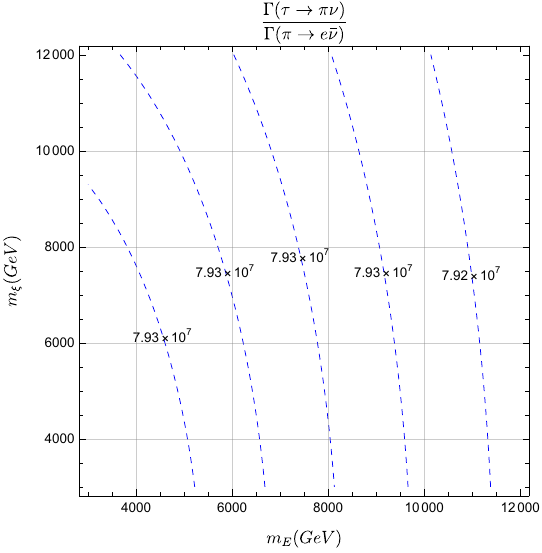} }
\caption{ Contour plot of the ratio $\frac{\Ga
		(\pi \rightarrow \mu \bar{\nu})}{\Ga(\pi \rightarrow e \bar{\nu})}$  Fig\ref{pimunubarFig} with experiment value $\Big[ \frac{\Ga
		(\pi \rightarrow \mu \bar{\nu})}{\Ga(\pi \rightarrow e \bar{\nu})}\Big]_{exp}=8.13(3)\times 10^3$  and contour plot of the ratio $\frac{\Gamma(\tau \rightarrow \pi \nu)}{\Gamma(\pi \rightarrow e \bar{\nu})}$ Fig\ref{taupinuFig} with experiment value $\Big[ \frac{\Ga
		(\tau \rightarrow \pi \nu)}{\Ga(\pi \rightarrow e \bar{\nu})}\Big]_{exp}=7.90(5)\times 10^7 $}
\label{d2uFig} 
\end{figure}

	These two ratios are represented as in FIG.\ref{pimunubarFig} and FIG.\ref{taupinuFig}.  The parameter range for  dark matter candidate $m_\xi$ is restricted to be $m_\xi \in [1.53,3.0]$TeV, while  no mass hierarchy is assumed,  the mass of the exotic lepton is $m_E \leq 8$TeV and the mass of exotic quark $m_U$, within the scale of the model, will give values of the two ratios align with experiment data.

\subsection{s to u  transitions}

We next will study the non-universality of flavor in the decays of $K^+\rightarrow \pi^0 l^+ \nu$,$K\rightarrow l \nu$, $\tau \rightarrow K \nu$. For our purpose, we consider ratios
\begin{align}
\frac{\Ga(K \rightarrow \mu \bar{\nu})}{\Ga(K \rightarrow e \bar{\nu})} &=\frac{\sum_k |\mathcal{C}^{us}_{2k}|^2}{\sum_k |\mathcal{C}^{us}_{1k}|^2} \times   \left[ \frac{\sum_k |\mathcal{C}^{us}_{1k}|^2}{\sum_k |\mathcal{C}^{us}_{2k}|^2} \right]_{SM} \times
\left[ \frac{\Ga(K \rightarrow \mu \bar{\nu})}{\Ga(K \rightarrow e \bar{\nu})} \right]_{SM}\,,   \\
\frac{\Ga(\tau \rightarrow K \nu)}{\Ga(K \rightarrow e \bar{\nu})} &=\frac{\sum_k |\mathcal{C}^{us}_{3k}|^2}{\sum_k |\mathcal{C}^{us}_{1k}|^2} \times   \left[ \frac{\sum_k |\mathcal{C}^{us}_{1k}|^2}{\sum_k |\mathcal{C}^{us}_{3k}|^2} \right]_{SM} \times
\left[ \frac{\Ga(\tau \rightarrow K \nu )}{\Ga(K \rightarrow e \bar{\nu})} \right]_{SM}\,,\\ 
\frac{\Ga(K^+ \rightarrow \pi^0\bar{\mu} \nu)}{\Ga(K^+ \rightarrow \pi^0\bar{e} \nu)} &=\frac{\sum_k |\mathcal{C}^{us}_{2k}|^2}{\sum_k |\mathcal{C}^{us}_{1k}|^2} \times   \left[ \frac{\sum_k |\mathcal{C}^{us}_{1k}|^2}{\sum_k |\mathcal{C}^{us}_{2k}|^2} \right]_{SM} \times
\left[ \frac{\Ga(K^+ \rightarrow \pi^0\bar{\mu} \nu)}{\Ga(K^+ \rightarrow \pi^0\bar{e} \nu)} \right]_{SM}\,.
\end{align}

The experimental values of these ratios are given as in \cite{ParticleDataGroup:2022pth}
\begin{align}
\left[ \frac{\Ga(K \rightarrow \mu \bar{\nu})}{\Ga(K \rightarrow e \bar{\nu})}\right]_{exp}&=4.018(3)\times 10^4,
\left[\frac{\Ga(K^+ \rightarrow \pi^0\bar{\mu} \nu)}{\Ga(K^+ \rightarrow \pi^0\bar{e} \nu)}  \right]_{exp}=0.660(3),\\
\left[\frac{\Ga(\tau \rightarrow K \nu)}{\Ga(K \rightarrow e \bar{\nu})} \right]_{exp}&=1.89(3)\times 10^7 \,,
\end{align}
and the SM predicted value are \cite{Boucenna:2016qad}
\begin{align}
\left[ \frac{\Ga(K \rightarrow \mu \bar{\nu})}{\Ga(K \rightarrow e \bar{\nu})}\right]_{SM}&=4.0037(2)\times 10^4,
\left[\frac{\Ga(K^+ \rightarrow \pi^0\bar{\mu} \nu)}{\Ga(K^+ \rightarrow \pi^0\bar{e} \nu)}  \right]_{SM}=0.663(2),\\
\left[\frac{\Ga(\tau \rightarrow K \nu)}{\Ga(K \rightarrow e \bar{\nu})} \right]_{SM}&=1.939(4)\times 10^7 \,.
\end{align}

\begin{figure}[ht!] 
\centering
\includegraphics[width=0.3\linewidth]{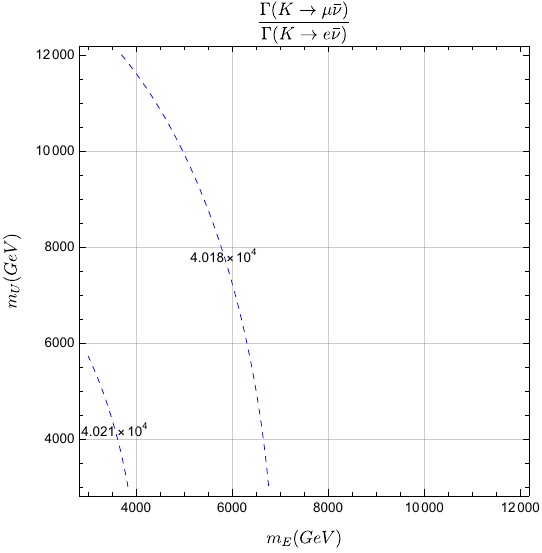} 
\includegraphics[width=0.3\linewidth]{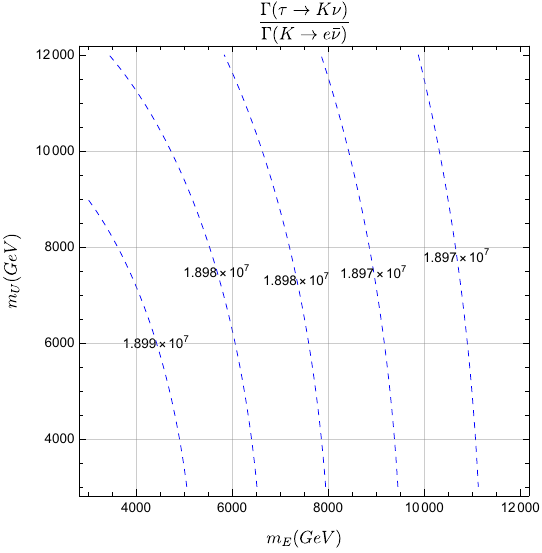} 
\includegraphics[width=0.3\linewidth]{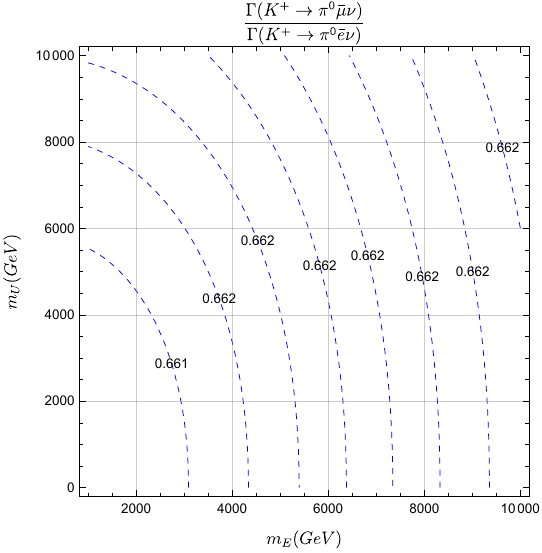}

\caption{Contour plot of the ratio $\frac{\Gamma(K \rightarrow \mu \bar{\nu})}{\Gamma(K \rightarrow e \bar{\nu})}$, $\frac{\Ga(\tau \rightarrow K \nu)}{\Ga(K \rightarrow e \bar{\nu})}$ and $\frac{\Ga(K^+ \rightarrow \pi^0\bar{\mu} \nu)}{\Ga(K^+ \rightarrow \pi^0\bar{e} \nu)}$ in $F3311$ model on $m_E-m_U$ plane. All contour lines are  consistent with experiment data }
\label{stouFig} 
\end{figure}

Similar to previous analysis, we examine these ratio to obtain the constraints on the mass of the new exotic particles Fig. \ref{stouFig}.
Our constrains on the mass of exotic leptons are consistent with previous allowed parameter space of the model. With $m_\xi \in [1.53,3.0]TeV$  and the mass of the exotic lepton $m_E \leq 7$ TeV and $m_U \in [6,12]TeV$  the $F3311$ model gives 
predictions  align with experiment results.

In summary, using  experiment data for $b\rightarrow c, d\rightarrow u, s\rightarrow u$ transitions, combine with the LUX-ZEPLIN (LZ) experiment data we have identify the working region  for  exotic quarks and leptons $U,E$ of the F3311 mdoel: $m_E \in [6.5, 9 ]TeV$ and $m_Q \in [6,11]TeV$, the mass of dark matter candidate in the F3311 model is narrowed to $m_\xi \in [1.53, 2]TeV$    

\section{Conclusion \label{Conclu}}

By extending the F331 model with an additional $U(1)_N$ gauge group, the F3311 model, after symmetry breaking, retains a residual symmetry characterized by a charge $W_P$. The new fermions carry odd $W_P$
charge, which enforces pairwise couplings in order to conserve $W_P$. The lightest of these new fermions are stabilized by this symmetry and thus can be considered as Dark Matter (DM) candidates. Since one generation of leptons transforms as a sextet, the F3311 model exhibits flavor-lepton violating (FLV) interactions at tree level, and because the lepton generations couple differently to the gauge bosons, the model has the potential to accommodate lepton universality violation.

In this work, we have studied FLV within the F3311 framework. We analyze FLV decays of the Z boson at tree level and the flavor-conserving Z decays at one loop. Additionally, we investigate the decays $l_i \rightarrow l_j \gamma$ and leptonic three-body decays. Using the most stringent experimental limits on these decays, we derive bounds on the Z-Z' mixing angle, finding $\sin \phi \leq 0.0015$ and $m_{Z'}\geq 3.2 TeV$.

Incorporating dark matter constraints, we have re-evaluated the constraints on the mass of DM candidate of the model using the LUX-ZEPLIN (LZ) experiment data. We then examine the charged current anomaly within the F3311 model. We compute the generic one-loop contribution to the decay process  $u_i\rightarrow d_j e_b \bar{\nu}_a$ and perform a detailed numerical analysis of observables in the transitions  $b\rightarrow c$,$s \rightarrow u$, $d\rightarrow u$.  We show that the model can address the 
$3.3 \sigma $ tension between the SM predictions and experimental observables  and identify the allowed region for the masses of exotic quarks and leptons. In the absence of a mass hierarchy, the DM candidate mass  range is narrowed down to 
$ 1.53 \leq m_\xi \leq 2.0$ TeV, while the mass of the  exotic leptons and quarks are of  several TeVs: $m_E \in [6.5, 9 ]TeV$ and $m_Q \in [6,11]TeV$.


\section*{Acknowledgments}
This research is funded by Vietnam National Foundation for Science and Technology Development (NAFOSTED) under grant number 103.01-2023.50.
\appendix

\section{Notations for Passarino-Veltman functions}

\subsection{General notation}

Package-X \cite{Patel:2015tea,PATEL201766} calculates dimensionally regulated $(d=4-2\epsilon)$ one loop tensor rank-P  up to 4-points integrals in the form

\begin{align}
&T_N^{\mu_1 ... \mu_P}(p1,...,p_N;m_0,m_1,...,m_N)=  \nonumber \\
& \mu^{2\epsilon}\int \frac{d^dk}{(2\pi)^d} \frac{k^{\mu_1}...k^{\mu_P}}{[k^2-m_0^2+i\epsilon][(k+p_1)^2-m_1^2+i\epsilon]...[(k+p_N)^2-m_N^2+i\epsilon]}
\end{align}

The scalar PV-functions are defined consistent with \cite{Denner:2005nn} where the evaluation and decomposition are :

\begin{align}
A_0(m_0)&=m_0^2\left[\Delta + \ln{\left(\frac{\mu^2}{m_0^2}\right)}   \right] \nonumber \\
B^\mu&=p_1^\mu B_1 \nonumber \\
B^{\mu \nu}&=p_1^\mu p_1^\nu B_{11} +g^{\mu \nu} B_{00} \nonumber \\
C^\mu &= \sum_{i_1=1}^2 p_{i_1}^\mu C_{i_1} \nonumber \\
C^{\mu \nu }&=\sum_{i_1=1, i_2=1}^2 p_{i_1}^\mu p_{i_2}^\mu C_{i_1 i_2} +g^{\mu \nu} C_{00} \nonumber \\
D^\mu &= \sum_{i_1=1}^2 p_{i_1}^\mu D_{i_1} \nonumber \\
D^{\mu \nu }&=\sum_{i_1=1, i_2=1}^2 p_{i_1}^\mu p_{i_2}^\mu D_{i_1 i_2} +g^{\mu \nu} D_{00} \nonumber \\	
\label{Reduction}
\end{align}

Here $\Delta$ is the one loop divergence

\begin{align}
\Delta=\frac{2}{4-d} -\gamma_E +\ln{(4\pi)}
\end{align}

where d is space-time dimensionality and $\gamma_E$ is the Euler's constant. Frampton:1992wt
For the purpose of the work we only use the decomposition of up to rank 2 tensor and up to 4 point integrals. All the PV functions then can be numerically evaluate using Collier \cite{Denner:2016kdg} or LoopTools \cite{HAHN1999153}.

\section{PV Calculation}

\subsection{Penguin diagrams}

\begin{figure}[ht] 
\includegraphics[width=0.3\linewidth]{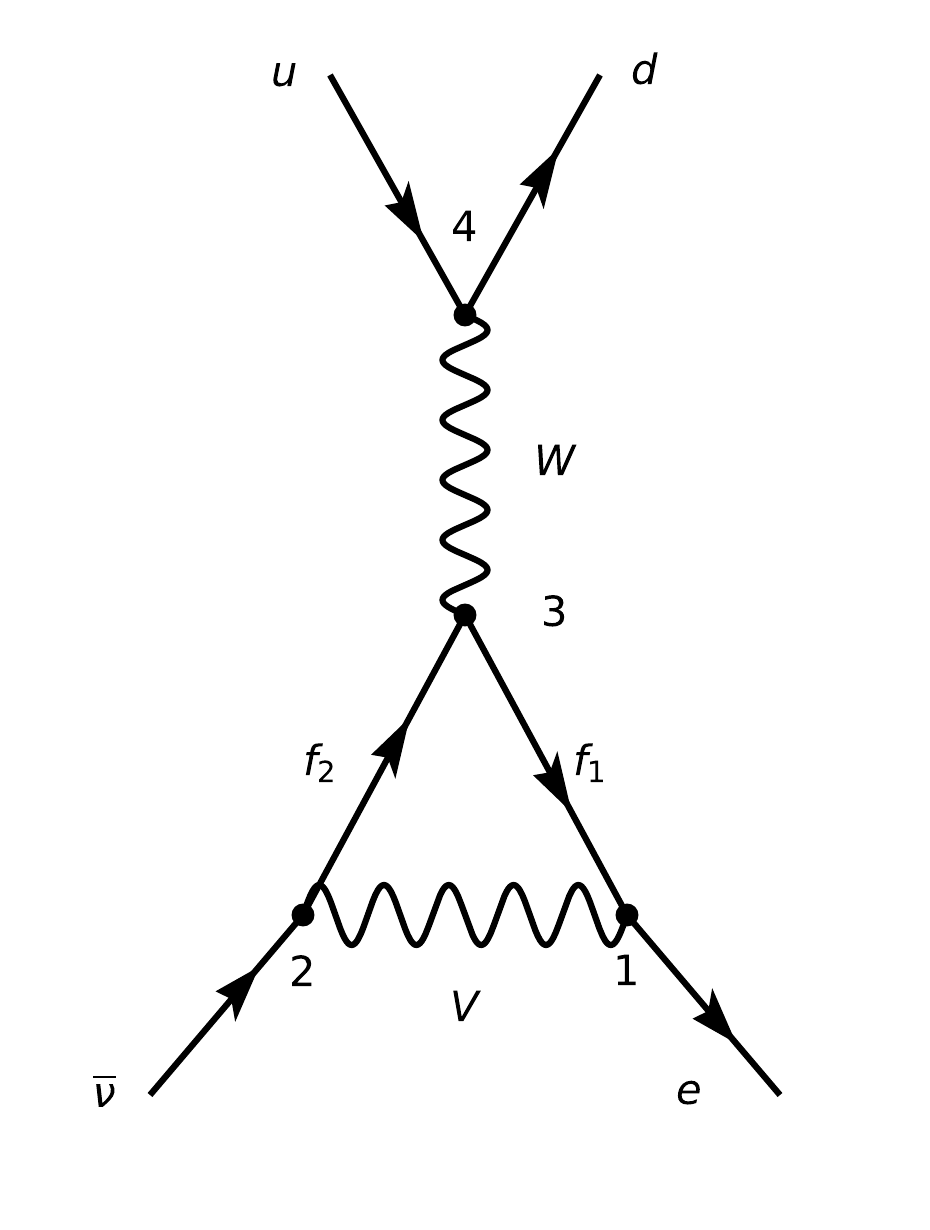} 
\includegraphics[width=0.3\linewidth]{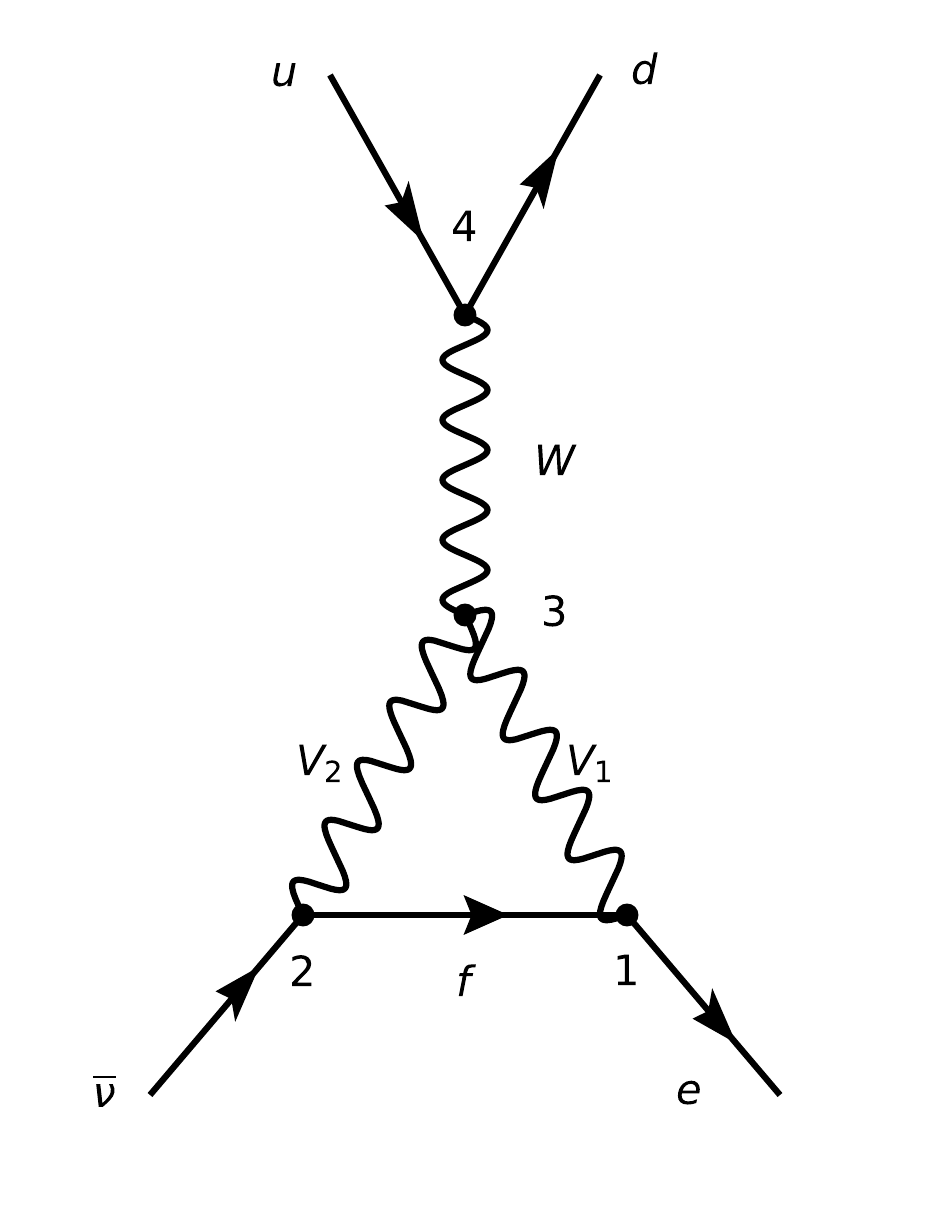} 
\includegraphics[width=0.3\linewidth]{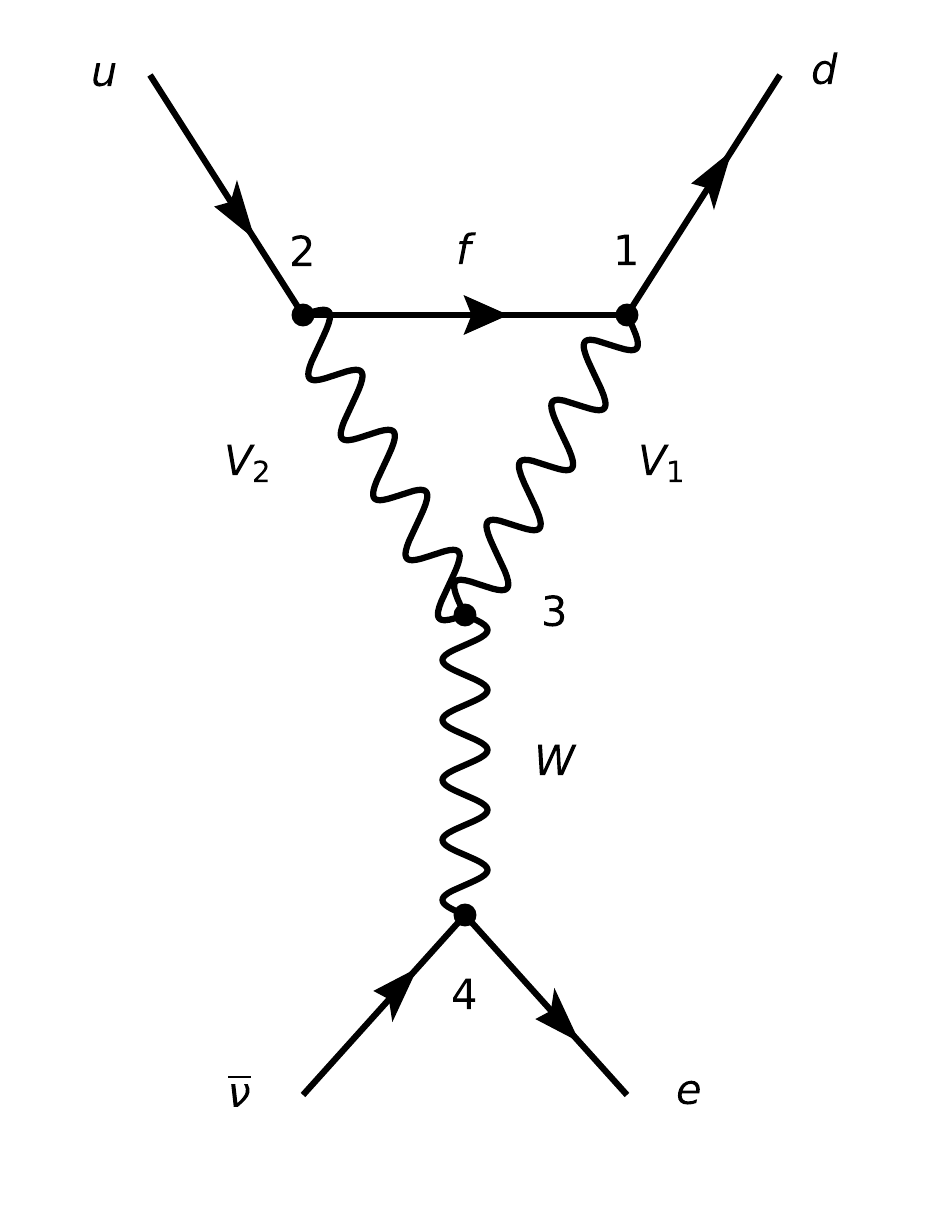} 

\caption{Generic penguin diagrams}
\label{genericpenguinfig} 
\end{figure}

The penguin diagrams in Fig.\ref{fig1}, Fig.\ref{fig2} can be classified into 3 generic diagrams as in Fig.\ref{genericpenguinfig}.
In the zero external momenta, internal momentum of vector boson $V_1$ is independent of the loop momentum. The propagator of the W boson can be approximated as $1/m_W^2$ The numerator of the amplitude of the diagrams can be written as:

\begin{align}
&I_{a}^{Num} =  \big( V_{CKM}\big)_{ij} \times\big[\bar{d}'^j  g_4\ga_{\mu} P_L u'^i \big]  \nonumber \\
&(V^l_L)_{bm}(V^{f_1}_L)^\dagger_{mc}(V^{f_1}_L)_{cn}(V^{f_2}_L)_{nd}^\dagger (V^{f_2}_L)_{dl}(V^{\nu}_L)_{la}^\dagger \left( g_{\al \beta} -\frac{k_\al k_\beta}{m_V^2} \right) \nonumber \\
&\big[\bar{e}'^b_L   \ga_{\al}(g_1^L P_L+g_1^R P_R)(\slashed{k}+ m_{f_{1}^c})
\ga^{\mu}(g_3^L P_L+g_3^R P_R)(\slashed{k}+ m_{f_{2}^d})\ga^{\al}(g_2^L P_L+g_2^R P_R) \nu_L'^a \big] \nonumber \\
\end{align}

\begin{align}
&I_{b}^{Num} =  \big( V_{CKM}\big)_{ij} \big( U_{PMNS}\big)_{ab} \times\big[\bar{d}'^j  g_4\ga_{\mu} P_L u'^i \big]  \left(g_{\al \al_1} -\frac{k_\al k_{\al_1}}{m_{V_1}^2}\right)    \left(g_{\beta \beta_1} -\frac{k_\beta k_{\beta_1}}{m_{V_2}^2}\right)  \nonumber \\
&\big[ \Ga^{VVV}(\mu,\beta_1,\al_1,0,-k,k) \times \bar{e}'^b_L   \ga_{\al}(g_1^L P_L+g_1^R P_R)(\slashed{k}+ m_{f})
\ga_{\beta}(g_2^L P_L+g_2^R P_R) \nu_L'^a \big] 
\end{align}

\begin{align}
& I_{c}^{Num} =  \big( V_{CKM}\big)_{ij} \big( U_{PMNS}\big)_{ab} \times\big[\bar{e}'^b  g_4\ga_{\mu} P_L \nu'^a \big]\left(g_{\al \al_1} -\frac{k_\al k_{\al_1}}{m_{V_1}^2}\right)    \left(g_{\beta \beta_1} -\frac{k_\beta k_{\beta_1}}{m_{V_2}^2}\right)  \nonumber \\
&\big[ \Ga^{VVV}(\mu,\beta_1,\al_1,0,-k,k) \times \bar{d}'^j_L   \ga_{\al}(g_1^L P_L+g_1^R P_R)(\slashed{k}+ m_{f})
\ga_{\beta}(g_2^L P_L+g_2^R P_R) u_L'^i \big] 
\end{align}

where the vetex of 3 gauge coupling is\\
$ \Gamma^{VVV} (\mu ,\alpha ,\beta ,k_1,k_2,k_3)= -i\frac{g}{\sqrt{2}}\big[( {k_1}-k_{2})_\beta g_{\alpha\mu }+ ({k_2}-k_{3})_\mu g_{\alpha\beta } + ({k_3}-k_{1})_\al g_{\beta\mu }\big]$ .

The amplitude is then be written as:

\begin{align}
M_{fi}=	\int \frac{d^4k}{(2\pi)^4}   \frac{1}{m_W^2}\frac{I_{1a}^{Num}}{(k^2-m_{V}^2)(k^2-m_{f_{1a}}^2)(k^2-m_{f_{2b}}^2)}
\end{align}

The $\bar{f}fV^\mu$ vertex interaction with left right coupling constants (Table \ref{coupling2}) is given as 
\begin{align}
L=\bar{f}\ga^\mu(g_L P_L + g_R P_R)fV^\mu
\end{align}
If $V^\mu$ is $Z,Z'$ boson we have $(V-A)$ current then
\begin{align}
g_L&= -\frac{g}{2c_W}(g_V^f+g_A^f) \\
g_R&= -\frac{g}{2c_W}(g_V^f-g_A^f) 
\end{align}

In case of charged current $V^\mu$ is the charged boson
we have
\begin{equation}
g_L=g_R, \hspace{1cm} g_L=\frac{g}{\sqrt{2}}
\end{equation}

The amplitude for penguin diagrams  can be written as:

\begin{align}
i \mathcal{ M}_V^{(a),(b)}&=\frac{g_4}{m_{W}^2}\Big( V_{CKM}\Big)_{ij}   \Big(U_{PMNS}\Big)^\dagger_{ab} [\bar{d}_j'\ga^\mu P_L u'_i]\times [\bar{e}'_b\Ga_\mu P_L \nu'_a] \\
i \mathcal{M}_V^{(c)}&=\frac{g_4}{m_{W}^2}\Big( V_{CKM}\Big)_{ij}   \Big(U_{PMNS}\Big)^\dagger_{ab}[\bar{d}_j'\ga^\mu P_L u'_i] \times [\bar{e}'_b\ga_\mu P_L \nu'_a] 
\end{align}

The effective Hamiltonian can be written as 

\begin{align}
\mathcal{H}=\mathcal{C}^{u_i d_j}_{e_a \nu_b}  [\overline{d}'_i\ga_\mu P_L u'_j]  [\bar{e}'_b\ga_\mu P_L \nu'_a]
\end{align}
The Wilson coefficients for each class of diagram are given as:

\begin{align}
\mathcal{C}^{u_i d_j}_{e_a \nu_b}(f_1f_2V)&= \frac{g_4^L}{m_{W}^2}\Big( V_{CKM}\Big)_{ij}   \Big(U_{PMNS}\Big)^\dagger_{ab} \Ga^{f_1^a f_2^b V} 
\label{penWC1}  \\
\mathcal{C}^{u_i d_j}_{e_a \nu_b}(V_1 V_2 f)&= \frac{g_4^L}{m_{W}^2}\Big( V_{CKM}\Big)_{ij}   \Big(U_{PMNS}\Big)^\dagger_{ab} \Ga^{V_1 V_2 f}   \label{penWC2}
\end{align}

where  the vertex functions $\Ga^{f_1^a f_2^b V}$,$\Ga^{V_1 V_2 f}$   are defined as in Appendix (\ref{ffVVertex}) and (\ref{VVfVertex}),(\ref{VgammafVertex}) respectively.

\subsection{Box diagrams}

\begin{figure}[ht!] 
\centering
\includegraphics[width=.5\linewidth]{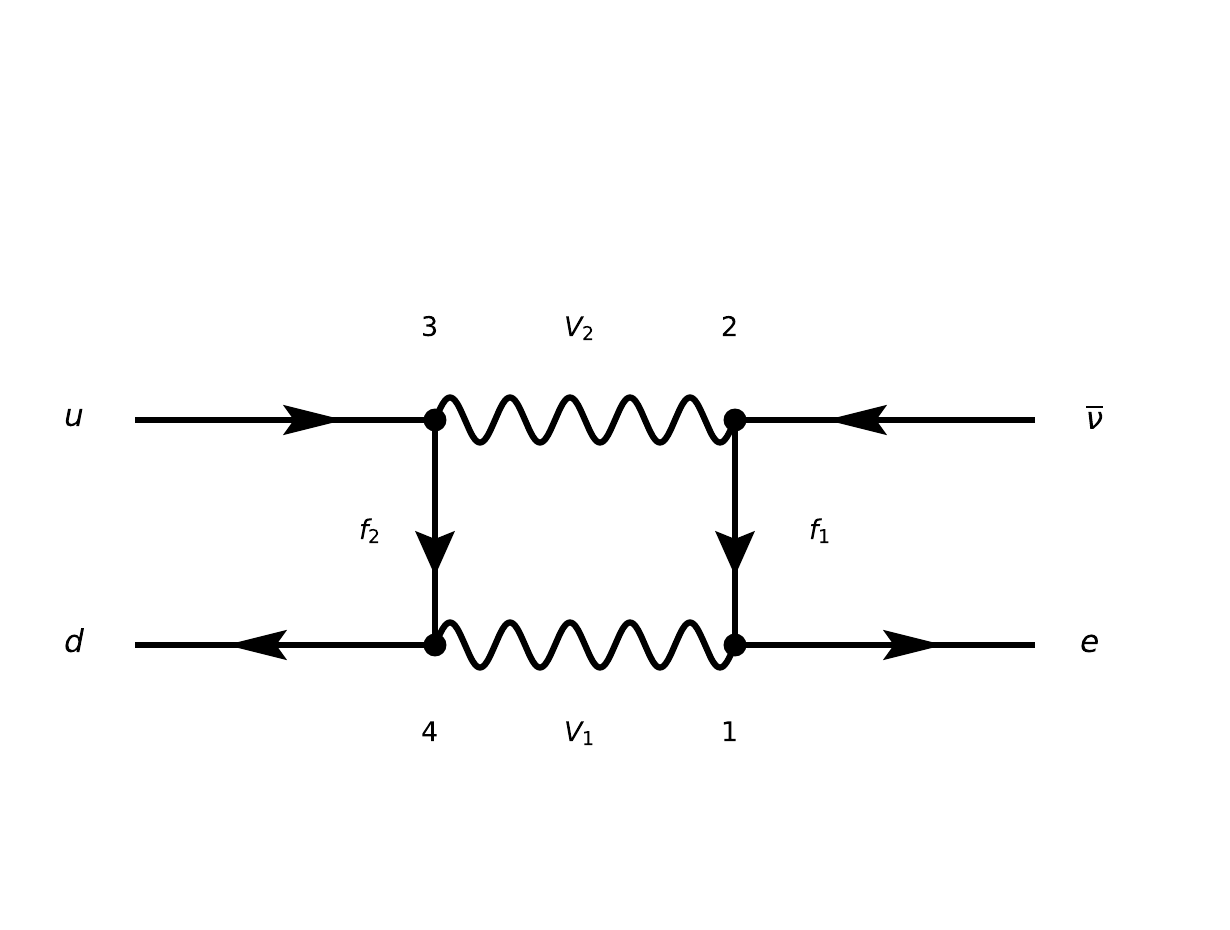} 
\caption{ Generic box diagram}
\label{Boxdiagrams} 
\end{figure}

\begin{align}
I_{box}^{Num} =&  (V^d_L)_{il} (V^{f_2}_L)^\dagger_{lk}  (V^{f2}_L)_{km}(V^{u}_L)^\dagger_{mj} \times (V^l_L)_{bd} (V^{f_1}_L)^\dagger_{dc}  (V^{f1}_L)_{ce}(V^{\nu}_L)^\dagger_{ea} \nonumber \\
&\big( V_{CKM}\big)_{ij} \big( U_{PMNS}\big)_{ab}  \left(g_{\mu \al} -\frac{k_\mu k_{\al}}{m_{V_1}^2}\right)    \left(g_{\nu \beta} -\frac{k_\nu k_{\beta}}{m_{V_2}^2}\right) \nonumber \\
&\times \big[\bar{d}'^j_L   \ga_{\mu}(g_4^L P_L+g_4^R P_R)(\slashed{k}+ m_{f_{1}})
\ga_{\nu}(g_3^L P_L+g_3^R P_R) u_L'^i \big] \nonumber \\
&\times\big[\bar{e}'^a_L   \ga_{\alpha}(g_1^L P_L+g_1^R P_R)(\slashed{k}+ m_{f_{2}})
\ga_{\beta}(g_2^L P_L+g_2^R P_R)\nu_L'^b \big] 
\end{align}

In the limit where external momenta small and set to zero  the box diagram Fig. \ref{Boxdiagrams} amplitude can be evaluated as:

\begin{align}
i\mathcal{M}_{box}&=\big( V_{CKM}\big)_{ij} \big( U_{PMNS}\big)_{ab}\Gamma^{f_1 V_1 f_2 V_2} [\overline{d}'_i\ga_\mu P_L u'_j]  [\bar{e}'_a\ga_\mu P_L \nu'_b]
\end{align}

The Wilson coefficient can be evaluated as:

\begin{align}
\big[\mathcal{C}^{u_id_j}_{e_a \nu_b}\big]_{box}^{f_1 V_1 f_2 V_2}&=\big( V_{CKM}\big)_{ij} \big( U_{PMNS}\big)_{ab} \Gamma^{f_1 V_1 f_2 V_2}(g_1^L,g_2^L,g_3^L,g_4^L,m_{f_1},m_{V_1},m_{f_2},m_{V_2})
\label{BoxWC}
\end{align}

where the vertex function $\Gamma^{f_1 V_1 f_2 V_2}$ are given as in Appendix (\ref{BoxVertex})

\section{Vertex function}
\label{Vertexfunction}
The vertex function defined above can be calculated using packageX \cite{Patel:2015tea,PATEL201766} and are give as followings:

\begingroup
\allowdisplaybreaks
\begin{align}
&\Gamma^{f_1f_2V}(g_1^L,g_2^L,g_3^L,g_3^R,m_{f_1},m_{f_2},m_V) =\frac{1}{16\pi^2}\frac{g_1^L g_2^L}{m_V^2} \nonumber \\
&\Big( 
2g_3^R m_{f_1} m_{f_2}m_V^2C_0(m_V,m_{f_1},m_{f_2})-2(g_3^R m_{f_1}m_{f_2}+2g_3^Lm_V^2)C_{00}(m_V,m_{f_1},m_{f_2}) \nonumber \\
&+ \frac{1}{3}g_3^L \big( A_0(m_{f_1}) +A_0(m_{f_2})+A_0(m_V) -m^2_{f_1} m^2_{f_1} C_0(m_{f_1},m_{f_2},m_V) \nonumber \\
&-m^2_{f_2}m_V^2C_0(m_{f_2},m_V,m_{f_1})
- m^2_{f_1}m^2_V C_0(m_V,m_{f_1},m_{f_2}) \nonumber \\
&+4(m^2_{f_1}+m^2_{f_2})C_{00}(m_{f_1},m_{f_2},m_V)  \nonumber \\
&+4(m^2_{f_2}+m_V^2)C_{00}(m_{f_2},m_V,m_{f_1})  
+4(m^2_{f_1}+m_V^2)C_{00}(m_V,m_{f_1},m_{f_2})
\big)
\Big)
\label{ffVVertex} 
\end{align}

\begin{align}
&\Gamma^{V_1V_2f}(g_1^L,g_2^L,g_3,m_{V_1},m_{V_2},m_f)=\frac{1}{64\pi^2}g_1^L g_2^L g_3  \nonumber \\
&\Big( -48C_{00}(m_f,m_{V_1},m_{V_2}) +  
\frac{1}{m_{V_1}^2}
\big(A_0(m_f) +A_0(m_{V_1}+A_0(m_{V_2})) \nonumber \\
&-m_f^2m_{V_1}^2C_0(m_f,m_{V_1},m_{V_2}) \nonumber \\
&-m_{V_1}^2 m_{V_2}^2C_0(m_{V_1},m_{V_2},m_f) - m_f^2m_{V_2}^2C_0(m_{V_2},m_f,m_{V_1})  \nonumber \\
&+ 4(m_f^2+m_{V_1}^2)C_{00}(m_f,m_{V_1},m_{V_2})  \nonumber \\
&+4(m_{V_1}^2+m_{V_2}^2)C_{00}(m_{V_1},m_{V_2},m_f) + 4(m_f^2+m_{V_2}^2)C_{00}(m_{V_2},m_f,m_{V_1}) 
\big) \nonumber \\
&+\frac{1}{m_{V_2}^2}
\big(A_0(m_f) +A_0(m_{V_1}+A_0(m_{V_2}))-m_f^2m_{V_1}^2C_0(m_f,m_{V_1},m_{V_2}) \nonumber \\
&-m_{V_1}^2 m_{V_2}^2C_0(m_{V_1},m_{V_2},m_f) - m_f^2m_{V_2}^2C_0(m_{V_2},m_f,m_{V_1})  \nonumber \\
&+ 4(m_f^2+m_{V_1}^2)C_{00}(m_f,m_{V_1},m_{V_2})  \nonumber \\
&+4(m_{V_1}^2+m_{V_2}^2)C_{00}(m_{V_1},m_{V_2},m_f) + 4(m_f^2+m_{V_2}^2)C_{00}(m_{V_2},m_f,m_{V_1}) 
\big)
\Big) 	
\label{VVfVertex} \\
&\Gamma^{V\gamma f}(g_1^L,g_2^L,g_3,m_{V},0,m_f)=\frac{1}{64\pi^2}\frac{g_1^L g_2^L g_3}{m_V^2}  \nonumber \\
&\Big(A_0(0)+A_0(m_f)+A_0(m_V)-m_f^2m_V^2C_0(m_f,m_V,0)+4m_f^2C_{00}(0,m_f,m_V) \nonumber \\
&+4(m_f^2-11m_V^2)C_{00}(m_f,m_V,0)+4m_V^2C_{00}(m_V,0,m_f)\Big)
\label{VgammafVertex} 
\end{align}

\begin{align}
&\Gamma^{f_1V_1f_2V_2}(g_1^L,g_2^L,g_3^L,g_4^L,m_{f_1},m_{V_1},m_{f_2},m_{V_2})=\frac{1}{256\pi^2}g_1^L g_2^L g_3^L g_4^L  \nonumber \\
&\Big( 
256D_{00}(m_{f_1},m_{V_1},m_{f_2},m_{V_2}) +  \frac{4}{m_{V_1}^2}\big(
B_0(m_{f_1},m_{V_1}) + B_0(m_{f_2},m_{V_2}) \nonumber \\
&+ B_0(m_{f_1},m_{V_2})  
+ B_0(m_{f_2},m_{V_1}) 
-m_{f_1}^2m_{V_1}^2D_0(m_{f_1},m_{V_1},m_{f_2},m_{V_2}) \nonumber \\	&-m_{f_2}^2m_{V_2}^2D_0(m_{f_2},m_{V_2},m_{f_1},m_{V_1}) 
-m_{f_2}^2m_{V_1}^2D_0(m_{V_1},m_{f_2},m_{V_2},m_{f_1}) \nonumber \\
&-m_{f_1}^2m_{V_2}^2D_0(m_{V_2},m_{f_1},m_{V_1},m_{f_2})  \nonumber \\
&+4(m_{f_1}^2+m_{V_1}^2)D_{00}(m_{f_1},m_{V_1},m_{f_2},m_{V_2})  \nonumber \\
&+4(m_{f_2}^2+m_{V_2}^2)D_{00}(m_{f_1},m_{V_1},m_{f_2},m_{V_2})
+4(m_{f_2}^2+m_{V_1}^2)D_{00}(m_{V_1},m_{f_2},m_{V_2},m_{f_1}) \nonumber \\
&+4(m_{f_1}^2+m_{V_2}^2)D_{00}(m_{V_2},m_{f_1},m_{V_1},m_{f_2})
\big)  
\nonumber \\
&+  \frac{4}{m_{V_2}^2}\big(
B_0(m_{f_1},m_{V_1}) + B_0(m_{f_2},m_{V_2}) + B_0(m_{f_1},m_{V_2})  \nonumber \\
&+ B_0(m_{f_2},m_{V_1}) 
-m_{f_1}^2m_{V_1}^2D_0(m_{f_1},m_{V_1},m_{f_2},m_{V_2}) \nonumber \\	&-m_{f_2}^2m_{V_2}^2D_0(m_{f_2},m_{V_2},m_{f_1},m_{V_1}) 
-m_{f_2}^2m_{V_1}^2D_0(m_{V_1},m_{f_2},m_{V_2},m_{f_1}) \nonumber \\
&-m_{f_1}^2m_{V_2}^2D_0(m_{V_2},m_{f_1},m_{V_1},m_{f_2}) +4(m_{f_1}^2+m_{V_1}^2)D_{00}(m_{f_1},m_{V_1},m_{f_2},m_{V_2})  \nonumber \\
&+4(m_{f_2}^2+m_{V_2}^2)D_{00}(m_{f_1},m_{V_1},m_{f_2},m_{V_2})
+4(m_{f_2}^2+m_{V_1}^2)D_{00}(m_{V_1},m_{f_2},m_{V_2},m_{f_1}) \nonumber \\
&+4(m_{f_1}^2+m_{V_2}^2)D_{00}(m_{V_2},m_{f_1},m_{V_1},m_{f_2})
\big)  \nonumber \\
&+\frac{1}{m_{V_1}^2m_{V_2}^2} \big(
A_0(m_{f_1}) + A_0(m_{f_2}) +A_0(m_{V_1}) + A_0(m_{V_2}) + m_{f_1}^2B_0(m_{f_1},m_{f_2})  \nonumber \\
&+ m_{f_1}^2B_0(m_{f_1},m_{V_1}) + m_{f_1}^2B_0(m_{f_1},m_{V_2}) + m_{f_2}^2B_0(m_{f_1},m_{f_2})  \nonumber \\
&+ m_{f_2}^2B_0(m_{f_2},m_{V_1}) + m_{f_2}^2B_0(m_{f_2},m_{V_2}) + m_{V_1}^2B_0(m_{V_1},m_{f_1})  \nonumber \\
&+ m_{V_1}^2B_0(m_{V_1},m_{f_2}) + m_{V_1}^2B_0(m_{V_1},m_{V_2}) + m_{V_2}^2B_0(m_{V_2},m_{f_1}) \nonumber \\
&+ m_{V_2}^2B_0(m_{V_2},m_{f_2}) + m_{V_2}^2B_0(m_{V_2},m_{V_1}) 
- 2m_{f_1}^2m_{f_2}^2m_{V_1}^2D_0(m_{f_1},m_{V_1},m_{f_2},m_{V_2}) \nonumber \\
&- 2m_{f_1}^2m_{f_2}^2m_{V_2}^2D_0(m_{f_2},m_{V_2},m_{f_1},m_{V_1}) 
- 2m_{f_2}^2m_{V_1}^2m_{V_2}^2D_0(m_{V_1},m_{f_2},m_{V_2},m_{f_1}) \nonumber \\
& - 2m_{f_1}^2m_{V_1}^2m_{V_2}^2D_0(m_{V_2},m_{f_1},m_{V_1},m_{f_2}) \nonumber \\
&+ 4(m_{f_2}^2m_{V_1}^2+m_{f_1}^2(m_{f_2}^2+m_{V_1}^2))D_{00}(m_{f_1},m_{V_1},m_{f_2},m_{V_2}) \nonumber \\
&+ 4(m_{f_2}^2m_{V_2}^2+m_{f_1}^2(m_{f_2}^2+m_{V_2}^2))D_{00}(m_{f_2},m_{V_2},m_{f_1},m_{V_1}) \nonumber \\
&+ 4(m_{f_1}^2m_{V_2}^2+m_{f_2}^2(m_{V_1}^2+m_{V_2}^2))D_{00}(m_{V_1},m_{f_2},m_{V_2},m_{f_1}) \nonumber \\
&+ 4(m_{f_1}^2m_{V_2}^2+m_{f_1}^2(m_{V_1}^2+m_{V_2}^2))D_{00}(m_{f_1},m_{V_1},m_{f_2},m_{V_2})
\big)
\Big)
\label{BoxVertex}
\end{align}
\endgroup
\section{\label{Coupling}Coupling constants}

\begin{table}[ht!]
\caption{The couplings of $Z$ with fermions ($f\ne\nu_R$), the coupling for $Z'$ are obtained by replacement: $c_\varphi\to s_\varphi$ and $s_\varphi\to -c_\varphi$.}
{	
\begin{tabular}{lcc}
	$f$ & $g^{Z}_V(f)$ & $g^{Z}_A(f)$ \\ \hline 
	$\nu_1$ & $\frac{s_\varphi c_{2W}+c_\varphi\sqrt{1+2c_{2W}}}{2\sqrt{1+2c_{2W}}}$ & $\frac{s_\varphi c_{2W}+c_\varphi\sqrt{1+2c_{2W}}}{2\sqrt{1+2c_{2W}}}$ \\
	$\nu_\alpha$ & $-\frac{s_\varphi-c_\varphi\sqrt{1+2c_{2W}}}{2\sqrt{1+2c_{2W}}}$ & $-\frac{s_\varphi-c_\varphi\sqrt{1+2c_{2W}}}{2\sqrt{1+2c_{2W}}}$ \\
	$e_1$ & $\frac{(s_\varphi \left(2c_{2W}-1 \right)-c_\varphi\sqrt{1+2c_{2W}})(2c_{2W}-1)}{2\sqrt{1+2c_{2W}}}$ & $\frac{s_\varphi-c_\varphi\sqrt{1+2c_{2W}}}{2\sqrt{1+2c_{2W}}}$ \\ 
	$e_\al$ & $\frac{s_\varphi-c_\varphi\sqrt{1+2c_{2W}}(2c_{2W}-1)}{2\sqrt{1+2c_{2W}}}$ & $-\frac{s_\varphi c_{2W}+c_\varphi\sqrt{1+2c_{2W}}}{2\sqrt{1+2c_{2W}}}$ \\ 
	$u_a$ & $\frac{s_\varphi(4-c_{2W})+c_\varphi\sqrt{1+2c_{2W}}(4c_{2W}-1)}{6\sqrt{1+2c_{2W}}}$ & $\frac{s_\varphi c_{2W}+c_\varphi\sqrt{1+2c_{2W}}}{2\sqrt{1+2c_{2W}}}$ \\
	$d_a$ &$\frac{(s_\varphi-c_\varphi \sqrt{1+2c_{2W}}) \sqrt{1+2c_{2W}}}{6}$ &  $\frac{s_\varphi-c_\varphi\sqrt{1+2c_{2W}}}{2\sqrt{1+2c_{2W}}}$\\
	$\xi^0$ & $-\frac{s_\varphi c_{W}^2}{\sqrt{1+2c_{2W}}}$ &  $-\frac{s_\varphi c_{W}^2}{\sqrt{1+2c_{2W}}}$ \\
	$\xi$ & $2c_\varphi c^2_W$ &  $-\frac{2s_\varphi c_{W}^2}{\sqrt{1+2c_{2W}}}$ \\
	$E_1$ & $\frac{2s_\varphi c_{2W}+c_\varphi\sqrt{1+2c_{2W}}(1-c_{2W})}{\sqrt{1+2c_{2W}}}$ & $\frac{2s_\varphi c_{W}^2}{\sqrt{1+2c_{2W}}}$\\
	$E_\al$ & $\frac{s_\varphi(3c_{2W}-1)-2c_\varphi\sqrt{1+2c_{2W}}(c_{2W}-1)}{2\sqrt{1+2c_{2W}}}$ & $\frac{s_\varphi c_{W}^2}{\sqrt{1+2c_{2W}}}$\\
	$U_a$ & $\frac{s_\varphi(1-7c_{2W})+4c_\varphi\sqrt{1+2c_{2W}}(c_{2W}-1)}{6\sqrt{1+2c_{2W}}}$ & $-\frac{s_\varphi c^2_W}{\sqrt{1+2c_{2W}}}$ \\
	\hline
\end{tabular}
\label{coupling1}
}
\end{table}

\begin{table}[ht!]
\centering
\caption{Gauge and fermion couplings }
{
\begin{tabular}{ll}
	\hline
	& Couplings \\
	\hline
	W boson & 
	\(\begin{aligned}
		g^{L}_{\nu_a e_b W} &= -\frac{g}{\sqrt{2}},\quad
		g^{L}_{u_i d_j W} =-\frac{g}{\sqrt{2}}
	\end{aligned}\) \\
	\hline
	X-boson  &
	\(\begin{aligned}{}
		g^{L}_{\nu_1 E_1 X} &= -g,\quad
		g^{L}_{\nu_\alpha E_\alpha X} &=&-\frac{g}{\sqrt{2}},\\
		g^{L}_{U_a d_a X} &= -\frac{g}{\sqrt{2}},\quad
		g^{L}_{\nu_1 \xi X} &=& -g,\\
		g^{L}_{e_1 \xi^0 X} &= -\frac{g}{\sqrt{2}}
	\end{aligned}\) \\
	\hline
	Y-boson  &
	\(\begin{aligned}
		g^{L}_{e_1 E_1 Y} &= -g,\quad
		g^{L}_{e_\alpha E_\alpha Y} &&= -\frac{g}{\sqrt{2}},\\
		g^{L}_{U_a u_a Y} &= -\frac{g}{\sqrt{2}},\quad
		g^{L}_{e_1 \xi Y} &&= -g,\\
		g^{L}_{\nu_1 \xi^0 Y} &= -\frac{g}{\sqrt{2}}
	\end{aligned}\) \\
	\hline
	Z/Z' boson & \(g^{L}_{\nu \nu Z, Z'} = -\frac{g}{2 c_W}\left[ g^V_{Z,Z'}(\nu) + g^A_{Z,Z'}(\nu) \right]\) \\
	\hline
	Z/Z' boson & \(g^{R}_{\nu \nu Z, Z'} = -\frac{g}{2 c_W}\left[ g^V_{Z,Z'}(\nu) - g^A_{Z,Z'}(\nu) \right]\) \\
	\hline
	Z/Z' boson & \(g^{L}_{e e Z, Z'} = -\frac{g}{2 c_W}\left[ g^V_{Z,Z'}(e) + g^A_{Z,Z'}(e) \right]\) \\
	\hline
	Z/Z' boson & \(g^{R}_{e e Z, Z'} = -\frac{g}{2 c_W}\left[ g^V_{Z,Z'}(e) - g^A_{Z,Z'}(e) \right]\) \\
	\hline
\end{tabular}
}
\label{coupling2}
\end{table}

\bibliographystyle{apsrev4-1}
\bibliography{3311_FFlip_Banomalies.bib}

@article{LZ:2024zvo,
    author = "Aalbers, J. and others",
    collaboration = "LZ",
    title = "{Dark Matter Search Results from 4.2{\,}{\,}Tonne-Years of Exposure of the LUX-ZEPLIN (LZ) Experiment}",
    eprint = "2410.17036",
    archivePrefix = "arXiv",
    primaryClass = "hep-ex",
    reportNumber = "FERMILAB-PUB-24-0796-V",
    doi = "10.1103/4dyc-z8zf",
    journal = "Phys. Rev. Lett.",
    volume = "135",
    number = "1",
    pages = "011802",
    year = "2025"
}

@article{RevModPhys.88.030501,
	title = {Nobel Lecture: Discovery of atmospheric neutrino oscillations},
	author = {Kajita, Takaaki},
	journal = {Rev. Mod. Phys.},
	volume = {88},
	issue = {3},
	pages = {030501},
	numpages = {7},
	year = {2016},
	month = {Jul},
	publisher = {American Physical Society},
	doi = {10.1103/RevModPhys.88.030501},
	url = {https://link.aps.org/doi/10.1103/RevModPhys.88.030501}
}

@article{RevModPhys.88.030502,
	title = {Nobel Lecture: The Sudbury Neutrino Observatory: Observation of flavor change for solar neutrinos},
	author = {McDonald, Arthur B.},
	journal = {Rev. Mod. Phys.},
	volume = {88},
	issue = {3},
	pages = {030502},
	numpages = {9},
	year = {2016},
	month = {Jul},
	publisher = {American Physical Society},
	doi = {10.1103/RevModPhys.88.030502},
	url = {https://link.aps.org/doi/10.1103/RevModPhys.88.030502}
}

@article{Bertone:2004pz,
	author = "Bertone, Gianfranco and Hooper, Dan and Silk, Joseph",
	title = "{Particle dark matter: Evidence, candidates and constraints}",
	eprint = "hep-ph/0404175",
	archivePrefix = "arXiv",
	reportNumber = "FERMILAB-PUB-04-047-A",
	doi = "10.1016/j.physrep.2004.08.031",
	journal = "Phys. Rept.",
	volume = "405",
	pages = "279--390",
	year = "2005"
}

@article{Arcadi:2017kky,
	author = "Arcadi, Giorgio and Dutra, Ma\'\i{}ra and Ghosh, Pradipta and Lindner, Manfred and Mambrini, Yann and Pierre, Mathias and Profumo, Stefano and Queiroz, Farinaldo S.",
	title = "{The waning of the WIMP? A review of models, searches, and constraints}",
	eprint = "1703.07364",
	archivePrefix = "arXiv",
	primaryClass = "hep-ph",
	doi = "10.1140/epjc/s10052-018-5662-y",
	journal = "Eur. Phys. J. C",
	volume = "78",
	number = "3",
	pages = "203",
	year = "2018"
}

@article{Canetti:2012zc,
	author = "Canetti, Laurent and Drewes, Marco and Shaposhnikov, Mikhail",
	title = "{Matter and Antimatter in the Universe}",
	eprint = "1204.4186",
	archivePrefix = "arXiv",
	primaryClass = "hep-ph",
	reportNumber = "TTK-12-04",
	doi = "10.1088/1367-2630/14/9/095012",
	journal = "New J. Phys.",
	volume = "14",
	pages = "095012",
	year = "2012"
}

@article{Singer:1980sw,
	author = "Singer, M. and Valle, J. W. F. and Schechter, J.",
	title = "{Canonical Neutral Current Predictions From the Weak Electromagnetic Gauge Group SU(3) X $u$(1)}",
	reportNumber = "SU-4217-162, COO-3533-162",
	doi = "10.1103/PhysRevD.22.738",
	journal = "Phys. Rev. D",
	volume = "22",
	pages = "738",
	year = "1980"
}

@article{Valle:1983dk,
	author = "Valle, J. W. F. and Singer, M.",
	title = "{Lepton Number Violation With Quasi Dirac Neutrinos}",
	reportNumber = "RL-83-018",
	doi = "10.1103/PhysRevD.28.540",
	journal = "Phys. Rev. D",
	volume = "28",
	pages = "540",
	year = "1983"
}

@article{Pisano:1992bxx,
	author = "Pisano, F. and Pleitez, V.",
	title = "{An SU(3) x U(1) model for electroweak interactions}",
	eprint = "hep-ph/9206242",
	archivePrefix = "arXiv",
	reportNumber = "IFT-P-008-92",
	doi = "10.1103/PhysRevD.46.410",
	journal = "Phys. Rev. D",
	volume = "46",
	pages = "410--417",
	year = "1992"
}

@article{Frampton:1992wt,
	author = "Frampton, P. H.",
	title = "{Chiral dilepton model and the flavor question}",
	reportNumber = "IFP-427-UNC",
	doi = "10.1103/PhysRevLett.69.2889",
	journal = "Phys. Rev. Lett.",
	volume = "69",
	pages = "2889--2891",
	year = "1992"
}

@article{PhysRevD.50.R34,
	title = {SU(3${)}_{\mathit{L}}$\ensuremath{\bigotimes}U(1${)}_{\mathit{N}}$ and SU(4${)}_{\mathit{L}}$\ensuremath{\bigotimes}U(1${)}_{\mathit{N}}$ gauge models with right-handed neutrinos},
	author = {Foot, Robert and Long, Hoang Ngoc and Tran, Tuan A.},
	journal = {Phys. Rev. D},
	volume = {50},
	issue = {1},
	pages = {R34--R38},
	numpages = {0},
	year = {1994},
	month = {Jul},
	publisher = {American Physical Society},
	doi = {10.1103/PhysRevD.50.R34},
	url = {https://link.aps.org/doi/10.1103/PhysRevD.50.R34}
}

@article{Tully:2000kk,
	author = "Tully, M. B. and Joshi, Girish C.",
	title = "{Generating neutrino mass in the 331 model}",
	eprint = "hep-ph/0011172",
	archivePrefix = "arXiv",
	reportNumber = "UM-P-2000-047",
	doi = "10.1103/PhysRevD.64.011301",
	journal = "Phys. Rev. D",
	volume = "64",
	pages = "011301",
	year = "2001"
}

@article{Dias:2005yh,
	author = "Dias, Alex G. and de S. Pires, C. A. and Rodrigues da Silva, P. S.",
	title = "{Naturally light right-handed neutrinos in a 3-3-1 model}",
	eprint = "hep-ph/0508186",
	archivePrefix = "arXiv",
	doi = "10.1016/j.physletb.2005.09.028",
	journal = "Phys. Lett. B",
	volume = "628",
	pages = "85--92",
	year = "2005"
}

@article{Chang:2006aa,
	author = "Chang, Darwin and Long, Hoang Ngoc",
	title = "{Interesting radiative patterns of neutrino mass in an SU(3)(C) x SU(3)(L) x U(1)(X) model with right-handed neutrinos}",
	eprint = "hep-ph/0603098",
	archivePrefix = "arXiv",
	doi = "10.1103/PhysRevD.73.053006",
	journal = "Phys. Rev. D",
	volume = "73",
	pages = "053006",
	year = "2006"
}

@article{Dong:2006mt,
	author = "Dong, P. V. and Long, Hoang Ngoc and Soa, D. V.",
	title = "{Neutrino masses in the economical 3-3-1 model}",
	eprint = "hep-ph/0610381",
	archivePrefix = "arXiv",
	doi = "10.1103/PhysRevD.75.073006",
	journal = "Phys. Rev. D",
	volume = "75",
	pages = "073006",
	year = "2007"
}

@article{Dong:2008sw,
	author = "Dong, P. V. and Long, Hoang Ngoc",
	title = "{Neutrino masses and lepton flavor violation in the 3-3-1 model with right-handed neutrinos}",
	eprint = "0801.4196",
	archivePrefix = "arXiv",
	primaryClass = "hep-ph",
	doi = "10.1103/PhysRevD.77.057302",
	journal = "Phys. Rev. D",
	volume = "77",
	pages = "057302",
	year = "2008"
}

@article{Dong:2010gk,
	author = "Dong, P. V. and Hue, L. T. and Long, H. N. and Soa, D. V.",
	title = "{The 3-3-1 model with $\mathrm{A}_4$ flavor symmetry}",
	eprint = "1001.4625",
	archivePrefix = "arXiv",
	primaryClass = "hep-ph",
	doi = "10.1103/PhysRevD.81.053004",
	journal = "Phys. Rev. D",
	volume = "81",
	pages = "053004",
	year = "2010"
}

@article{Dong:2010zu,
	author = "Dong, P. V. and Long, H. N. and Soa, D. V. and Vien, V. V.",
	title = "{The 3-3-1 model with $S_4$ flavor symmetry}",
	eprint = "1009.2328",
	archivePrefix = "arXiv",
	primaryClass = "hep-ph",
	doi = "10.1140/epjc/s10052-011-1544-2",
	journal = "Eur. Phys. J. C",
	volume = "71",
	pages = "1544",
	year = "2011"
}

@article{Dong:2011vb,
	author = "Dong, P. V. and Long, H. N. and Nam, C. H. and Vien, V. V.",
	title = "{The $S_3$ flavor symmetry in 3-3-1 models}",
	eprint = "1111.6360",
	archivePrefix = "arXiv",
	primaryClass = "hep-ph",
	doi = "10.1103/PhysRevD.85.053001",
	journal = "Phys. Rev. D",
	volume = "85",
	pages = "053001",
	year = "2012"
}

@article{Boucenna:2014ela,
	author = "Boucenna, Sofiane M. and Morisi, Stefano and Valle, Jose W. F.",
	title = "{Radiative neutrino mass in 3-3-1 scheme}",
	eprint = "1405.2332",
	archivePrefix = "arXiv",
	primaryClass = "hep-ph",
	doi = "10.1103/PhysRevD.90.013005",
	journal = "Phys. Rev. D",
	volume = "90",
	number = "1",
	pages = "013005",
	year = "2014"
}

@article{Boucenna:2014dia,
	author = "Boucenna, Sofiane M. and Fonseca, Renato M. and Gonzalez-Canales, Felix and Valle, Jose W. F.",
	title = "{Small neutrino masses and gauge coupling unification}",
	eprint = "1411.0566",
	archivePrefix = "arXiv",
	primaryClass = "hep-ph",
	doi = "10.1103/PhysRevD.91.031702",
	journal = "Phys. Rev. D",
	volume = "91",
	number = "3",
	pages = "031702",
	year = "2015"
}

@article{Long:2003hht,
	author = "Long, Hoang Ngoc and Lan, Nguyen Quynh",
	title = "{Selfinteracting dark matter and Higgs bosons in the SU(3)(C) x SU(3)(L) x U(1)(N) model with right-handed neutrinos}",
	eprint = "hep-ph/0309038",
	archivePrefix = "arXiv",
	reportNumber = "IC-2003-33",
	doi = "10.1209/epl/i2003-00267-5",
	journal = "EPL",
	volume = "64",
	pages = "571",
	year = "2003"
}

@article{Filippi:2005mt,
	author = "Filippi, Simonetta and Ponce, William A. and Sanchez, Luis A.",
	title = "{Dark matter from the scalar sector of 3-3-1 models without exotic electric charges}",
	eprint = "hep-ph/0509173",
	archivePrefix = "arXiv",
	doi = "10.1209/epl/i2005-10349-x",
	journal = "EPL",
	volume = "73",
	pages = "142--148",
	year = "2006"
}

@article{Mizukoshi:2010ky,
	author = "Mizukoshi, J. K. and de S. Pires, C. A. and Queiroz, F. S. and Rodrigues da Silva, P. S.",
	title = "{WIMPs in a 3-3-1 model with heavy Sterile neutrinos}",
	eprint = "1010.4097",
	archivePrefix = "arXiv",
	primaryClass = "hep-ph",
	doi = "10.1103/PhysRevD.83.065024",
	journal = "Phys. Rev. D",
	volume = "83",
	pages = "065024",
	year = "2011"
}

@article{Ruiz-Alvarez:2012nvg,
	author = "Ruiz-Alvarez, J. D. and de S. Pires, C. A. and Queiroz, Farinaldo S. and Restrepo, D. and Rodrigues da Silva, P. S.",
	title = "{On the Connection of Gamma-Rays, Dark Matter and Higgs Searches at LHC}",
	eprint = "1206.5779",
	archivePrefix = "arXiv",
	primaryClass = "hep-ph",
	reportNumber = "FERMILAB-PUB-12-873-AE",
	doi = "10.1103/PhysRevD.86.075011",
	journal = "Phys. Rev. D",
	volume = "86",
	pages = "075011",
	year = "2012"
}

@article{Dong:2013ioa,
	author = "Dong, P. V. and Nguyen, T. Phong and Soa, D. V.",
	title = "{3-3-1 model with inert scalar triplet}",
	eprint = "1308.4097",
	archivePrefix = "arXiv",
	primaryClass = "hep-ph",
	doi = "10.1103/PhysRevD.88.095014",
	journal = "Phys. Rev. D",
	volume = "88",
	number = "9",
	pages = "095014",
	year = "2013"
}

@article{Profumo:2013sca,
	author = "Profumo, Stefano and Queiroz, Farinaldo S.",
	title = "{Constraining the $Z'$ mass in 331 models using direct dark matter detection}",
	eprint = "1307.7802",
	archivePrefix = "arXiv",
	primaryClass = "hep-ph",
	reportNumber = "CETUP2013-012",
	doi = "10.1140/epjc/s10052-014-2960-x",
	journal = "Eur. Phys. J. C",
	volume = "74",
	number = "7",
	pages = "2960",
	year = "2014"
}

@article{Pisano:1996ht,
	author = "Pisano, Felice",
	title = "{A Simple solution for the flavor question}",
	eprint = "hep-ph/9609358",
	archivePrefix = "arXiv",
	reportNumber = "IFT-P-031-96",
	doi = "10.1142/S0217732396002630",
	journal = "Mod. Phys. Lett. A",
	volume = "11",
	pages = "2639--2647",
	year = "1996"
}

@article{Doff:1998we,
	author = "Doff, A. and Pisano, F.",
	title = "{Charge quantization in the largest leptoquark bilepton chiral electroweak scheme}",
	eprint = "hep-ph/9812303",
	archivePrefix = "arXiv",
	doi = "10.1142/S0217732399001218",
	journal = "Mod. Phys. Lett. A",
	volume = "14",
	pages = "1133--1142",
	year = "1999"
}

@article{deSousaPires:1998jc,
	author = "de Sousa Pires, Carlos Antonio and Ravinez, O. P.",
	title = "{Charge quantization in a chiral bilepton gauge model}",
	eprint = "hep-ph/9803409",
	archivePrefix = "arXiv",
	reportNumber = "IFT-P-014-98",
	doi = "10.1103/PhysRevD.58.035008",
	journal = "Phys. Rev. D",
	volume = "58",
	pages = "035008",
	year = "1998"
}

@article{deSousaPires:1999ca,
	author = "de Sousa Pires, Carlos Antonio",
	title = "{Remark on the vector - like nature of the electromagnetism and the electric charge quantization}",
	eprint = "hep-ph/9902406",
	archivePrefix = "arXiv",
	doi = "10.1103/PhysRevD.60.075013",
	journal = "Phys. Rev. D",
	volume = "60",
	pages = "075013",
	year = "1999"
}

@article{Dong:2005ebq,
	author = "Dong, Phung Van and Long, Hoang Ngoc",
	title = "{Electric charge quantization in SU(3)(C) x SU(3)(L) x U(1)(X) models}",
	eprint = "hep-ph/0507155",
	archivePrefix = "arXiv",
	doi = "10.1142/S0217751X06035191",
	journal = "Int. J. Mod. Phys. A",
	volume = "21",
	pages = "6677--6692",
	year = "2006"
}

@article{Duy:2022qhy,
	author = "Duy, N. T. and Thu, P. N. and Huong, D. T.",
	title = "{New physics in $\text {b} \rightarrow \text {s}$ transitions in the MF331 model}",
	eprint = "2205.02995",
	archivePrefix = "arXiv",
	primaryClass = "hep-ph",
	doi = "10.1140/epjc/s10052-022-10916-7",
	journal = "Eur. Phys. J. C",
	volume = "82",
	number = "10",
	pages = "966",
	year = "2022"
}

@article{Huong:2019vej,
	author = "Huong, D. T. and Dinh, D. N. and Thien, L. D. and Van Dong, Phung",
	title = "{Dark matter and flavor changing in the flipped 3-3-1 model}",
	eprint = "1906.05240",
	archivePrefix = "arXiv",
	primaryClass = "hep-ph",
	doi = "10.1007/JHEP08(2019)051",
	journal = "JHEP",
	volume = "08",
	pages = "051",
	year = "2019"
}

@article{Dinh:2019jdg,
	author = "Dinh, D. N. and Huong, D. T. and Duy, N. T. and Nhuan, N. T. and Thien, L. D. and Van Dong, Phung",
	title = "{Flavor changing in the flipped trinification}",
	eprint = "1901.07969",
	archivePrefix = "arXiv",
	primaryClass = "hep-ph",
	doi = "10.1103/PhysRevD.99.055005",
	journal = "Phys. Rev. D",
	volume = "99",
	number = "5",
	pages = "055005",
	year = "2019"
}

@article{Thu:2023xai,
	author = "Thu, P. N. and Duy, N. T. and Carcamo Hernandez, A. E. and Huong, D. T.",
	title = "{Lepton universality violation in the minimal flipped 331 model}",
	eprint = "2304.03003",
	archivePrefix = "arXiv",
	primaryClass = "hep-ph",
	doi = "10.1093/ptep/ptad135",
	journal = "PTEP",
	volume = "2023",
	number = "12",
	pages = "123B01",
	year = "2023"
}

@article{Abu-Ajamieh:2025vxw,
	author = "Abu-Ajamieh, Fayez and Ahriche, Amine and Okada, Nobuchika",
	title = "{Novel and Updated Bounds on Flavor-violating Z Interactions in the Lepton Sector}",
	eprint = "2503.07236",
	archivePrefix = "arXiv",
	primaryClass = "hep-ph",
	month = "3",
	year = "2025"
}

@article{Dong:2013wca,
	author = "Dong, P. V. and Hung, H. T. and Tham, T. D.",
	title = "{3-3-1-1 model for dark matter}",
	eprint = "1305.0369",
	archivePrefix = "arXiv",
	primaryClass = "hep-ph",
	doi = "10.1103/PhysRevD.87.115003",
	journal = "Phys. Rev. D",
	volume = "87",
	number = "11",
	pages = "115003",
	year = "2013"
}

@article{Dong:2015yra,
	author = "Dong, P. V.",
	title = "{Unifying the electroweak and B-L interactions}",
	eprint = "1505.06469",
	archivePrefix = "arXiv",
	primaryClass = "hep-ph",
	doi = "10.1103/PhysRevD.92.055026",
	journal = "Phys. Rev. D",
	volume = "92",
	number = "5",
	pages = "055026",
	year = "2015"
}

@article{Dong:2014wsa,
	author = "Dong, P. V. and Huong, D. T. and Queiroz, Farinaldo S. and Thuy, N. T.",
	title = "{Phenomenology of the 3-3-1-1 model}",
	eprint = "1405.2591",
	archivePrefix = "arXiv",
	primaryClass = "hep-ph",
	doi = "10.1103/PhysRevD.90.075021",
	journal = "Phys. Rev. D",
	volume = "90",
	number = "7",
	pages = "075021",
	year = "2014"
}

@article{Alves:2016fqe,
	author = "Alves, Alexandre and Arcadi, Giorgio and Dong, P. V. and Duarte, Laura and Queiroz, Farinaldo S. and Valle, Jos\'e W. F.",
	title = "{Matter-parity as a residual gauge symmetry: Probing a theory of cosmological dark matter}",
	eprint = "1612.04383",
	archivePrefix = "arXiv",
	primaryClass = "hep-ph",
	reportNumber = "IFIC-16-XX",
	doi = "10.1016/j.physletb.2017.07.056",
	journal = "Phys. Lett. B",
	volume = "772",
	pages = "825--831",
	year = "2017"
}

@article{BaBar:2012obs,
	author = "Lees, J. P. and others",
	collaboration = "BaBar",
	title = "{Evidence for an excess of $\bar{B} \to D^{(*)} \tau^-\bar{\nu}_\tau$ decays}",
	eprint = "1205.5442",
	archivePrefix = "arXiv",
	primaryClass = "hep-ex",
	reportNumber = "BABAR-PUB-12-012, SLAC-PUB-15028",
	doi = "10.1103/PhysRevLett.109.101802",
	journal = "Phys. Rev. Lett.",
	volume = "109",
	pages = "101802",
	year = "2012"
}

@article{BaBar:2013mob,
	author = "Lees, J. P. and others",
	collaboration = "BaBar",
	title = "{Measurement of an Excess of $\bar{B} \to D^{(*)}\tau^- \bar{\nu}_\tau$ Decays and Implications for Charged Higgs Bosons}",
	eprint = "1303.0571",
	archivePrefix = "arXiv",
	primaryClass = "hep-ex",
	reportNumber = "BABAR-PUB-13-001, SLAC-PUB-15381",
	doi = "10.1103/PhysRevD.88.072012",
	journal = "Phys. Rev. D",
	volume = "88",
	number = "7",
	pages = "072012",
	year = "2013"
}

@article{Belle:2016dyj,
	author = "Hirose, S. and others",
	collaboration = "Belle",
	title = "{Measurement of the $\tau$ lepton polarization and $R(D^*)$ in the decay $\bar{B} \to D^* \tau^- \bar{\nu}_\tau$}",
	eprint = "1612.00529",
	archivePrefix = "arXiv",
	primaryClass = "hep-ex",
	reportNumber = "KEK-PREPRINT-2016-53, BELLE-PREPRINT-2016-14",
	doi = "10.1103/PhysRevLett.118.211801",
	journal = "Phys. Rev. Lett.",
	volume = "118",
	number = "21",
	pages = "211801",
	year = "2017"
}

@article{Belle:2015qfa,
	author = "Huschle, M. and others",
	collaboration = "Belle",
	title = "{Measurement of the branching ratio of $\bar{B} \to D^{(\ast)} \tau^- \bar{\nu}_\tau$ relative to $\bar{B} \to D^{(\ast)} \ell^- \bar{\nu}_\ell$ decays with hadronic tagging at Belle}",
	eprint = "1507.03233",
	archivePrefix = "arXiv",
	primaryClass = "hep-ex",
	reportNumber = "KEK-REPORT-2015-18",
	doi = "10.1103/PhysRevD.92.072014",
	journal = "Phys. Rev. D",
	volume = "92",
	number = "7",
	pages = "072014",
	year = "2015"
}

@article{Belle:2019rba,
	author = "Caria, G. and others",
	collaboration = "Belle",
	title = "{Measurement of $\mathcal{R}(D)$ and $\mathcal{R}(D^*)$ with a semileptonic tagging method}",
	eprint = "1910.05864",
	archivePrefix = "arXiv",
	primaryClass = "hep-ex",
	reportNumber = "Belle-2019-18, KEK-2019-40",
	doi = "10.1103/PhysRevLett.124.161803",
	journal = "Phys. Rev. Lett.",
	volume = "124",
	number = "16",
	pages = "161803",
	year = "2020"
}

@article{LHCb:2015gmp,
	author = "Aaij, Roel and others",
	collaboration = "LHCb",
	title = "{Measurement of the ratio of branching fractions $\mathcal{B}(\bar{B}^0 \to D^{*+}\tau^{-}\bar{\nu}_{\tau})/\mathcal{B}(\bar{B}^0 \to D^{*+}\mu^{-}\bar{\nu}_{\mu})$}",
	eprint = "1506.08614",
	archivePrefix = "arXiv",
	primaryClass = "hep-ex",
	reportNumber = "CERN-PH-EP-2015-150, LHCB-PAPER-2015-025, CERN-PH-EP-2015-150, LHCb-PAPER-2015-025",
	doi = "10.1103/PhysRevLett.115.111803",
	journal = "Phys. Rev. Lett.",
	volume = "115",
	number = "11",
	pages = "111803",
	year = "2015",
	note = "[Erratum: Phys.Rev.Lett. 115, 159901 (2015)]"
}

@article{LHCb:2017smo,
	author = "Aaij, R. and others",
	collaboration = "LHCb",
	title = "{Measurement of the ratio of the $B^0 \to D^{*-} \tau^+ \nu_{\tau}$ and $B^0 \to D^{*-} \mu^+ \nu_{\mu}$ branching fractions using three-prong $\tau$-lepton decays}",
	eprint = "1708.08856",
	archivePrefix = "arXiv",
	primaryClass = "hep-ex",
	reportNumber = "LHCB-PAPER-2017-017, CERN-EP-2017-212",
	doi = "10.1103/PhysRevLett.120.171802",
	journal = "Phys. Rev. Lett.",
	volume = "120",
	number = "17",
	pages = "171802",
	year = "2018"
}

@article{LHCb:2017rln,
	author = "Aaij, R. and others",
	collaboration = "LHCb",
	title = "{Test of Lepton Flavor Universality by the measurement of the $B^0 \to D^{*-} \tau^+ \nu_{\tau}$ branching fraction using three-prong $\tau$ decays}",
	eprint = "1711.02505",
	archivePrefix = "arXiv",
	primaryClass = "hep-ex",
	reportNumber = "CERN-EP-2017-256, LHCB-PAPER-2017-027",
	doi = "10.1103/PhysRevD.97.072013",
	journal = "Phys. Rev. D",
	volume = "97",
	number = "7",
	pages = "072013",
	year = "2018"
}

@article{LHCb:2017vlu,
	author = "Aaij, R. and others",
	collaboration = "LHCb",
	title = "{Measurement of the ratio of branching fractions $\mathcal{B}(B_c^+\,\to\,J/\psi\tau^+\nu_\tau)$/$\mathcal{B}(B_c^+\,\to\,J/\psi\mu^+\nu_\mu)$}",
	eprint = "1711.05623",
	archivePrefix = "arXiv",
	primaryClass = "hep-ex",
	reportNumber = "LHCB-PAPER-2017-035, CERN-EP-2017-275",
	doi = "10.1103/PhysRevLett.120.121801",
	journal = "Phys. Rev. Lett.",
	volume = "120",
	number = "12",
	pages = "121801",
	year = "2018"
}

@article{KATRIN:2021uub,
	author = "Aker, M. and others",
	collaboration = "KATRIN",
	title = "{Direct neutrino-mass measurement with sub-electronvolt sensitivity}",
	eprint = "2105.08533",
	archivePrefix = "arXiv",
	primaryClass = "hep-ex",
	doi = "10.1038/s41567-021-01463-1",
	journal = "Nature Phys.",
	volume = "18",
	number = "2",
	pages = "160--166",
	year = "2022"
}

@article{ParticleDataGroup:2022pth,
	author = "Workman, R. L. and others",
	collaboration = "Particle Data Group",
	title = "{Review of Particle Physics}",
	doi = "10.1093/ptep/ptac097",
	journal = "PTEP",
	volume = "2022",
	pages = "083C01",
	year = "2022"
}

@article{HFLAV:2022esi,
	author = "Amhis, Yasmine Sara and others",
	collaboration = "HFLAV",
	title = "{Averages of b-hadron, c-hadron, and \ensuremath{\tau}-lepton properties as of 2021}",
	eprint = "2206.07501",
	archivePrefix = "arXiv",
	primaryClass = "hep-ex",
	doi = "10.1103/PhysRevD.107.052008",
	journal = "Phys. Rev. D",
	volume = "107",
	number = "5",
	pages = "052008",
	year = "2023"
}

@article{Boucenna:2016qad,
	author = "Boucenna, Sofiane M. and Celis, Alejandro and Fuentes-Martin, Javier and Vicente, Avelino and Virto, Javier",
	title = "{Phenomenology of an $SU(2) \times SU(2) \times U(1)$ model with lepton-flavour non-universality}",
	eprint = "1608.01349",
	archivePrefix = "arXiv",
	primaryClass = "hep-ph",
	reportNumber = "LMU-ASC-34-16, IFIC-16-62",
	doi = "10.1007/JHEP12(2016)059",
	journal = "JHEP",
	volume = "12",
	pages = "059",
	year = "2016"
}

@article{Fonseca:2016tbn,
	author = "Fonseca, Renato M. and Hirsch, Martin",
	title = "{A flipped 331 model}",
	eprint = "1606.01109",
	archivePrefix = "arXiv",
	primaryClass = "hep-ph",
	reportNumber = "IFIC-16-33",
	doi = "10.1007/JHEP08(2016)003",
	journal = "JHEP",
	volume = "08",
	pages = "003",
	year = "2016"
}

@article{VanLoi:2020xcq,
	author = "Van Loi, Duong and Nam, Cao H. and Van Dong, Phung",
	title = "{Dark matter in the fully flipped 3-3-1-1 model}",
	eprint = "2012.10979",
	archivePrefix = "arXiv",
	primaryClass = "hep-ph",
	doi = "10.1140/epjc/s10052-021-09374-4",
	journal = "Eur. Phys. J. C",
	volume = "81",
	number = "7",
	pages = "591",
	year = "2021"
}

@article{Gross:1972pv,
	author = "Gross, David J. and Jackiw, R.",
	title = "{Effect of anomalies on quasirenormalizable theories}",
	doi = "10.1103/PhysRevD.6.477",
	journal = "Phys. Rev. D",
	volume = "6",
	pages = "477--493",
	year = "1972"
}

@article{Georgi:1972bb,
	author = "Georgi, Howard and Glashow, Sheldon L.",
	title = "{Gauge theories without anomalies}",
	reportNumber = "PRINT-72-2263",
	doi = "10.1103/PhysRevD.6.429",
	journal = "Phys. Rev. D",
	volume = "6",
	pages = "429",
	year = "1972"
}

@article{Banks:1976yg,
	author = "Banks, J. and Georgi, H.",
	title = "{Comment on Gauge Theories Without Anomalies}",
	doi = "10.1103/PhysRevD.14.1159",
	journal = "Phys. Rev. D",
	volume = "14",
	pages = "1159--1160",
	year = "1976"
}

@article{Okubo:1977sc,
	author = "Okubo, Susumu",
	title = "{Gauge Groups Without Triangular Anomaly}",
	reportNumber = "UR-628, COO-3065-178",
	doi = "10.1103/PhysRevD.16.3528",
	journal = "Phys. Rev. D",
	volume = "16",
	pages = "3528",
	year = "1977"
}

@article{Kim:2016bdu,
	author = "Kim, C. S. and Yuan, Xing-Bo and Zheng, Ya-Juan",
	title = "{Constraints on a Z' boson within minimal flavor violation}",
	eprint = "1602.08107",
	archivePrefix = "arXiv",
	primaryClass = "hep-ph",
	doi = "10.1103/PhysRevD.93.095009",
	journal = "Phys. Rev. D",
	volume = "93",
	number = "9",
	pages = "095009",
	year = "2016"
}

@article{Esteban:2020cvm,
	author = "Esteban, Ivan and Gonzalez-Garcia, M. C. and Maltoni, Michele and Schwetz, Thomas and Zhou, Albert",
	title = "{The fate of hints: updated global analysis of three-flavor neutrino oscillations}",
	eprint = "2007.14792",
	archivePrefix = "arXiv",
	primaryClass = "hep-ph",
	reportNumber = "IFT-UAM/CSIC-112, YITP-SB-2020-21",
	doi = "10.1007/JHEP09(2020)178",
	journal = "JHEP",
	volume = "09",
	pages = "178",
	year = "2020"
}

@article{Patel:2015tea,
	author = "Patel, Hiren H.",
	title = "{Package-X: A Mathematica package for the analytic calculation of one-loop integrals}",
	eprint = "1503.01469",
	archivePrefix = "arXiv",
	primaryClass = "hep-ph",
	doi = "10.1016/j.cpc.2015.08.017",
	journal = "Comput. Phys. Commun.",
	volume = "197",
	pages = "276--290",
	year = "2015"
}

@article{PATEL201766,
	title = {Package-X 2.0: A Mathematica package for the analytic calculation of one-loop integrals},
	journal = {Computer Physics Communications},
	volume = {218},
	pages = {66-70},
	year = {2017},
	issn = {0010-4655},
	doi = {https://doi.org/10.1016/j.cpc.2017.04.015},
	url = {https://www.sciencedirect.com/science/article/pii/S0010465517301297},
	author = {Hiren H. Patel}
}

@article{Denner:2005nn,
	author = "Denner, Ansgar and Dittmaier, S.",
	title = "{Reduction schemes for one-loop tensor integrals}",
	eprint = "hep-ph/0509141",
	archivePrefix = "arXiv",
	reportNumber = "MPP-2005-84, PSI-PR-05-08",
	doi = "10.1016/j.nuclphysb.2005.11.007",
	journal = "Nucl. Phys. B",
	volume = "734",
	pages = "62--115",
	year = "2006"
}

@article{Denner:2016kdg,
	author = "Denner, Ansgar and Dittmaier, Stefan and Hofer, Lars",
	title = "{Collier: a fortran-based Complex One-Loop LIbrary in Extended Regularizations}",
	eprint = "1604.06792",
	archivePrefix = "arXiv",
	primaryClass = "hep-ph",
	reportNumber = "FR-PHENO-2016-003, ICCUB-16-016",
	doi = "10.1016/j.cpc.2016.10.013",
	journal = "Comput. Phys. Commun.",
	volume = "212",
	pages = "220--238",
	year = "2017"
}

@article{HAHN1999153,
	title = {Automated one-loop calculations in four and D dimensions},
	journal = {Computer Physics Communications},
	volume = {118},
	number = {2},
	pages = {153-165},
	year = {1999},
	issn = {0010-4655},
	doi = {https://doi.org/10.1016/S0010-4655(98)00173-8},
	url = {https://www.sciencedirect.com/science/article/pii/S0010465598001738},
	author = {T. Hahn and M. Pérez-Victoria}
}

\end{document}